\documentclass[11pt, spanish]{article}
\usepackage{cite}
\usepackage{amsmath,amsfonts,amssymb}
\usepackage[small,bf,hang]{caption}
\usepackage{slashed}
\usepackage{color}
\usepackage{multirow}
\usepackage{extarrows}
\usepackage{hyperref}

\usepackage{graphicx}
\usepackage{bbold}
\usepackage{stackrel}
\usepackage{mathrsfs}
\usepackage{amsmath}
\usepackage[makeroom]{cancel}
\usepackage{amssymb}

\usepackage{empheq}
\usepackage{tensor}
\usepackage{ulem}

\usepackage{geometry}
\def\hybrid{
        \topmargin -20pt
        \oddsidemargin 0pt
        \headheight 0pt \headsep 0pt
        \textwidth 6.25in 
        \textheight 9.5in 
        \marginparwidth .875in
        \parskip 5pt plus 1pt \jot = 1.5ex}

\hybrid

 \csname
@addtoreset\endcsname{equation}{section}

\usepackage{epsfig}
\usepackage{color}
\usepackage{graphics}
\usepackage{graphicx}
\usepackage{slashed}

\usepackage{amsthm}
\usepackage{amssymb}
\usepackage{amsmath}
\usepackage{amsfonts}

\numberwithin{equation}{section}

\newcommand{\be}{\begin{equation}}
\newcommand{\ee}{\end{equation}}
\newcommand{\ba}{\begin{eqnarray}}
\newcommand{\ea}{\end{eqnarray}}

\newcommand{\nn}{\nonumber}

\newcommand{\un}{\underline}
\newcommand{\ov}{\overline}

\begin{document}
\begin{titlepage}
\rightline{}
\begin{center}
\vskip 1.5cm
 {\Large \bf{ Hidden symmetries from extra dimensions}}
\vskip 1.7cm

{\large\bf {Marco Ciafardini$^1$, Diego Marqu\'es$^{1,2} $, Carmen A. N\'u\~nez$^1$} \\
\vskip 0,5cm
and Agustina Pereyra Grau$^{1,2}$}
\vskip 1cm
$^1$ {\it   Instituto de Astronom\'ia y F\'isica del Espacio}, (CONICET-UBA)\\
{\it Ciudad Universitaria, Pabell\'on IAFE, CABA, C1428ZAA, Argentina}\\

\vskip .3cm

$^2$ {\it Departamento de F\'isica, Universidad de Buenos Aires.}\\
{\it Ciudad Universitaria, Pabell\'on 1, CABA, C1428ZAA, Argentina}\\

\vskip .4cm

\{mciafardini, diegomarques, carmen, agustinapereyra\}@iafe.uba.ar\\

\vskip .4cm

\vskip .4cm

\end{center}
\bigskip\bigskip

\begin{center} 
\textbf{Abstract}

\end{center} 
\begin{quote}  

In Kaluza-Klein compactifications, some symmetries of the higher dimensional theory are preserved in lower dimensions, others are broken, and occasionally, there are symmetry enhancements. The symmetries that are enhanced by toroidal compactifications were recently shown to define a symmetry principle with constrained parameters that fixes the action {\it  before} dimensional reduction. Here we show the opposite: symmetries of the  higher dimensional theory that are broken in the reduction process, can actually be  realized  {\it  after} dimensional reduction as a global symmetry principle with constrained parameters that fixes couplings in the lower dimensional theory. We implement this principle in pure gravity, half-maximal supergravity and the circle reduction of 11 dimensional supergravity to Type IIA superstring theory. As a further application, we show that it can be used to constrain the quartic Ramond-Ramond couplings in Type IIA superstring theory from the four-point $\zeta(3)\, \alpha'{}^3\,  t_8 t_8 R^{(-) 4}$ interactions.

\end{quote} 
\vfill
\setcounter{footnote}{0}
\end{titlepage}

\tableofcontents

\section{Introduction}

Classical field theories are usually defined  by action principles with some interactions fixed by symmetries, like general coordinate transformations, gauge or global symmetries, supersymmetries,  etc. When the action is compactified, some of these symmetries are preserved, others are partially broken and, occasionally, there are symmetry enhancements mostly related to the isometries and structure of the compact space. 

Recently, it was shown in \cite{Beta1,Beta2} that the symmetry enhancement due to toroidal compactifications  of supergravities can be used  as a symmetry principle that fixes all the couplings in the higher dimensional theory. Strictly, the  symmetries that are enhanced in the lower dimensional theory are not symmetries of the parent theory, because the latter has no information on the compact space that connects both theories and produces the enhancement. However, a symmetry principle can still be defined in the parent theory through {\it constrained} parameters, with constraints that encode the information on the would be compact space. Specifically, $\beta$ symmetry \cite{Beta1} is the realization in the higher dimensional theory of the (non-geometric part of the) $O(d,d)$ enhancement of toroidal compactifications of supergravity. The parameter is a constant bivector $\beta^{\mu \nu}$ and the constraint  $\beta^{\mu \nu} \partial_\nu \dots = 0$ states that it is orthogonal to the coordinates of the  reduced theory. 

Here we consider the opposite situation: symmetries of the higher dimensional theories that are partially broken by toroidal compactifications,  can still be realized in the lower dimensional theory.  Explicitly, we show that the higher dimensional diffeomorphisms can  be realized in the compactified theory through  constrained parameters $\alpha_m{}^i \partial_i \dots = 0$, $m$ labeling internal directions. The idea is sketched below on general grounds, and in the forthcoming sections we apply it to pure gravity  and half-maximal supergravities  in generic dimensions, as well as to the circle reduction of $11$ dimensional supergravity to type IIA. 

The following notation is used throughout the paper. The coordinates of the uncompactified $D$ dimensional directions are labeled with Greek letters $\mu,\nu,\rho, \dots = 0,\dots,D-1$. They  split into $n$ external coordinates carrying Latin indices $i,j,k, \dots = 0,\dots,n-1$  and $d=D-n$  compact directions conveniently identified  in each section.

Kaluza-Klein (KK) reductions  of $D$ dimensional general coordinate transformations  on $T^d$ break down to three types of symmetries in $n = D-d$ lower dimensions  when only the massless modes are kept. Indeed, consider the Lie derivative of a tensorial density $V_\mu{}^\nu (x,y)$ of weight $\omega$   in $D$ dimensions with vector parameter  $\xi^\mu(x,y)$,  where $x$ are external $n$ dimensional coordinates and $y$ are internal $d$ dimensional coordinates, namely  
\be
\delta_\xi V_\mu{}^\nu = L^\omega_\xi V_\mu{}^\nu = \xi^\rho \partial_\rho V_\mu{}^\nu + \partial_\mu \xi^\rho V_\rho{}^\nu - \partial_\rho \xi^\nu V_\mu{}^\rho + \omega \partial_\rho \xi^\rho\, V_\mu{}^\nu \ . \label{LieDerivative}
\ee
The reduction ansatz
\be
V_\mu{}^\nu (x,y) = \left(\begin{matrix} v_i{}^j(x) & v_i{}^n(x) \\ v_m{}^j(x) & v_m{}^n(x)\end{matrix} \right) \ , \ \ \ \xi^\mu (x,y) = \left(\begin{matrix} \xi^i(x)  \\ \lambda^m(x) + y^p \, \alpha_p{}^m\end{matrix}\right)\, , \label{KKansatz}
\ee
with $m,n,p,q,\dots$ internal $d$ dimensional indices,
includes generators of lower dimensional diffeomorphisms  $\xi^{i}(x)$, gauge parameters $\lambda^m(x)$, and constant $GL(d)$ parameters $\alpha_m{}^n$. Their impact in the lower dimensional theory is obtained by replacing (\ref{KKansatz}) into (\ref{LieDerivative}), which leads to
\begin{eqnarray}
\delta v_i{}^j &=& L^\omega_\xi v_i{}^j + \partial_i \lambda^p v_p{}^j + \omega \alpha_p{}^p v_i{}^j \, ,\\
\delta v_i{}^n &=& L^\omega_\xi v_i{}^n + \partial_i \lambda^p v_p{}^n - \partial_k \lambda^n v_i{}^k - \alpha_p{}^n v_i{}^p + \omega \alpha_p{}^p v_i{}^n \, ,\\
\delta v_m{}^j &=& L^\omega_\xi v_m{}^j + \alpha_m{}^p v_p{}^j +\omega \alpha_p{}^p v_m{}^j \, ,\\
\delta v_m{}^n &=& L^\omega_\xi v_m{}^n - \partial_k \lambda^n v_m{}^k + \alpha_m{}^p v_p{}^n - \alpha_p{}^n v_m{}^p + \omega \alpha_p{}^p v_m{}^n \, .
\end{eqnarray}
Consistency of the ansatz is due to the elimination of the internal coordinates. When $\alpha_p{}^p = 0$, the global symmetry is broken to $SL(d)$. 

In this paper  we observe that there is an extra symmetry with a consistent reduction to lower dimensions, given by the following parameter
\be
\xi^\mu (x,y) = \left(\begin{matrix} y^p \, \alpha_p{}^i  \\ 0 \end{matrix}\right) \ , \label{NewParameter}
\ee
where $\alpha_m{}^i$ is constant and constrained by
\be
\alpha_m{}^i \partial_i \dots = 0 \ . \label{constraint}
\ee
Indeed, plugging \eqref{NewParameter}  into (\ref{LieDerivative}) leads to the following $\alpha$ transformations 
\begin{eqnarray}
\delta_\alpha v_i{}^j &=&  - \alpha_p{}^j v_i{}^p\, ,\\
\delta_\alpha v_i{}^n &=& 0\, ,\\
\delta_\alpha v_m{}^j &=& \alpha _m{}^k v_k{}^j - \alpha_p{}^j v_m{}^p\, ,\\
\delta_\alpha v_m{}^n &=& \alpha_m{}^k v_k{}^n\, .
\end{eqnarray}
Due to the constraint (\ref{constraint}), the dependence on the internal coordinates $y$ proposed in (\ref{NewParameter}) has disappeared.

Importantly, this transformation mixes all the components of $V_\mu{}^\nu$. Hence it will exchange their respective curvatures. Moreover, notice that any $D$ dimensional scalar, e.g. the Lagrangian $\delta L = \partial_\mu \left(\xi^\mu L\right)$, will compactify to a strict invariant due to the constraint (\ref{constraint}). As such, this is a good candidate to be a hidden symmetry principle with constrained parameters that fixes couplings in the lower dimensional theory. We  call it $\alpha$ symmetry.  

Note that a solution to (\ref{constraint}) would require truncating the external coordinate dependence in the directions of non-vanishing $\alpha_m{}^i$, or equivalently compactifying more dimensions. This would consequently enhance the global symmetry group $GL(d)$, with contributions from $\alpha$ symmetry and external diffeomorphisms. In this sense, $\alpha$ symmetry can be regarded as the realization  in the lower dimensional theory of the {\it would be} $GL(D)$ symmetry if all directions were compactified.

Let us emphasize that we are  partially truncating the internal coordinates, as we are looking for hidden symmetries in the lower dimensional theory. 
In this sense, our work stands  between two extreme cases:
\begin{itemize}
\item The case in which all the internal coordinate dependence is truncated. This is the standard massless KK reduction, where the higher symmetries are broken to the manifest symmetries of the lower dimensional theory, typically diffeomorphisms and gauge symmetries, that are not enough to fix the lower dimensional couplings. 

\item The case in which there is no truncation at all, but the action and the transformations of the higher dimensional theory are rewritten so that they {\it look like} those in the compactified theory. This is the spirit behind the Kaluza-Klein formulation of double field theory \cite{KKdft} and exceptional field theory \cite{EFT}, and the same procedure could be applied in standard (super)gravity \cite{Aulakh:1985un}. In this case the symmetry principle fixes the action completely.
\end{itemize}

What is missing, and we provide here, is a symmetry principle that fixes the couplings in the truncated theory. Even if in the second item the couplings can be fixed {\it prior} to the truncation, once it is performed there are not enough symmetries left to fix all the couplings. The key observation  introduced in this paper is that the truncation can be slightly relaxed as in (\ref{NewParameter}), so as to end with a new symmetry principle that fixes the lower dimensional action.

We implement this idea in the theories obtained through KK reduction of pure Einstein gravity in Section \ref{Einstein} and of half-maximal 10 dimensional supergravity  in Section \ref{Sugra}. The dimensional reduction of the effective action of M-theory to Type IIA superstring theory in 10 dimensions is considered in Sections \ref{4} and \ref{5}. In the former we explore the impact of $11$ dimensional general coordinate transformations  in the two derivative action of Type IIA supergravity,  and  in the latter we deal with the higher derivative corrections.  In particular, we discuss the restrictions imposed by $\alpha$ symmetry on the Ramond-Ramond (R-R) sector completeness of the four-point $\zeta(3)\, \alpha'{}^3\,  t_8 t_8 R^{(-) 4}$ Neveu Schwarz-Neveu Schwarz (NS-NS) interactions of Type IIA superstring theory.  Section \ref{6} contains a summary of the results and comments on possible further applications of our work.

\section{Pure Einstein gravity}\label{Einstein}

\subsection{Metric formulation}

In this section we consider the theory obtained through KK reduction of pure Einstein gravity in $D$ dimensions. The metric tensor $G_{\mu \nu}$ is parameterized as 
\be
G_{\mu \nu}(x,y) = \left(\begin{matrix} g_{i j} + A_i{}^p g_{pq} A_j{}^q & A_i{}^p g_{p n} \\ g_{m p} A_j{}^p & g_{m n} \end{matrix} \right) (x)\, .
\ee 
While the standard lower dimensional symmetries given by (\ref{KKansatz}) lead to 
\ba
\delta g_{ij} &=& L_\xi g_{i j} \, , \label{SymmetriesGravity} \\
\delta A_i{}^n &=& L_\xi A_i{}^n + \partial_i \lambda^n  - \alpha_p{}^n A_i{}^p\, ,  \\
\delta g_{m n} &=& L_\xi g_{m n} + 2 \alpha_{(m}{}^p g_{n)p} \, , \label{SymmetriesGravity1}
\ea
the Lie derivative with respect to the parameter (\ref{NewParameter}) leads to the $\alpha$ transformations of the lower dimensional fields
\ba
\delta_\alpha g_{ij} &=& - 2 \alpha_m{}^k g_{k(i} A_{j)}{}^m\, , \label{AlphaSymmetryGravity} \\
\delta_\alpha A_i{}^n &=& \alpha_p{}^k (g^{pn} g_{k i} - A_k{}^n A_i{}^p)\, , \\
\delta_\alpha g_{m n} &=& 2 \alpha_{(m}{}^k g_{n) p} A_k{}^p \, . \label{AlphaSymmetryGravity1}
\ea

The most general  Lagrangian that is invariant under the symmetries (\ref{SymmetriesGravity})-\eqref{SymmetriesGravity1} is
\be
L = R + a\, g_{m n} F_{i j}{}^m F^{ij\, n} + b\, \nabla_i g_{m n} \nabla^i g^{m n} + c\, g^{m n} \nabla_i g_{m n} g^{p q} \nabla^i g_{p q} + d\,  g^{m n} \nabla_i \nabla^i g_{m n}\, ,
\ee
where $R$ is the Ricci scalar and $F_{ij}{}^m = 2 \partial_{[i}A_{j]}{}^m$ are the gauge curvatures.  
The $\alpha$ transformations (\ref{AlphaSymmetryGravity})-\eqref{AlphaSymmetryGravity1}  lead to 
\begin{eqnarray}
    \delta_\alpha R &=& 2 \nabla^i \left(\alpha_m{}^j F_{i j}{}^m\right) + F_{i j}{}^m \nabla^i \alpha_m{}^j\, ,\\
    \delta_\alpha \left( g_{m n} F_{i j}{}^m F^{ij\, n}\right) &=& 4 \alpha_m{}^i g^{m n} F_{i j}{}^p \nabla^j g_{n p} + 4 F_{i j}{}^m \nabla^i \alpha_m{}^j\, ,\\
    \delta_\alpha \left( \nabla_i g_{m n} \nabla^i g^{m n}\right) &=& 4 \alpha_m{}^i g^{m n} F_{i j}{}^p \nabla^j g_{n p}\, , \\
    \delta_\alpha \left( g^{m n} \nabla_i g_{m n} g^{p q} \nabla^i g_{p q} \right) &=& - 4 \alpha_m{}^i F_{i j}{}^m g^{n p} \nabla^j g_{n p}\, ,\\
    \delta_\alpha \left(g^{m n} \nabla_i \nabla^i g_{m n}\right) &=& \alpha_m{}^iF_{i j}{}^m  g^{n p} \nabla^j g_{n p} - 4 \alpha_m{}^i g^{m n} F_{i j}{}^p \nabla^j g_{n p } + 2 \nabla^i\left(\alpha_m{}^j F_{i j}{}^m \right) \, , \ \ \ \ \ \ \ 
\end{eqnarray}
which imply the  following variation of the Lagrangian
\begin{eqnarray}
\delta_\alpha L &=&   (3 + 4 a + 2 d)  \nabla^i \alpha_m{}^j F_{i j}{}^m  + (2 + 2 d) \alpha_m{}^j \nabla^i F_{i j}{}^m  \nonumber\\
&& +(4 a + 4 b - 4 d) \alpha_m{}^i g^{m n} F_{i j}{}^p \nabla^j g_{n p}  + (-4 c + d)\alpha_m{}^iF_{i j}{}^m  g^{n p} \nabla^j g_{n p} \, . 
\end{eqnarray}
Hence, $\alpha$ symmetry determines
\be
a = -\frac 1 4 \ , \ \ \ b = - \frac 3 4 \ , \ \ \ c =  - \frac 1 4 \ , \ \ \ d = - 1 \ .
\ee

To construct an invariant action, note that
\be
\delta_\alpha g = - g\,  \text g^{-1} \delta_\alpha \text g = -2 g \alpha_m{}^i A_i{}^m \ ,
\ee
where we have defined the determinants
\be
g = \det (g_{ij}) \ , \ \ \ \text g = \det (g_{m n}) \ .
\ee
Then $\sqrt{|g|}$ is not an invariant density, and therefore it is not a proper measure for an action. Since $\delta (g \text g) = 0$, the $\alpha$ invariant action is 
\begin{eqnarray}
S &=& \int d^nx \sqrt{|g|}\sqrt{|\text g|} \left(R - \frac 1 4 \, g_{m n} F_{i j}{}^m F^{ij\, n} -\frac 3 4\, \nabla_i g_{m n} \nabla^i g^{m n} \right.\nonumber \\ && \ \ \ \ \ \ \ \ \ \ \ \ \ \ \ \ \ \ \ \ \ \ \ \ \ \ \ \left.-\frac 1 4\, g^{m n} \nabla_i g_{m n} g^{p q} \nabla^i g_{p q} -\,  g^{m n} \nabla_i \nabla^i g_{m n}\right) \, .
\end{eqnarray}
One can get rid of the non-standard $\sqrt{|\text g|}$ factor by re-scaling the metric as
\be
\widetilde g_{i j} = \text g^{\frac 1 n} g_{ i j} \ , \ \ \ \widetilde g = \det(\widetilde g_{i j}) = g \,\text g \ \ \ \Rightarrow \ \ \ \delta_\alpha \widetilde g = 0 \ .
\ee

We close this section recalling an interesting observation recently made in \cite{Gomez-Fayren:2024cpl}. There is another hidden symmetry in the KK reduction of pure gravity that does not descend from higher dimensional diffeomorphisms. It can however  be realized in higher dimensions assuming a non-trivial topology on the background. In the approach we have followed here instead,  the hidden symmetries originate from standard diffeomorphims in higher dimensions.

\subsection{Frame formulation}
Now we perform the KK reduction of pure Einstein gravity
in the frame formulation. The reason why frame formulations  are interesting in this context is that curvatures and connections are contracted with flat metrics, which are invariant under $\alpha$ transformations. Then, the role of $\alpha$ symmetry in fixing the couplings is easier to envision. 

The  $n$ and $d$ dimensional metrics can be written in terms of the external and internal frames $e_i{}^a$ and $\nu_m{}^{\bar a}$, respectively as 
\be
g_{i j} = e_i{}^a g_{a b} e_j{}^b \ , \ \ \ g_{m n} = \nu_m{}^{\bar a} g_{\bar a \bar b} \nu_n{}^{\bar b} \ ,
\ee
where $g_{ab}$ and $g_{\bar a \bar b}$ are the $\alpha$ invariant flat  metrics with indices $a,b$ and $\bar a, \bar b$ labeling tangent external and internal  directions. We refer to Appendix \ref{a} for more details on the conventions. 

The frames have the following $\alpha$ transformations
(up to Lorentz rotations $\delta_\Lambda e_i{}^a = e_i{}^b \Lambda_b{}^a$ and $\delta_\Lambda \nu_m{}^{\bar a} = \nu_m{}^{\bar b} \Lambda_{\bar b}{}^{\bar a}$)
\ba
\delta_\alpha  e_i{}^a=-\alpha_{\bar c}{}^aA_b{}^{\bar c}e_i{}^b, \qquad
\delta_\alpha \nu_m{}^{\bar a}=\alpha_{\bar c}{}^{ b}A_b{}^{\bar a}\nu_m{}^{\bar c}\, , \label{transfframe}
\ea
where $\alpha_{\bar c}{}^a=\nu^m{}_{\bar c} e_k{}^a\alpha_m{}^k$ and  $A_b{}^{\bar c}=e^k{}_b\nu_m{}^{\bar c}A_k{}^m$. Notice that due to the constraint (\ref{constraint}), flat derivatives $D_a = e_a{}^i \partial_i$ turn out to be $\alpha$ invariant  $[\delta_\alpha\, ,\, D_a] = 0$.

Defining the  flat version of the gauge curvature $F_{a b}{}^{\bar c} =  e^i{}_a e^j{}_b  F_{i j}{}^m \nu_m{}^{\bar c}$,  the external spin connection $w_{abc}$ (\ref{spinconnection}), and an internal scalar connection $\Omega_{a\bar b\bar c} = \nu_{m\bar b}D_a\nu^m{}_{\bar c}$ \eqref{om}, we readily find that $\alpha$ transformations mix them linearly
\ba
\delta_\alpha w_{a}{}^{bc}&=&\alpha_{\bar c}{}^{[b}F_{a}{}^{c]\bar c}+\frac12\alpha_{\bar c a}F^{bc\bar c}\, ,\label{aTransf1}\\
\delta_\alpha F_{ab}{}^{\bar c}&=&-2\left(\alpha^{\bar cc}w_{cab} +2\alpha_{\bar a[a}\Omega_{b]}{}^{(\bar a\bar c)}\right)\equiv-2\nabla_{[a}\alpha^{\bar c}{}_{b]}\, ,\\
\delta_\alpha\Omega_{a}{}^{\bar b}{}_{\bar c}&=&-\alpha_{\bar c}{}^bF_{ab}{}^{\bar b}\, .\label{aTransf2}
\ea

The most general diffeomorphism, gauge and Lorentz invariant $n$ dimensional Lagrangian is
\ba
L=R+a\ F_{ab}{}^{\bar c}F^{ab\bar d}g_{{\bar c\bar d}}+  b\ \Omega_{a (\bar b\bar c)}\Omega^{a\bar b\bar c}+c\ \Omega^{a\bar a}{}_{\bar a}\Omega_{a\bar c}{}^{\bar c}+d\ \nabla_a\Omega^{a\bar c}{}_{\bar c}\, .\label{mg}
\ea
Each term $\alpha$ transforms as
\ba
\delta_\alpha R&=&2 {\alpha}{_{\overline{c}}{}^b}\nabla^a {F}{_{ab}{}^{\overline{c}}}-3{\alpha}{_{\overline{c}}}{}^c{w}_c{}{^{ab}}{F}{_{ab}{}^{\overline{c}}}\, ,\label{tf}\\
\delta_\alpha (F_{ab}{}^{\bar c}F^{ab\bar d}g_{\bar c\bar d})&=&-4F_{ab}{}^{\bar c}\nabla^{a}\alpha_{\bar c}{}^{b}\, ,\\
 \delta_\alpha (\Omega^{a(\bar b\bar c)}\Omega_{a\bar b\bar c})&=&-2\alpha_{\bar c}{}^b\Omega^{a}{}^{(\bar b\bar c)}F_{ab\bar b}\, ,\\
\delta_\alpha(\Omega_{a\bar b}{}^{\bar b}\Omega^{a\bar c}{}_{\bar c})&=&-2\alpha_{\bar b}{}^bF_{ab}{}^{\bar b}\Omega^{a\bar c}{}_{\bar c}\, ,\\
\delta_\alpha(\nabla_a\Omega^{a\bar c}{}_{\bar c})&=&\ -\alpha_{\bar c}{}^b\nabla^aF_{ab}{}^{\bar c}+\alpha_{\bar c}{}^cw_{c}{}^{ab}F_{ab}{}^{\bar c}+\alpha_{\bar c}{}^aF_{ab}{}^{\bar c}\Omega^{b\bar a}{}_{\bar a}\, ,
\ea
 which imply the following variation of the Lagrangian \eqref{mg}
\ba
\delta_\alpha L&=&(2-d){\alpha}{_{\overline{a}}{}^b}\nabla^a {F}{_{ab}{}^{\overline{a}}}+(-4a+b){\alpha}{_{\overline{b}}{}^a}{\Omega}{^{b\overline{b}}{}_{\overline{a}}}{F}{_{ab}{}^{\overline{a}}}-(2c+d)\alpha_{\bar c}{}^{b}{}F_{ab}{}^{\bar c}\Omega^{a\bar b}{}_{\bar b}\nn\\
&&-(3+4a-d){\alpha}{_{\overline{a}}{}^{c}}{w}_c{}{^{ab}}{F}{_{ab}{}^{\overline{a}}}+(4a-b)\alpha_{\bar a}{}^b\Omega^{a}{}_{\bar b}{}^{\bar a}F_{ab}{}^{\bar b}\, .
\ea
Hence, invariance of the Lagrangian under \eqref{aTransf1}-\eqref{aTransf2} requires $a=-\frac14, b=-1, c=-1, d=2$, leading to the unique $\alpha$ invariant expression
\ba
L=R-\frac14F_{ab}{}^{\bar c}F^{ab\bar d}g_{{\bar c\bar d}}-\Omega_{a\bar b\bar c}\Omega^{a(\bar c\bar b)}-\Omega^{a\bar a}{}_{\bar a}\Omega_{a\bar c}{}^{\bar c}+2\nabla_a\Omega^{a\bar c}{}_{\bar c}\, .
\ea

\section{Half-maximal Supergravity} \label{Sugra}

\subsection{Metric formulation}
In this section we consider KK reductions of half-maximal $N = 1$ supergravity in $D = 10$ dimensions. In order to reproduce the structure of Section \ref{Einstein}, it is convenient to use the formulation of double field theory \cite{Siegel,DFT1}, where the two-form is geometerized and supergravity can be considered a generalized gravitational theory that is invariant under generalized diffeomorphisms. 

The generalized Ricci scalar is defined in terms of a (gauged supergravity adapted) generalized metric \cite{GenMet1,GenMet2}
\be
{\cal H}_{\widehat M \widehat N} = \left( \begin{matrix} g^{i j} & - g^{i k} c_{k j} & - g^{i k} A_{k N} \\ - g^{j k} c_{k i} & g_{i j} + c_{k i} c_{l j} g^{k l} + A_i{}^P A_j{}^Q {\cal M}_{P Q} & c_{k i} g^{k l} A_{l N} + A_i{}^P {\cal M}_{P N} \\ - g^{j k} A_{k M} & c_{k j} g^{k l} A_{l M} + A_j{}^P {\cal M}_{P M} & {\cal M}_{M N} + A_{k M} A_{l N} g^{k l} \end{matrix} \right) \, ,
\ee
where $c_{i j} = b_{i j}  + \frac 1 2 A_i{}^M A_{jM}$, $A_i{}^M$ are $2d$ Abelian gauge fields with $M=1,...,2d$, $b_{ij}$ is the external two-form, and ${\cal M}_{M N}$ is the scalar matrix. $O(D,D)$ indices $\widehat M, \widehat N$  and $O(d,d)$ indices $M, N$ are   raised and lowered with their corresponding $\eta$ invariant metrics
\be
\eta_{\widehat M \widehat N} = \left( \begin{matrix} 0 & \delta^i{}_j & 0 \\ \delta_i{}^j & 0 \\ 0 & 0& \eta_{M N}\end{matrix} \right)\, \qquad {\text {and}}\qquad \eta_{M N} = \left( \begin{matrix} 0 & \delta^m{}_n  \\ \delta_m{}^n & 0 \end{matrix} \right)\, .
\ee
The scalar matrix is a symmetric element of $O(d,d)$, namely ${\cal M}_{M P} \eta^{P Q} {\cal M}_{Q N} = \eta_{M N}$. In addition, there is a generalized dilaton, defined as $e^{-2d} = \sqrt{|g|} e^{-2\phi}$, where $\phi$ is the standard dilaton. 

The local symmetries are governed by infinitesimal generalized diffeomorphisms acting as follows\cite{DFT2}
\begin{eqnarray}
{\cal L}_\xi {\cal H}_{\widehat M\widehat N} &=& \xi^{\widehat P} \partial_{\widehat P} {\cal H}_{\widehat M\widehat N} + \left( \partial_{\widehat M} \xi^{\widehat P} - \partial^{\widehat P} \xi_{\widehat M}\right) {\cal H}_{\widehat P \widehat N}  + \left( \partial_{\widehat N} \xi^{\widehat P} - \partial^{\widehat P} \xi_{\widehat N}\right) {\cal H}_{\widehat M \widehat P} \, , \label{gendiffs} \\
{\cal L}_\xi e^{-2d} &=& \partial_{\widehat M} \left(\xi^{\widehat M} e^{-2d}\right)\ . \label{measure}
\end{eqnarray}
There exists an invariant Lagrangian $L({\cal H},d)$ whose details can be found in \cite{DFT2}. Here we just mention a crucial fact about it: there is a proper action principle because the combination $e^{-2d} L$ transforms as a total derivative
\be
\delta_\xi \left(e^{-2d} L\right) = \partial_{\widehat M} \left(\xi^{\widehat M} e^{-2d} L\right) \ , \label{InvariantL}
\ee
provided a ``strong constraint'' holds
\be
\partial_{\widehat M} \dots \partial^{\widehat M} \dots = 0 \ , \ \ \ \ \partial_{\widehat M} \partial^{\widehat M} \dots = 0 \ , \label{strongconstraint}
\ee
where  dots represent any field or gauge parameter.

Under a $GL(n)\times O(d,d)$ decomposition,  $O(D,D)$ coordinates split as $X^{\widehat M} = \left( \tilde x_i\, , \ x^i \, , \ y^M \right)$. While the fields only depend on the external coordinates $x$ under KK reduction, the vector parameterizing infinitesimal generalized diffeomorphisms additionally admits the following linear dependence on $y$ 
\be
\xi^{\widehat M} = \left(\begin{matrix} \lambda_i (x) \\ \xi^i (x) \\ \Lambda^M(x) + \frac 1 2 y^N \alpha_{N}{}^M\end{matrix}\right) \ .
\ee
This generates the standard diffeomorphisms, local transformations of the gauge fields and two-form (including the Green-Schwarz transformation), and infinitesimal global $O(d,d)$ transformations parameterized by $\alpha_{M N} = \alpha_{[M N]}$, specifically
\ba
\delta g_{ij} &=& L_\xi g_{i j}\, , \label{GaugeTransfSugra}\\
\delta A_i{}^M &=& L_\xi A_i{}^M + \partial_i \Lambda^M - A_i{}^N \alpha_N{}^M\, , \\
\delta b_{i j} &= & L_\xi b_{i j} + 2 \partial_{[i} \lambda_{j]} + A_{[i}{}^M \partial_{j]}\Lambda_M \, , \\
\delta \phi &=& L_\xi \phi \, ,\\
\delta {\cal M}_{M N} &=& L_\xi {\cal M}_{M N} + 2 \alpha_{(M}{}^P {\cal M}_{N) P}  \ . \label{GaugeTransfSugra1}
\ea
In turn, these transformations  determine the covariant curvatures of the gauge fields
\begin{eqnarray}
F_{i j}{}^M = 2 \partial_{[i} A_{j]}{}^M \ , \ \ \  \qquad H_{i j k} = 3 \partial_{[i} b_{jk]} - 3 A_{[i}{}^M \partial_j A_{k]M}  \ .
\end{eqnarray}

Now we propose the following additional diffeomorphism 
\be
\xi^{\widehat M} = \left(\begin{matrix} 0 \\ y^N \alpha_N{}^i \\ 0 \end{matrix}\right)\, , \label{paramalpha1}
\ee
which produces the $\alpha$ transformations
\ba
\delta_\alpha g_{ij} &=& - 2 \alpha_M{}^k g_{k(i} A_{j)}{}^M\, , \label{AlphaTransfSugra1} \\
\delta_\alpha A_i{}^M &=& \alpha_N{}^j \left(g_{i j} {\cal M}^{M N}- A_i{}^N A_j{}^M\right) + \alpha^{Mj} \left(b_{i j} + \frac 1 2 A_i{}^N A_{j N}\right)\, ,   \\
\delta_\alpha b_{i j} &=& \alpha_M{}^k \left({\cal M}^{M N} g_{k[i} A_{j]N} + \frac 1 2 A_{[i}{}^N A_{j]}{}^M A_{kN} + A_{[i}{}^M b_{j]k}\right)\, ,\\
\delta_\alpha \phi &=& - \frac 1 2 \alpha_M{}^i A_i{}^M \, ,\\
\delta_\alpha {\cal M}_{M N} &=& 2 \alpha_{(M}{}^k {\cal M}_{N)P} A_k{}^P - 2 \alpha_P{}^k A_{k(M} {\cal M}_{N)}{}^P \, ,\label{AlphaTransfSugra2}
\ea
and leads to $y^M$ independent transformations in the $n$ dimensional  theory, provided the following constraint holds 
\be
\alpha_M{}^i \partial_i \dots = 0 \ . \label{constalpha}
\ee
The parameter (\ref{paramalpha1}) together with the truncated fields trivially satisfy the strong constraint (\ref{strongconstraint}). 
Moreover, using (\ref{paramalpha1}) as the diffeomorphism parameter in (\ref{measure}) and (\ref{InvariantL}) implies that both the measure $e^{-2d}$ and the Lagrangian $L$ are strictly invariant under $\alpha$ transformations (\ref{AlphaTransfSugra1})-\eqref{AlphaTransfSugra2} as long as (\ref{constalpha}) holds.

Curiously,  the same transformations (\ref{AlphaTransfSugra1})-\eqref{AlphaTransfSugra2} can be  obtained from a different generalized diffeomorphism parameterized by
\be
\xi^{\widehat M} = \left(\begin{matrix} 0 \\ 0 \\ -\alpha_M{}^i \tilde x_i\end{matrix}\right)\, . \label{paramalpha2}
\ee
In this case, the $\alpha$ transformations obtained by inserting (\ref{paramalpha2}) into (\ref{gendiffs}) are automatically independent of $\widetilde x_i$ and $y^M$, and the restriction (\ref{constalpha}) arises instead imposing the strong constraint  $\partial^{\widehat P}\xi^{\widehat M}\partial_{\widehat P}f(x)=0$, for arbitrary $f(x)$. 

The most general Lagrangian that is invariant under the standard symmetries (\ref{GaugeTransfSugra})-(\ref{GaugeTransfSugra1}) is\footnote{Other possible kinetic terms for the scalar fields, like ${\cal M}^{M N} \nabla_i\nabla^i {\cal M}_{M N}$, ${\cal M}^{P Q} \nabla_i{\cal M}_P{}^N \nabla^i{\cal M}_{N Q}$, ${\cal M}^{P Q} {\cal M}^{M N} \nabla_i{\cal M}_{PM} \nabla^i{\cal M}_{QN}$, etc.  can be shown to be either zero, or linearly dependent of the one considered here.}
\begin{eqnarray}
L &=& R + a \nabla_i \phi \nabla^i \phi + b \nabla_i \nabla^i \phi + c F_{i j}{}^M F^{i j}{}_M + d F_{i j}{}^M {\cal M}_{M N} F^{i j N} \nonumber \\
&& + e H_{i j k} H^{i j k} + f \nabla_i {\cal M}_{M N} \nabla^i {\cal M}^{M N}  \ . \label{mgl}
\end{eqnarray}
Each of these terms transforms with respect to (\ref{AlphaTransfSugra1})-\eqref{AlphaTransfSugra2} as follows 
\ba
\delta_\alpha R &=& 2 \nabla^i \left(\alpha_M{}^j F_{i j}{}^M\right) + F_{i j}{}^M \nabla^i \alpha_M{}^j \, ,\label{AlphaTransfSugraCurva}\\
\delta_\alpha \left( \nabla_i \phi \nabla^i \phi \right) &=& \alpha_M{}^i \nabla^j \phi F_{i j}{}^M\, , \\
\delta_\alpha  \left( \nabla_i \nabla^i \phi  \right) &=& \alpha_M{}^i \nabla^j \phi F_{i j}{}^M - \frac 1 2 \nabla^i \left(\alpha_M{}^j F_{i j}{}^M\right) \, ,\\
\delta_\alpha \left(F_{i j}{}^M F^{i j}{}_M  \right)  &=& 2 \alpha_M{}^i F^{j k M} H_{i j k} + 4 F_{i j N} \nabla^i \left(\alpha_M{}^j {\cal M}^{M N}\right) \, ,\\
\delta_\alpha \left(F_{i j}{}^M {\cal M}_{M N} F^{i j N}  \right)  &=& 2 \alpha_M{}^i F^{j k}{}_N H_{i j k} {\cal M}^{M N} - 4 \alpha_M{}^i F_{i j}{}^N {\cal M}_{N P} \nabla^j {\cal M}^{P M} \, \nn\\
&& \ + 4 F_{i j}{}^M \nabla^i \alpha_M{}^j \, ,\\
\delta_\alpha \left( H_{i j k} H^{i j k}  \right)  &=& - 6 \alpha_M{}^i F^{j k}{}_N H_{i j k} {\cal M}^{M N}\, , \label{AlphaTransfSugraCurvh}\\
\delta_\alpha \left(\nabla_i {\cal M}_{M N} \nabla^i {\cal M}^{M N}  \right)  &=& - 8 \alpha_M{}^i F_{i j}{}^N {\cal M}_{N P} \nabla^j {\cal M}^{P M} \ .
\ea
It is then immediate to see that requiring $\alpha$ symmetry, namely $\delta_\alpha L = 0$, uniquely fixes the couplings of the  Lagrangian to the values
\be
a = -4 \ , \ \ \ b= 4 \ , \ \ \ c = 0 \ , \ \ \ d = - \frac 1 4 \ , \ \ \ e = - \frac 1 {12} \ , \ \ \ f = \frac 1 8 \  .
\ee
Regarding the measure, we find
\be
\delta_\alpha \sqrt{|g|} = - \alpha_M{}^i A_i{}^M \sqrt{|g|} \ , \ \ \ \delta_\alpha e^{-2\phi} = \alpha_M^i A_i{}^M e^{- 2\phi}\ , \ \ \  \delta_\alpha \left(\sqrt{|g|} e^{-2\phi}\right) = 0\, ,
\ee
so that the $\alpha$ invariant action is finally
\begin{eqnarray}
S &=& \int d^nx \, \sqrt{|g|} e^{-2\phi} \left(R -4 \nabla_i \phi \nabla^i \phi + 4 \nabla_i \nabla^i \phi - \frac 1 4  F_{i j}{}^M {\cal M}_{M N} F^{i j N} \right. \nonumber \\
&& \ \ \ \ \ \ \ \ \ \ \ \ \ \ \ \ \ \ \ \ \ \ \ \ \ \ \  \left. - \frac 1 {12} H_{i j k} H^{i j k} + \frac 1 8 \nabla_i {\cal M}_{M N} \nabla^i {\cal M}^{M N}  \right)\, .
\end{eqnarray}

\subsubsection{Gauged supergravity}
The extension of the formalism to the non-Abelian case can be found in \cite{Gauged1,Gauged2}. The generalized diffeomorphims (\ref{gendiffs}) are now deformed by the gaugings $f_{\widehat M \widehat N \widehat P}$ 
\be
{\cal L}_\xi^f {\cal H}_{\widehat M \widehat N} = f_{\widehat M \widehat P}{}^{\widehat Q} \xi^{\widehat P} {\cal H}_{\widehat Q \widehat N} + f_{\widehat N \widehat P}{}^{\widehat Q} \xi^{\widehat P} {\cal H}_{\widehat M \widehat Q} \ . \label{gaugedgendiffs}
\ee
This deformation breaks the global symmetry of the ungauged theory to a local subgroup, and requires extra consistency constraints, in particular
\be
f_{\widehat M \widehat N}{}^{\widehat P} \partial_{\widehat P} \dots = 0  \ ,\label{gaugedconstraint}
\ee
where again,  dots represent generalized fields or parameters.
Additionally, we are interested in the case in which the gaugings contain only constant internal $O(d,d)$ components $f_{M N P}$ satisfying linear and quadratic constraints
\be
f_{MNP}=f_{[MNP]}\, ,\qquad\qquad  f_{[MN}{}^{Q}f_{P]Q}{}^{ R}=0\, .
\ee

Introducing the  parameter (\ref{paramalpha1}) in (\ref{gaugedgendiffs})  produces no additional contributions to the $\alpha$ transformations (\ref{AlphaTransfSugra1})-\eqref{AlphaTransfSugra2}, which remain ungauged. However, insertion of (\ref{paramalpha1}) into (\ref{gaugedconstraint}) requires an extra constraint on $\alpha_M{}^i$\footnote{Again, the other choice of parameter (\ref{paramalpha2}) produces the opposite situation: (\ref{constalpha2}) arises by demanding that (\ref{gaugedgendiffs}) are $\widetilde x_i$ independent, while (\ref{gaugedconstraint}) is automatically satisfied.}
\be
f_{M N}{}^P \alpha_P{}^i = 0 \ . \label{constalpha2}
\ee
The local symmetries (\ref{GaugeTransfSugra}) are now given by
\ba
\delta g_{ij} &=& L_\xi g_{i j}\, , \\
\delta A_i{}^M &=& L_\xi A_i{}^M + \partial_i \Lambda^M - f_{P Q}{}^M \Lambda^P A_i{}^Q \, , \\
\delta b_{i j} &=& L_\xi b_{i j} + 2 \partial_{[i} \lambda_{j]} + A_{[i}{}^M \partial_{j]}\Lambda_M \, , \\
\delta \phi &=& L_\xi \phi \, ,\\
\delta {\cal M}_{M N} &=& L_\xi {\cal M}_{M N}  + 2 f_{P(M}{}^Q {\cal M}_{N) Q} \Lambda^P \ ,
\ea
and the gauge covariant curvatures take the form
\ba
        {F}_{ij}{}^{ M}&=&2\partial_{[i}{A}_{j]}{}^{M}+ f_{NP}{}^{ M} {A}_i{}^{ N}{A}_j{}^{ P} \, ,\label{fg}\\
       {H}_{ijk}&=&3 \partial_{[i}b_{jk]}-3 {A}_{[i}{}^{ M}\partial_j {A}_{k]M}- f_{MNP} {A}_i{}^{ M}{A}_j{}^{ N}{A}_k{}^{ P}\,  , \label{hg} \\
       \nabla_i \mathcal{M}_{MN}&= &\partial_i \mathcal{M}_{MN}+ f_{MP}{}^{Q} {A}_{i}{}^{ P} \mathcal{M}_{QN}+ f_{NP}{}^{Q} {A}_i{}^{ P} \mathcal{M}_{MQ}\, .
    \ea
The most general Lagrangian  compatible with these symmetries is 
\ba
           {L}&=    &R+ c_1 \,  \nabla_i \phi \nabla^i\phi+c_2  \,  \nabla_i \nabla^i \phi +c_3\, {F}_{ij}{}^{ M} \eta_{MN}{F}^{ijN}+c_4\, {F}_{ij}{}^{ M} \mathcal{M}_{MN}{F}^{ijN}+c_5 \, {H}_{ijk}H^{ijk}\nn\\
    &&+c_6 \, \nabla_i \mathcal{M}_{MN}\nabla^i \mathcal{M}^{MN}+c_7\,  f_{MP}^{\, \, \, \, \, \, \, \, \, R} f_{NQ}^{\, \, \, \, \, \, \, \, \, S} \mathcal{M}^{MN}\mathcal{M}^{PQ} \mathcal{M}_{RS}+ c_8 \,  f_{MPQ} f_{NR}^{\, \, \, \, \, \, \, \, \, Q} \mathcal{M}^{MN}\mathcal{M}^{PR}\nn\\
    &&+ c_9 \,  f_{MP}^{\, \, \, \, \, \, \, \, \, Q} f_{NQ}^{\, \, \, \, \, \, \, \, \, P} \mathcal{M}^{MN}+c_{10} \, f_{MNP} f^{MNP} \ .
    \label{lag gauging}
\ea
The transformations \eqref{AlphaTransfSugraCurva} -\eqref{AlphaTransfSugraCurvh} keep the same form, with the replacement of the curvatures by \eqref{fg} and \eqref{hg}. In addition, we get
\ba
        \delta_\alpha \left( \nabla_i \mathcal{M}_{MN}\nabla^i \mathcal{M}^{MN}\right)
     &=&-8 \alpha^{M i} {F}_{ij}{}^{N} \mathcal{M}_{N}{}^{ P}  \nabla^{j}\mathcal{M}_{MP}+ 4   \alpha_M{}^{i}  {A}_i^{\, \, \, N}   f_{N}^{ \, \, \, \, \,PQ}   f_{RPQ}   \mathcal{M}^{M R}\nn\\
    && - 4  \alpha_M{}^{i}  \mathcal{A}_i^{\, \, \, P}   f_{PRT}   f_{QSN}   \mathcal{M}^{M Q}
     \mathcal{M}^{RS}   \mathcal{M}^{TN}\, ,\\
    \delta_\alpha \left( f_{MN}{}^{P}f_{QR}{}^{S} \mathcal{M}^{MQ}\mathcal{M}^{NR} \mathcal{M}_{PS}\right)&=&-6  \alpha_{T}{}^{i} {A}_i^{\, \, \, M}  f_{MNP}f_{QRS}\mathcal{M}^{NR} \mathcal{M}^{PS} \mathcal{M}^{TQ}\, ,\\
     \delta_\alpha \left( f_{MPQ}f_{NR}{}^Q \mathcal{M}^{MN}\mathcal{M}^{PR} \right)&=&-4\alpha_S{}^kA_k{}^{M}f_{MPQ}f_{NR}{}^Q{\cal M}^{NS}\mathcal{M}^{PR}\, ,\\
    \delta_\alpha \left( f_{MPQ}f_{N}{}^{PQ} \mathcal{M}^{MN}\right)&=&-2   \alpha_R{}^{i} {A}_i^{\, \, \,M}  f_N^{\, \, \, \, \, PQ} f_{MPQ}\mathcal{M}^{NR} \, ,\\
    \delta_\alpha \left(f_{MNP} f^{MNP}\right)&=&0\, ,
    \ea
where we used
\begin{equation}
    \alpha_M{}^{i} \partial_i=0 \,\, \, \, \,\,\,\,\ {\rm and } \,\, \, \, \,\,\,\,\,  \alpha_M{}^{ i} f^{M}_{\, \,\, \, \, NP}=0\, .\label{cond}
\end{equation}
Then, $\alpha$ invariance of the Lagrangian requires
\begin{equation}
    \begin{split}
         c_1=-c_2=-4, \, \, \, \quad c_3=c_8=0\, , \, \, \, \quad c_4=c_9=-\frac{1}{4}, \, \, \, \quad c_5=c_7=-\frac{1}{12}, \, \, \, \quad c_6=\frac{1}{8}, 
    \end{split}
\end{equation}
leading to
\begin{equation}
    \begin{split}
            {L}=
    &R-4 \nabla_i \phi\nabla^i+4  \,  \Box \phi-\frac{1}{4}\, {F}_{ij}^{\, \, \, M} \mathcal{M}_{MN}{F}^{ijN}-\frac{1}{12} \, {H}^{ijk}H_{ijk}+\frac{1}{8} \, \nabla_i \mathcal{M}_{MN}\nabla^i \mathcal{M}^{MN}+\\
    &-\frac{1}{12} \,  f_{MP}^{\, \, \, \, \, \, \, \, \, R} f_{NQ}^{\, \, \, \, \, \, \, \, \, S} \mathcal{M}^{MN}\mathcal{M}^{PQ} \mathcal{M}_{RS}-\frac{1}{4} \, f_{MP}^{\, \, \, \, \, \, \, \, \, Q} f_{NQ}^{\, \, \, \, \, \, \, \, \, P} \mathcal{M}^{MN}+c_{10} \,  f_{MNP} f^{MNP}\, .
    \end{split}
\end{equation}
Note that $\alpha$ symmetry does not fix the coefficient $c_{10}$. When $c_{10}=0$ the theory is a truncation of  maximal supergravity  \cite{amnr,dgr}, while a non-vanishing $c_{10}$ implies that there is no possible uplift due to the presence of sources.

\subsection{Frame formulation }

We now consider the frame formulation of double field theory \cite{Siegel} (see also \cite{Frame,Exploring}). There are two frames, an external  $e_i{}^a$ and an internal  $\nu_M{}^A$, and we maintain all the conventions of the previous section for the external one. The difference with standard gravity is that the internal local symmetry consists of a pair Lorentz groups, with invariant metrics $\eta_{A B}$ and ${\cal H}_{A B}$
\ba
\eta_{AB}=\nu^M{}_A\ \eta_{MN}\ \nu^N{}_B\, ,\qquad\qquad {\cal H}_{AB}=\nu^M{}_A\ {\cal M}_{MN}\ \nu^N{}_B\, .
\ea
Flat indices are raised and lowered with $\eta_{AB}$.

It is convenient to define a complete set of projectors  
\ba
P_{AB}=\frac12(\eta_{AB}-{\cal H}_{AB})\, ,\qquad\qquad \ov P_{AB}=\frac12(\eta_{AB}+{\cal H}_{AB})\, .
\ea
Under gauge and Lorentz transformations, the external and internal frames transform as
\be
\delta_\Lambda e_i{}^a = e_i{}^b \Lambda_b{}^a \ , \ \ \ \delta_\Lambda \nu_M{}^A =f^A{}_{BC}\Lambda^B\nu_M{}^C+ \nu_M{}^B \Lambda_B{}^{A} \ ,
\ee
with a constrained double Lorentz parameter
\be
\delta_\Lambda \eta_{A B} = 2 \Lambda_{(A B)} = 0 \ , \ \ \ \delta_\Lambda {\cal H}_{A B} = 2 {\cal H}_{C(A} \Lambda^C{}_{B)} = 0 \ , \ \ \ \implies \ \  \  P_{A}{}^C \ov P_B{}^D \Lambda_{C D} = 0 \ . \label{projLambda}
\ee

We define the following curvatures and connections
\ba
F_{ab}{}^C&=&e^i{}_ae^j{}_b F_{i j}{}^M \nu_M{}^C\, ,\\
H_{abc}&=& e^i{}_ae^j{}_{b}e^k{}_c H_{ijk}\, ,\\
\Omega_{a B C}&=&\nu_{N B}D_a\nu^N{}_{ C}-f_{ B CD}A_a{}^D\, ,\\
f_{ABC}&=&f_{MNP}\, \nu^M{}_A\, \nu^N{}_B\, \nu^P{}_C\, ,
\ea
which are all gauge invariant and scalars under diffeomorphisms. These objects are Lorentz covariant, except for $\Omega_{aBC}$ (\ref{noncovariantOmega}). Hence we define the projection $P_A{}^C \ov P_{B}{}^D \Omega_{a C D}$, which is Lorentz covariant 
due to (\ref{projLambda}).

We take the $\alpha$ variations of the frames to be
\begin{subequations} \label{ta}
\begin{align}
\delta_\alpha e_i{}^a&= \ -\alpha_A{}^aA_b{}^Ae_i{}^b\, , \label{te}\\
\delta_\alpha\nu_M{}^A&=\ 2 A_a{}^{[A} \alpha^{B]a} \nu_{MB}\, ,\label{tanu}
\end{align}
\end{subequations}
such that  $[\delta_\alpha\, ,\, D_a] = 0$. The curvatures mix linearly among each other under $\alpha$ transformations 
\ba
\delta_\alpha w_{a}{}^{bc}&=&\alpha_{A}{}^{[b}F_{a}{}^{c]A}+\frac12\alpha_{A a}F^{bcA}\, ,\label{aTransf}\\
\delta_\alpha H_{abc}&=& -3\alpha^A{}_{[a}F_{bc]}{}^B{\cal H}_{AB}\, ,\\
\delta_\alpha F_{ab}{}^{C}&=&\alpha^{Cd}H_{abd}+2\alpha^A{}_{[b}\Omega_{a]}{}^{CB}{\cal H}_{AB}+2\nabla_{[a}\left(\alpha^A{}_{b]}{\cal H}_A{}^C\right)\ ,\\
\delta_\alpha\Omega_{aBC}&=& 2F_{ab[C}\alpha_{B]}{}^b-\alpha_{Aa}f_{BCD}{\cal H}^{AD}\, , \\
\delta_\alpha f_{ABC}&=&-3\alpha_{[A}{}^{a}f_{BC]D}A_{a}{}^{D}\, .
\ea

The most general $n$ dimensional Lagrangian that is invariant under the local symmetries is given by
\ba
L&=&R+a_1 D_a\phi D^a\phi+a_2\nabla_a\nabla^a \phi+a_3F_{ab}{}^CF^{ab}{}_C+a_4F_{ab}{}^CF^{abD}{\cal H}_{CD}+a_5H_{abc}H^{abc}\nn\\
&&+a_6\Omega_{aBC}\Omega^{a}{}_{DE}P^{BD}\ov P^{CE}+a_7f_{ABC}f_{DEF}{\cal H}^{AD}{\cal H}^{BE}{\cal H}^{CF}+a_8f_{ABC}f_{DE}{}^{C}{\cal H}^{AD}{\cal H}^{BE}\nn\\
&&+a_9f_{ABC}f_{D}{}^{BC}{\cal H}^{AD}+a_{10}f_{ABC}f^{ABC}\, .
\ea

The variation of each term produces
\ba
\delta_\alpha R&=& 2\nabla_a\left(\alpha_A{}^cF^a{}_c{}^A\right)+F^a{}_b{}^A\nabla_a\alpha_A{}^b+F_{abC}\alpha_B{}^b\Omega^{aBC}\, ,\\
\delta_\alpha(\nabla_a\phi\nabla^a\phi)&=&\alpha_A{}^aF_{ab}{}^A\nabla^b\phi \ ,\\
\delta_\alpha(\nabla^2\phi)&=&\alpha_A{}^a\nabla^b\phi F_{ab}{}^A-\frac12\nabla^a(\alpha_A{}^bF_{ab}{}^A) \ ,\\
\delta_\alpha(F_{ab}{}^CF^{ab}{}_C)&=& 4\nabla_{a}(\alpha^A{}_{b}{\cal H}_{A}{}^{C})F^{ab}{}_{C}+2\alpha^C{}_d{H}_{ab}{}^dF^{ab}{}_{C}\nn\\
&&  +\ 2\alpha_A{}^bF_{abC}\Omega^{aC}{}_{B}{\cal H}^{AB}\, ,\\
\delta_\alpha(F_{ab}{}^CF^{abD}{\cal H}_{CD})&=& 4\nabla_{a}(\alpha^A{}_{b}{\cal H}_{A}{}^{C})F^{abD}{\cal H}_{CD}+2F^{abD}{\cal H}_{CD}\alpha^C{}_d{H}_{ab}{}^d\nn\\
&&  +\ 4\alpha_A{}^bF_{ab}{}^{C}\Omega^{aD}{}_{B}{\cal H}^{AB}{\cal H}_{CD}\, ,\\
\delta_\alpha(H_{abc}H^{abc})&=& -6\alpha^A{}_aF_{bc}{}^B{\cal H}_{AB}H^{abc}\, ,\\
\delta_\alpha(\Omega_{aBC}\Omega^{a}{}_{DE}P^{BD}\ov P^{CE})&= & F_{abC}\alpha_B{}^b\Omega^{aB C}+\frac12\alpha_A{}^af_{BCD}f^{ B C}{}_EA_a{}^E{\cal H}^{AD}\\
&&-\left(F_{abC}\alpha_B{}^b\Omega^{a}{}_{DE}+\frac12\alpha_A{}^af_{BCF}f_{ DEG}A_a{}^G{\cal H}^{AF}\right){\cal H}^{BD}{\cal H}^{CE}\ , \nn\\
\delta_\alpha(f_{ABC}f_{DEF}{\cal H}^{AD}{\cal H}^{BE}{\cal H}^{CF})&=& -6\alpha_A{}^{a}f_{BCG}f_{DEF}{\cal H}^{AD}{\cal H}^{BE}{\cal H}^{CF}A_a{}^{G}\, ,\\
\delta_\alpha(f_{ABC}f_{DE}{}^{C}{\cal H}^{AD}{\cal H}^{BE})&=& -4\alpha_B{}^af_{CAD}f^A{}_{EF}{\cal H}^{BE}{\cal H}^{CF}A_a{}^{D}\, ,\\
\delta_\alpha(f_{ABC}f_{D}{}^{BC}{\cal H}^{AD})&=& -2\alpha_A{}^af_{BCD}f^{BC}{}_EA_a{}^{E}{\cal H}^{AD} \ ,\\
\delta_\alpha(f_{ABC}f^{ABC})&=&0\, ,
\ea
and leads to the following variation of the Lagrangian 
\ba
\delta_\alpha L&=& (2-\frac12a_2)\nabla_a\left(\alpha_A{}^cF^a{}_c{}^A\right)+(1+4a_4)F^a{}_b{}^A\nabla_a\alpha_A{}^b\nn\\
&&+(1-a_6)F_{abC}\alpha_B{}^b\Omega^{aBC}+(a_1+a_2)\alpha_A{}^aF_{ab}{}^A\nabla^b\phi\nn\\
&&+a_3\left( 4\nabla_{a}(\alpha^A{}_{b}{\cal H}_{A}{}^{C})F^{ab}{}_{C}+2\alpha^C{}_d{H}_{ab}{}^dF^{ab}{}_{C}+2\alpha_A{}^bF_{abC}\Omega^{aC}{}_{B}{\cal H}^{AB}\right)\nn\\
&&-(6a_5-2a_4) \alpha^A{}_aF_{bc}{}^B{\cal H}_{AB}H^{abc}-(a_6+4a_4)F_{abC}\alpha_B{}^b\Omega^{a}{}_{ DB}{\cal H}^{AD}{\cal H}^{BC}\nn\\
&&-(6a_7 +\frac12a_6)\alpha_A{}^{a}f_{BCG}f_{DEF}{\cal H}^{AD}{\cal H}^{BE}{\cal H}^{CF}A_a{}^{G}\nn\\
&&-4a_8\alpha_B{}^af_{CAD}f^A{}_{EF}{\cal H}^{BE}{\cal H}^{CF}A_a{}^{D}-(2a_9-\frac12a_6)\alpha_C{}^af_{ABD}f^{AB}{}_E{\cal H}^{CE}A_a{}^{D}\, . \ \  \
\ea
Hence, $\alpha$ invariance requires \be a_1=-a_2=-4, \quad a_3=a_8=0,\quad  a_4=-a_9=-\frac14,\quad   a_5=a_7=-\frac1{12},\quad  a_6=1,\ee and fixes the action uniquely up to a single parameter $a_{10}$
\ba
L&=&R-4D_a\phi D^a\phi+4\nabla_a\nabla^a \phi-\frac14F_{ab}{}^CF^{abD}{\cal H}_{CD}-\frac1{12}H_{abc}H^{abc}+\Omega_{a  AB  }\Omega^{a\un A\ov B} \nn\\
&&-\frac1{12}f_{ABC}f_{DEF}{\cal H}^{AD}{\cal H}^{BE}{\cal H}^{CF}+\frac14f_{ABC}f_{D}{}^{BC}{\cal H}^{AD}+a_{10}f_{ABC}f^{ABC}\, .\qquad
\ea

\subsection{$\beta$ symmetry of gauged supergravities}

Interestingly, the procedure outlined above allows to compute the $\beta$ symmetry \cite{Beta1,Beta2} of gauged supergravities through the following choice of parameter
\be
\xi^{\widehat M} = \left(\begin{matrix} 0 \\ - \frac 1 2  \beta^{i j} \widetilde x_j \\0 \end{matrix}\right) \ .
\ee
Plugging this specific generalized diffeomorphism into (\ref{gendiffs}), gauged by (\ref{gaugedgendiffs}), leads to the following $\beta$ transformation rules
\begin{subequations} \label{AlphaTransfSugra}
\begin{align}
\delta_\beta g_{ij} &= -2 \beta^{k}{}_{(i}  c_{j)k}  \, ,  \\
\delta_\beta A_i{}^M &= - g_{i j} \beta^{j k} A_{kP} {\cal M}^{P M} - c_{i j} \beta^{j k} A_{k}{}^{M}\, ,   \\
\delta_\beta b_{i j} &=  - \beta_{i j} - c_{k[i} \beta^{k l} c_{j]l} - \beta_{[i}{}^k A_{j]}{}^M {\cal M}_{M N} A_k{}^N\, ,\\
\delta_\beta \phi &=  \frac 1 2 \beta^{i j} b_{i j}\, ,\\
\delta_\beta {\cal M}_{M N} &= 2 \beta^{i j} A_{i(M}  {\cal M}_{N) P} A_j{}^P\, ,
\end{align}
\end{subequations}
where we remind that $c_{i j} = b_{i j} + \frac 1 2 A_i{}^M A_{jM}$. These reproduce the $\beta$ transformations found in \cite{Beta1} for the specific case $A_i{}^M = 0$ and ${\cal M}_{M N} = \eta_{M N}$, which are given by
\be
\delta_\beta E_{i j} = - E_{i k} \beta^{k l} E_{l j} \ ,  \ \ \ \delta_\beta \phi = \frac 1 2 \beta^{i j} E_{i j} \ ,
\ee
with $E_{i j} = g_{i j} + b_{i j}$. Interestingly, taking $n = 10$, setting the internal indices in the adjoint of the heterotic gauge group, and truncating ${\cal M}_{M N}$ to coincide with the killing metric of the heterotic group, these reproduce the $\beta$ transformations of heterotic supergravity \cite{bnr}.

\section{The symmetries of type IIA from 11d supergravity}\label{4}

\subsection{Metric formulation}

Maximal $11$ dimensional supergravity  on a circle reduces to Type IIA supergravity in $10$ dimensions. Following the approach in this paper, we expect that the  invariance under $11$ dimensional general coordinate transformations descends to  $\alpha$ symmetry in Type IIA. We explore this issue in this section, starting with the action of $11$ dimensional supergravity
\begin{equation}
S_{11}=\int d^{11}x \sqrt{-G} \left[ {\cal R} -\frac{1}{2 \cdot 4!}{\cal F}_{\mu \nu\rho\sigma}{\cal F}^{\mu \nu\rho\sigma} + \left(\frac{2}{4!}\right)^4 \epsilon^{\mu_1\dots \mu_{11}} {\cal A}_{\mu_1 \mu_2\mu_3} {\cal F}_{\mu_4 \dots \mu_7} {\cal F}_{\mu_8 \dots \mu_{11}} \right] \ . \nn
\end{equation}
The fields are a metric $G_{\mu \nu}$  and a three-form  ${\cal A}_{\mu \nu \rho}$ with the following gauge transformation and curvature
\begin{equation}
    \delta_\Lambda {\cal A}_{\mu \nu\rho} = 3 \partial_{[\mu}\Lambda_{\nu \rho]}  
        \ , \ \ \ \ \ \mathcal{F}_{\mu\nu\rho\sigma}=4 \partial_{[\mu}{\cal A}_{\nu\rho\sigma]} \ .
\end{equation}
Diffeomorphisms act as usual with the Lie derivative
\be
\delta_\xi G_{\mu \nu} = \xi^\rho \partial_\rho G_{\mu \nu} + 2 \partial_{(\mu} \xi^\rho G_{\nu)\rho} \ , \ \ \ \delta_\xi {\cal A}_{\mu \nu \rho} = \xi^\sigma \partial_\sigma{\cal A}_{\mu \nu\rho} + 3 \partial_{[\mu}\xi^\sigma {\cal A}_{\nu \rho] \sigma} \ . \label{diffs11}
\ee

We consider the following KK ansatz for the fields on the circle with coordinate $y$,
\begin{eqnarray}
&& G_{\mu \nu}(x, y)=\begin{pmatrix}
e^{-2\Phi/3}g_{ij}+A_i A_j e^{4\Phi/3} &  A_i e^{4\Phi/3} \\
A_j e^{4\Phi/3} & e^{4\Phi/3}
\end{pmatrix}(x)\ ,\label{ansatz11} \\ && \mathcal{A}_{ijk}(x, y)\equiv A_{ijk}(x), \, \ \ \  \mathcal{A}_{ij10}(x,y)\equiv b_{ij}(x) \ ,\nonumber 
\end{eqnarray}
in terms of the Type IIA NS-NS fields: the metric $g_{i j}$, dilaton $\Phi$ and two-form $b_{i j}$, and the R-R fields: the one-form $A_i$ and three-form $A_{i j k}$. We also propose an ansatz for the local parameters
\be
\xi^{\mu} (x,y) = \begin{pmatrix} \xi^i(x) + \alpha^i\, y\\  \lambda(x) + \rho \, y\end{pmatrix} \ , \ \ \ \Lambda_{ijk}(x,y) = \lambda_{i j k}(x) \ , \ \ \ \Lambda_{ij 10}(x,y) = \lambda_{i j}(x) \ .\label{alpharho}
\ee
These descend in Type IIA to diffeomorphisms parameterized by $\xi^i(x)$, gauge transformations 
\be
 \delta_\lambda A_{i} = \partial_i \lambda \ , \ \ \ \delta_\lambda b_{i j} = 2 \partial_{[i}\lambda_{j]} \ , \ \ \ \delta_\lambda A_{i j k} = 3 (\partial_{[i} \lambda_{j k]} + \partial_{[i}\lambda \, b_{jk]}) \, ,
\ee
plus additional global symmetries parameterized by the constants $\alpha^i$ and $\rho$, to be discussed soon. The gauge symmetries have their associated invariant curvatures
\begin{equation}
    H_{ijk}\equiv3\partial_{[i}b_{jk]} \ , \ \ \ F_{ij}\equiv 2\partial_{[i}A_{j]}\ , \ \ \  \tilde{F}_{ijkl}\equiv 4 \left(\partial_{[i}A_{jkl]}+A_{[i}H_{jkl]}\right) \ .
\end{equation}

We can now write the most general two-derivative diffeomorphism and gauge invariant action in $10$ dimensions
\begin{equation}
\begin{split}
        &S_{10}=
         \int 
        d^{10}x \sqrt{-g} \, f(\Phi)\left[f_0(\Phi) R +f_1(\Phi)\left(\partial\Phi\right)^2+f_2(\Phi) \Box \Phi
        \right.\\
        &\ \ \ \ \ \ \ \ \ \ \ \ \ \  \left.
        +f_3(\Phi) H_{i j k} H^{i j k} +f_4(\Phi)F_{i j}F^{i j}+\, f_5(\Phi) \tilde{F}_{i j k l}\tilde{F}^{i j k l} \right.\\
        &\ \ \ \ \ \ \ \ \ \ \ \ \ \    \left. + \gamma_6 f^{-1}(\Phi) \epsilon^{i_1\dots i_{10}} b_{i_1 i_2} F_{i_3 \dots i_6} F_{i_7 \dots i_{10}}+ \gamma_7 f^{-1}(\Phi) \epsilon^{i_1\dots i_{10}} A_{i_1 i_2 i_3} H_{i_4 i_5  i_6} F_{i_7 \dots i_{10}} \right]. 
\end{split}\label{ActionTypeIICoefs}
\end{equation}
Note that gauge invariance requires the Chern-Simons interactions to be defined in terms of the closed form
\be
F_{i j k l} = 4 \partial_{[i} A_{j k l]} \ , \ \ \ \delta_\lambda F_{i j k l} = - 4 \partial_{[i} \left(\lambda \, H_{j k l]}\right) \ ,
\ee
not to be confused with $\tilde F_{i j k l}$.

We now intend to restrict $f_i(\Phi)$, $\gamma_6$ and $\gamma_7$ through $\rho$ and $\alpha$ symmetries. For this we plug (\ref{alpharho}) and (\ref{ansatz11}) into (\ref{diffs11}), and see that $y$ independence is obtained by the expected constraint on $\alpha$
\be
\alpha^i \partial_i \dots = 0 \ ,
\ee
which we impose from now on. The effect on the Type IIA fields is the following for $\rho$ transformations
\begin{eqnarray}
&&\delta_\rho \Phi=\frac{3}{2} \rho \ , \ \ \ \delta_\rho g_{ij} = \rho \, g_{ij} \ , \ \ \ \delta_\rho b_{ij}=\rho\, b_{ij} \ ,\label{rho1} \\
 &&  \delta_\rho A_i=-\rho\, A_{i} \ , \ \ \ \delta_\rho A_{ijk}=0\ ,  \ \ \ \delta_\rho \epsilon^{i_1\dots i_{10}} = - 5 \rho \epsilon^{i_1\dots i_{10}} \ ,
\end{eqnarray}
and  for $\alpha$ transformations
\begin{eqnarray}
   && \delta_\alpha \Phi= \frac{3}{2} \alpha^i A_i \ , \ \ \ \delta_\alpha b_{ij}=\alpha^k A_{ijk} \ , \ \ \ \delta_\alpha g_{ij}=\alpha^k \left( A_k g_{ij}- 2 A_{(i} g_{j)k} \right) \ , \\
    &&\delta_\alpha A_i=\alpha^j \left( e^{-2 \Phi} g_{ij}-A_i A_j\right) \ , \ \ \  \delta_\alpha A_{ijk}=0 \  , \ \ \ \delta_\alpha \epsilon^{i_1\dots i_{10}} = - 4 \alpha^k A_k \epsilon^{i_1\dots i_{10}} \ . \label{al2}
\end{eqnarray}

These in turn imply the following $\rho$ and $\alpha$ transformations 
\begin{eqnarray}
    \delta_{\alpha, \rho}  F_{ij} &=&\alpha^{k} \left(2{A}_{[i}{F}_{j]k} -{A}_k {F}_{ij}\right)+2g_{k[j}\nabla_{i]} \left( \alpha^{k}e^{-2 \Phi}\right) - \rho \, F_{i j}\ ,\\
    \delta_{\alpha, \rho}  H_{ijk}&=& \alpha^l F_{ijkl} + \rho \, H_{i j k} \ ,\\
    \delta_{\alpha, \rho}  \tilde{F}_{ijkl} &=& 4 \alpha^m\left( e^{-2\Phi}g_{m[i}H_{jkl]}+A_{[i}\tilde{F}_{jkl]m}\right) \ , \\
    \delta_{\alpha, \rho} F_{i j k l} &=& 0 \ , \\
    \delta_{\alpha, \rho}  \sqrt{-g} &=& \left( 4 \alpha^i A_i+5 \rho \right)\, \sqrt{- g} \ ,  \label{transfdetg}\\
    \delta_{\alpha, \rho} e^{\gamma \Phi} &=& \frac 3 2 \gamma \left(\alpha^i A_i+\rho \right) e^{\gamma \Phi} \label{transfgammaphi} \ . 
\end{eqnarray}
From (\ref{transfdetg}) and (\ref{transfgammaphi}) we see that it is not possible to define simultaneously a $\rho$ and $\alpha$ invariant measure. We choose to maintain $\alpha$ invariance simultaneausly for the measure $\sqrt{-g} f(\Phi)$ and the rest of the Lagrangian. This implies 
\be
f(\Phi) = e^{- \frac 8 3 \Phi} \ , \label{fs1}
\ee
which is consistent with the fact that $\sqrt{-G} = \sqrt{-g} e^{- \frac 8 3 \Phi}$. 

Demanding that each term inside the brackets in (\ref{ActionTypeIICoefs}) be $\rho$ invariant fixes
\begin{eqnarray}
&&f_0(\Phi) = \gamma_0 e^{\frac 2 3 \Phi} \ , \ \ \ f_1(\Phi) = \gamma_1 e^{\frac 2 3 \Phi} \ , \ \ \ f_2(\Phi) = \gamma_2 e^{\frac 2 3 \Phi} \ , \ \ \ f_3(\Phi) = \gamma_3 e^{\frac 2 3 \Phi}\ , \nonumber \\ && f_4(\Phi) = \gamma_4 e^{\frac 8 3 \Phi} \ , \ \ \ f_5(\Phi) = \gamma_5 e^{\frac 8 3 \Phi} \ . \label{fs2}
\end{eqnarray}

Plugging (\ref{fs1}) and (\ref{fs2}) in (\ref{ActionTypeIICoefs}) gives 
\begin{eqnarray}
&& S_{10}=
         \int 
        d^{10}x \sqrt{-g} e^{-\frac 8 3 \Phi}\left\{ \, e^{\frac 2 3 \Phi}\left[\gamma_0  R + \gamma_1 \left(\partial\Phi\right)^2+\gamma_2 \Box \Phi
        +\gamma_3 H_{i j k} H^{i j k} \right]\right.\\
        && \quad \left.
          + e^{\frac 8 3 \Phi} \left[\gamma_4 F_{i j}F^{i j}+\, \gamma_5 \tilde{F}_{i j k l}\tilde{F}^{i j k l}  + \epsilon^{i_1\dots i_{10}} \left(  \gamma_6\,  b_{i_1 i_2} F_{i_3 \dots i_6} F_{i_7 \dots i_{10}}+ \gamma_7  A_{i_1 i_2 i_3} H_{i_4 i_5  i_6} F_{i_7 \dots i_{10}} \right) \right]\right\} \nonumber
\end{eqnarray}
The coefficients $\gamma_a$ with $a= 0,\dots, 7$ are further restricted by $\alpha$ symmetry. We split the $\alpha$ transformation of each term in the Lagrangian as follows:
\begin{itemize}
\item Terms descending from the $11$ dimensional Ricci scalar
\begin{eqnarray}        
\delta_\alpha \left(e^{2\Phi/3}R\right)
        &=& e^{2\Phi/3} \left[-7 \nabla^i \left( \alpha^j F_{ij}\right)+F_{ij}\nabla^i \alpha^{j}\right] \, ,\\   \delta_\alpha \left(e^{2\Phi/3} (\partial \Phi)^2 \right)&=&-3 e^{2\Phi/3} \alpha^i F_{ij} \nabla^j \Phi\, ,
        \\
        \delta_\alpha \left( e^{2\Phi/3}\Box \Phi\right)&=& e^{2\Phi/3} \left[-3 \alpha^i F_{ij} \nabla^j \Phi +\frac{3}{2} \nabla^i\left(\alpha^j F_{i j}\right)\right]\, ,\\ 
\delta_\alpha \left( e^{8\Phi/3} F_{ij} F^{ij}\right)
    &=&e^{2\Phi/3} \left(8 \alpha^i F_{ij} \nabla^j \Phi +4 F_{i j} \nabla^i \alpha^j     \right)\, ,
\end{eqnarray}
fix the following coefficients 
\begin{equation} \gamma_1 = - \frac {16} 3 \gamma_0 \ , \ \ \gamma_2 = \frac {14} 3 \gamma_0 \ , \ \ \ \gamma_4 = -\frac 1 4 \gamma_0 \ .\end{equation}

\item Terms descending from the $11$ dimensional ${\cal F}_4^2$ 
\begin{eqnarray}
    \delta_\alpha \left( e^{2\Phi/3}H_{i j k} H^{i j k}\right)&=&-2e^{2\Phi/3}\alpha^i \tilde{F}_{ijkl}H^{jkl}\, ,
\\    \delta_\alpha \left(e^{8\Phi/3}\tilde F_{i j k l} \tilde F^{i j k l}\right)&=& 8e^{2\Phi/3}  \alpha^i\tilde F_{ijkl}H^{jkl}\, ,
\end{eqnarray}
fix
\begin{eqnarray} \gamma_5 = \frac 1 4 \gamma_3 \ .\end{eqnarray}

 \item Terms descending from the $11$ dimensional Chern-Simons term
\begin{eqnarray}
     \delta_\alpha \left(\sqrt{-g}\epsilon^{i_1 i_2...i_{10}} b_{i_1 i_2} F_{i_3 \dots i_6} F_{i_7 \dots i_{10}} \right)&=& \alpha^k\sqrt{-g}  \epsilon^{i_1 \dots i_{10}}  A_{i_1 i_2 k} F_{i_3...i_6} F_{i_7...i_{10}} \, ,\\
      \delta_\alpha \left( \sqrt{-g}\epsilon^{i_1 i_2...i_{10}} A_{i_1 i_2 i_3} H_{i_4 i_5 i_6} F_{i_7 \dots i_{10}} \right)&=& \alpha^k \sqrt{-g}  \epsilon^{i_1 \dots i_{10}}A_{i_1 i_2 i_3} F_{i_4 i_5 i_6 k} F_{i_7 \dots i_{10}} \, ,\ 
\end{eqnarray}
fix
\be
\gamma_7 = \frac 8 3 \gamma_6 \ ,
\ee
where we used  the identity
\be 0 = 11 A_{[i_1 i_2 i_3} F_{i_4 \dots i_7} F_{i_8 i_9 i_{10} k]} = 8 A_{[i_1 i_2 i_3} F_{i_4 \dots i_7} F_{i_8 i_9 i_{10}] k } + 3 A_{k[i_1 i_2} F_{i_3 \dots i_6} F_{i_7 \dots i_{10}]}\, .\ee
\end{itemize}
Then we  end with an action containing three $\alpha$ invariant terms
\begin{eqnarray}
S_{10} &=& \int d^{10} x \sqrt{-g} e^{-2\Phi} \left[\gamma_0 \left(R - \frac {16} 3 (\partial \Phi)^2 + \frac {14} 3 \Box \Phi - \frac 1 4 e^{2\Phi} F_{i j} F^{i j} \right) \right.  \label{S10aux} \\ && \ \ \ \ \ \ \ \ \ \ \ \ \ \ \ \ \ \ \ \ \ \ \  + \,\gamma_3 \left(H_{i j k} H^{i j k} + \frac 1 4 e^{2\Phi} {\tilde F}_{i j k l}{\tilde F}^{i j k l}\right)  \nonumber
\\ && \ \ \ \ \ \ \ \ \ \ \ \ \ \ \ \ \  \ \ \ \ \ \ \left. + \, \gamma_ 6\, e^{2 \Phi} \epsilon^{i_1 \dots i_{10}} \left(    b_{i_1 i_2} F_{i_3 \dots i_6} F_{i_7 \dots i_{10}}+ \frac 8 3 A_{i_1 i_2 i_3} H_{i_4 i_5  i_6} F_{i_7 \dots i_{10}} \right) \right] \ . \nonumber
\end{eqnarray} Without loss of generality we take $\gamma_0 = 1$. This is the most general action in $10$ dimensions that is invariant under diffeomorphisms, gauge transformations of the $p$-forms and $\alpha$ transformations, and that re-scales homogeneously under $\rho$ transformations. It is defined up to two parameters: $\gamma_3$ and $\gamma_6$. 

It is known that the universal NS-NS sector of string theory is invariant under the so-called $\beta$ symmetry \cite{Beta1, Beta2}. This symmetry was studied in the democratic formulation of the R-R sector in \cite{wn}. Ignoring here the R-R fields, $\beta$ symmetry fixes $\gamma_3 = - \frac 1 {12}$. With this in mind, and up to boundary terms and a re-scaling of the unique free parameter left to $\tilde \gamma$, we find the following final action
\begin{eqnarray}
S_{10} &=& \int d^{10} x \sqrt{-g} \left[ e^{-2\Phi} \left( R + 4 (\partial \Phi)^2 
-\frac 1 {12} H_{i j k} H^{i j k} \right)
- \frac 1 4  F_{i j} F^{i j} - \frac 1 {48}  {\tilde F}_{i j k l}{\tilde F}^{i j k l}  \right.  \nonumber
\\ &&\left.  \vphantom{\frac 1 2 }\ \ \ \ \ \ \ \ \ \ \ \ \ \ \ \ \  + \, \tilde \gamma\, \epsilon^{i_1 \dots i_{10}}     b_{i_1 i_2} F_{i_3 \dots i_6} F_{i_7 \dots i_{10}}  \right] \ . \label{IIA}
\end{eqnarray}
This precisely agrees with the action of Type IIA supergravity when $\tilde \gamma = - \frac {(4!)^2} 3$ \cite{Polchi}.

\subsection{Frame formulation}

We now move to the frame formulation of Type IIA supergravity. To this end, we compute the $\rho$ and $\alpha$ transformations of the vielbein from those of the metric, and get
\begin{equation}
        \delta_{\rho,\alpha} \tensor{e}{_{i}^{a}}
   =\frac{1}{2} (\rho + \alpha^d A_d) \tensor{e}{_i^a}-\alpha^a  A_i \ , \ \ \ \delta_{\rho,\alpha} \tensor{e}{^{i}_{a}}
   =- \frac{1}{2} (\rho + \alpha^d A_d) \tensor{e}{^i_a} +\alpha^i  A_a \ .
\end{equation}
We  define a flat derivative with a dilaton twist that commutes with $\rho$ and $\alpha$ transformations
\begin{equation}
    {\textup D}_{a}= e^{\frac \Phi 3} D_a \ , \ \ \ \left[\delta_\rho \, , \, {\textup D}_a\right] = 0 \ , \ \ \ \left[\delta_\alpha \, , \, {\textup D}_a\right] = 0 \ .
\end{equation}
An analogous definition in terms of the covariant flat derivative 
\begin{equation}
    {\textup D}^\nabla_{a}= e^{\frac \Phi 3} \nabla_a \ , \ \ \ \left[\delta_\rho \, , \, {\textup D}^\nabla_a\right] = 0 \ ,
\end{equation} commutes with $\rho$ transformations, but not with $\alpha$ transformations.

It is also useful to define $\rho$ invariant twisted connections and curvatures as follows
\begin{eqnarray}
&& \omega_{abc} = e^{\frac \Phi 3} w_{abc} \ , \ \ \ {\textup H}_{abc} = e^{\frac \Phi 3} H_{a b c} \ , \ \ \ {\textup F}_{ab} = e^{\frac 4 3 \Phi} F_{a b} \ , \nonumber \\ &&  \tilde {\textup F}_{abcd} = e^{\frac 4 3 \Phi} \tilde F_{a b cd} \ ,  \ \ \ {\textup R}_{abcd} = e^{\frac 2 3 \Phi} R_{abcd} \ .
\end{eqnarray}

In terms of these, the flat derivatives acting on the parameter $\alpha$ with tangent space indices satisfy the following relations
\begin{equation}
  {\textup D}_a \tensor{\alpha}{^c}= 2 \tensor{\alpha}{^d} \omega{_{[ad]}{}^c} \ , \  \ \ {\textup D}^\nabla_a \tensor{\alpha}{^c}= \alpha^d \, \omega_{ad}{}^c \ .
\end{equation}

We can now present the $\alpha$ transformations of the following $\rho$ invariant quantities
\begin{eqnarray}
        \delta_\alpha \omega{_{abc}}
        &=&e^{- \Phi }\left(\alpha^d g_{a[b}{\textup F}_{c]d}+\tensor{\alpha}{_{[c}} {\textup F}{_{b]a}}+ \frac{1}{2} {\alpha}{_{a}}{\textup F}{_{bc}}\right)
\ , \\        \delta_\alpha {\textup F}_{ab}&=&-2e^{-\Phi} \left( \alpha^c \omega_{cab}+2 \alpha_{[b}{\textup D}_{a]}\Phi\right)
\ , \\    \delta_\alpha {\textup H}_{abc}&=& e^{- \Phi}\alpha^d \tilde{\textup F}_{abcd}
\ , \\    \delta_\alpha \tilde{\textup{F}}_{abcd}&=& 4e^{- \Phi}\alpha_{[a}{\textup H}_{bcd]} \ , \\
\delta_\alpha {\textup{R}}_{abcd} &=& e^{- \Phi}\ \left(\alpha_{[d|}\textup {D}^\nabla_a {\textup{F}}_{|c]b}+\alpha_{[b}{\textup {D}}^\nabla_{a]}{\textup{F}}_{cd}-\alpha_{[d|}{\textup {D}}^\nabla_{b} {\textup{F}}_{|c]a}+\alpha^e {\textup D}^\nabla_a {\textup{F}}_{e[c}g_{d]b}-\alpha^e {\textup D}^\nabla_b  {\textup{F}}_{e[c}g_{d]a} \right. \nonumber\\
&& \ \  \  \  \  \  \ \ \left.-\alpha^e \omega_{ecd} {\textup{F}}_{ab} +\alpha^e \omega_{ea[c}{\textup{F}}_{d]b}-\alpha^e \omega_{eb[c}{\textup{F}}_{d]a}-\alpha^e \omega_{eab} {\textup{F}}_{cd}+\right.\nonumber\\
&& \ \  \  \  \  \  \ \ \left.+\alpha^e \omega{_{ea}{}^f}{\textup{F}}_{f[d}g_{c]b}-\alpha^e \omega{_{eb}{}^f}{\textup{F}}_{f[d}g_{c]a} +\frac{4}{3} \alpha^e {\textup{F}}_{e[d}g_{c]b} {\textup D}_a \Phi+\frac{4}{3} \alpha_{[c} {\textup{F}}_{d]b} {\textup D}_a\Phi-\right.\nonumber\\
&& \ \  \  \  \  \  \ \ \left.-\frac{4}{3} \alpha_{[c}{\textup{F}}_{d]a} {\textup D}_b \Phi-\frac{4}{3} \alpha_{[b|} {\textup{F}}_{cd} {\textup D}_{|a]}\Phi-\frac{4}{3}\alpha^e {\textup{F}}_{e[d} g_{c]a} {\textup D}_b\Phi
\right)        \ ,
\\
\delta_\alpha {\textup D}_a \Phi &=& - \frac 3 2 e^{-\Phi} \alpha^b {\textup F}_{b a} \ , \\ \delta_\alpha  {\textup D}^\nabla_{a} {\textup D}_b \Phi &=& e^{- \Phi} \alpha^{d}\, \left(\frac{3}{2}  \mathcal{\nabla}_a \textup{F}_{bd} -\frac{3}{2} \omega{_{da}{}^c} \textup{F}_{bc}+\frac{1}{2} g_{ab}\textup{F}_{cd} \textup{D}^c\Phi -2\textup{F}_{bd} \textup{D}_a \Phi+g_{d(a}\textup{F}_{b)c} \textup{D}^c \Phi\right)  \ . \ \ \ \ \ \ \ \ \ 
 \end{eqnarray}

Then  the following scalars transform as
\begin{eqnarray}
        \delta_\alpha {\textup R}
        &=& e^{-\Phi} \alpha^a \left( -7 {\cal \nabla}^{c}{\textup F}_{ca} - \frac {28} 3  {\textup F}_{a b}\, {\textup D}^b\Phi +6{\textup F}_{bc}\,  \omega_a{}^{bc}\right) \ ,  \\
    \delta_\alpha  ({\textup D} \Phi)^2&=&- 3 e^{- \Phi} \alpha^a {\textup F}_{ab}\, {\textup D}^b\Phi \ , \\
        \delta_\alpha\left({\textup D}^\nabla_a{\textup D}^a \Phi \right)&=& e^{- \Phi}\alpha^a \left(\frac{3}{2} {\cal \nabla}^c{\textup F}_{ca}-2 {\textup F}_{ab} {\cal \nabla}^b\Phi-\frac{3}{2} {\textup F}_{bc} \, \omega{_a{}^{bc}}\right) \,  \ , \\
          \delta_\alpha \left( {\textup F}_{ab} {\textup F}^{a b}\right)&=&-4 e^{- \Phi} \alpha^a\left( {\textup F}_{bc} \omega_a{}^{bc}  - 
   2  {\textup F}_{ab} {\textup D}^b \Phi\right) \ ,
    \end{eqnarray}
    yielding the $\alpha$ invariant expression
    \begin{equation}
     {\textup R} - \frac {62} 9 ({\textup D} \Phi)^2 + \frac {14} 3 {\textup D}^{\nabla}_a {\textup D}^a \Phi - \frac 1 4 {\textup F}_{a b} {\textup F}^{a b} \ , \label{R11invariant}
    \end{equation}
   that reproduces the combination with coefficient $\gamma_0$ in (\ref{S10aux}), as can be easily verified using the identity ${\textup D}^\nabla_a {\textup D}^a \Phi = e^{\frac 2 3 \Phi} \left(\Box \Phi + \frac 1 3 (\partial \Phi)^2\right)$.

On the other hand, the  transformations 
\begin{eqnarray}
 \delta_\alpha \left(  {\textup H}_{abc} {\textup H}^{abc}\right)&=&- 2 e^{- \Phi}\alpha^a {\textup H}^{bcd} \tilde {\textup F}_{abcd}\ , \\
    \delta_\alpha \left( \tilde {\textup F}_{abcd}\tilde {\textup F}^{abcd}\right)&=&  8 e^{-\Phi}\alpha^a {\textup H}^{bcd} \tilde{\textup F}_{abcd}\ , \label{F211invariant}
\end{eqnarray}
produce another $\alpha$ invariant combination
\be
 {\textup H}_{abc} {\textup H}^{abc} + \frac 1 4 \tilde {\textup F}_{abcd}\tilde {\textup F}^{abcd} \ ,
\ee
corresponding to the terms with coefficient $\gamma_3$ in (\ref{S10aux}).

The relative coefficient between (\ref{R11invariant}) and (\ref{F211invariant}) is finally fixed by $\beta$ symmetry to $-\frac 1 {12}$, rendering the $\alpha$ and $\rho$ invariant Lagrangian, up to CS terms,
\be
{\cal L}_{IIA} = {\textup R} - \frac {62} 9 ({\textup D} \Phi)^2 + \frac {14} 3 {\textup D}^{\nabla}_a {\textup D}^a \Phi - \frac 1 4 {\textup F}_{a b} {\textup F}^{a b} -\frac 1 {12} {\textup H}_{abc} {\textup H}^{abc} - \frac 1 {48} \tilde {\textup F}_{abcd}\tilde {\textup F}^{abcd} + C.S.
\ee

\section{Predicting quartic R-R couplings in IIA at order $\zeta(3)\alpha'^3$} \label{5}

In (\ref{IIA}) we displayed the $\rho$ and $\alpha$ completion of the universal NS-NS two derivative sector of Type IIA 
 supergravity. We saw that these symmetries   predict  the two derivative R-R interactions unambiguously, when the NS-NS sector is known.  In this section we extend the analysis to the next order in derivatives, to asses how $\alpha$ symmetry constrains the R-R couplings in Type IIA at order $\zeta(3)\alpha'^3$, starting from the well known interactions of the NS-NS fields.
 
 To follow the procedure implemented in the previous sections, we should take a generic combination of all possible diffeomorphism and gauge invariant eight-derivative terms in 10 dimensions, act on them with the transformations \eqref{rho1}$-$\eqref{al2} and determine the $\alpha$ invariant action.  For simplicity, instead, we restrict the analysis to quartic couplings and take advantage of known results on four-point interactions in the effective actions of M-theory and the NS-NS sector of Type II theories.

We will see that, contrary to what happens in the two-derivative case, at the eight-derivative level there are $\alpha$ invariants that vanish when the R-R fields are set to zero, and then these terms are not directly accessible from the pure NS-NS sector. 
Furthermore, the effective action of M-theory computed from four-point correlators  has ambiguities due to combinations of terms  with vanishing four-particle amplitudes. 

To control these two sources of ambiguities, we proceed as follows. We start with a generic linear combination of all independent diffeomorphism and gauge invariant eight-derivative four-field terms of 11 dimensional supergravity, compactify on the circle, set the R-R fields to zero and constrain the coefficients by requiring that the 10 dimensional expression agrees with the well known eight-derivative quartic interactions $\zeta(3) \alpha'{}^3 t_8t_8R^{(-)4}$ of NS-NS fields in Type II theories. This procedure allows us to classify the intrinsic redundancies of our method.

In order to eliminate them, we demand that the resulting 11 dimensional couplings agree with the effective action that produces the four-point amplitudes in M-theory \cite{Peeters}. We mentioned that these terms are also ambiguous in the sense that they are defined up to terms that vanish at the four-point level. However, we get rid of these ambiguities by requiring that the 11 dimensional action compatible with the results in \cite{Peeters}, reduces {\it exactly} to $\zeta(3) \alpha'{}^3 t_8t_8R^{(-)4}$, as opposed to {\it up to} five-point contributions. This procedure allows us to finally end with the $\alpha$ invariant completion of $\zeta(3) \alpha'{}^3 t_8t_8R^{(-)4}$ in Type IIA.

The following two observations are crucial:
\begin{enumerate}
\item Modulo the overall $e^{-2\phi}$ factor in the string frame, the structure of the tree-level and one-loop quartic NS-NS interactions  is identical in Type II theories \cite{schwarz}-\cite{Liu}.

\item The circle reduction of the one-loop quartic Riemann interactions in 11 dimensional supergravity, leads to the quartic Riemann terms at one-loop in the string coupling in 10 dimensional Type IIA theory  \cite{green}.
\end{enumerate}

Therefore, although many dilaton dependent terms can be redefined away with field redefinitions, in order to gain simplicity and also to put the tree-level and one-loop corrections on the same footing, from now on we ignore the dilaton $\phi \to 0$ and its $\alpha$ transformation $\alpha^i A_i \to 0$. 

Before turning to the explicit calculations, it is useful to recall the effective Lagrangian produced by the four-point amplitudes of NS-NS fields at order $\alpha'^3$ in  Type II theories \cite{schwarz}-\cite{sloan}
\be
{\cal L}_4 = \zeta(3)\alpha'^3 t_8^{i_1\dots i_8} t_{8\, j_1 \dots j_8} \, R^{(-)}_{i_1 i_2}{}^{j_1 j_2} R^{(-)}_{i_3 i_4}{}^{j_3 j_4} R^{(-)}_{i_5 i_6}{}^{j_5 j_6} R^{(-)}_{i_7 i_8}{}^{j_7 j_8} \ .  \label{t8t8}
\ee
 Here $R^{(-)}_{ijkl}$ is the torsional Riemann tensor, which to linear order reads
 \be
 R^{(-)}_{ijkl} = R_{ijkl} - \nabla_{[i}H_{j]kl} \ ,
 \ee
and the $t_8$ tensor can be defined through its action on generic antisymmetric matrices $M^a_{ij}$, with $a = 1 \dots 4$, as follows
\begin{eqnarray}
&& t_8^{i_1 \dots i_8} M^1_{i_1i_2} M^2_{i_3i_4} M^3_{i_5i_6} M^4_{i_7i_8} = 8 {\rm Tr}[M^1 M^2 M^3 M^4 + M^1 M^3 M^2 M^4 + M^1 M^3 M^4 M^2 ] \nonumber\\
&& \quad \quad -2 \left({\rm Tr}[M^1 M^2] {\rm Tr}[M^3 M^4] + {\rm Tr}[M^1 M^3]  {\rm Tr}[M^2 M^4] + {\rm Tr}[M^1 M^4]  {\rm Tr}[M^2 M^3]  \right) \ .
\end{eqnarray}

The NS-NS sector of Type II theories is invariant under the exchange of sign  of the two-form. To make this symmetry manifest, it is instructive to define new indices $\alpha = [ij]$. In terms of these, we can write $R^{(-)}_{\alpha\beta} = R_{\alpha \beta} - \nabla H_{\alpha\beta}$, where $R_{\alpha \beta}$ is symmetric and $\nabla H_{\alpha\beta}$ is antisymmetric under the exchange of $\alpha \leftrightarrow \beta$, due to the Bianchi identities $R_{\mu\nu\rho\sigma} = R_{\rho\sigma\mu\nu} $  and $\nabla_{[\mu} H_{\nu\rho\sigma]} = 0$.
 This, together with the fact that $t_8^{\alpha_1 \dots \alpha_4}$ is totally symmetric under the exchange of any pair of indices $\alpha_i$, implies that (\ref{t8t8}) can be rewritten as
\be
\frac 1 {\alpha'^3 \zeta(3)} {\cal L}_4 = R^4 + 6 R^2 \nabla H^2  +\nabla H^4\ , \label{nsns}
\ee
where
\begin{eqnarray}
R^4 &=& t_8^{\alpha_1 \dots \alpha_4} t_8^{\beta_1 \dots \beta_4} \, R_{\alpha_1 \beta_1} \dots R_{\alpha_4 \beta_4}\, , \label{R4}\\
R^2 \nabla H^2 &=& t_8^{\alpha_1 \dots \alpha_4} t_8^{\beta_1 \dots \beta_4} \,  R_{\alpha_1 \beta_1}  R_{\alpha_2 \beta_2} \nabla H_{\alpha_3 \beta_3}\nabla H_{\alpha_4 \beta_4} \, ,\label{Alpha3s}\\ 
\nabla H^4 &=& t_8^{\alpha_1 \dots \alpha_4} t_8^{\beta_1 \dots \beta_4} \, \nabla H_{\alpha_1 \beta_1} \dots \nabla H_{\alpha_4 \beta_4} \ ,  \label{NH4}
\end{eqnarray}
and the terms linear and cubic in powers of $\nabla H$ vanish. The three terms above are not related by $\alpha$ symmetry. Instead, their symmetric completion will induce additional terms with R-R couplings.

We also recall that quartic R-R interactions in Type IIA theories in 10 dimensions have been computed using different methods in  \cite{Policastro:2006vt}-\cite{Bakhtiarizadeh:2017ojz}.

\subsection{$\alpha$ symmetric completion of $R^4$ terms }

The quartic Riemann interactions (\ref{R4}) take the following simple form when decomposed in terms of explicit index contractions
\begin{eqnarray}
        {R^4}
        &=& 3\cdot 2^7 \left(\tensor{R}{_{i}  _{j}  _{k}  _{l}} \tensor{R}{^{j}  ^{m}  ^{l}  ^{n}} \tensor{R}{^{p}  ^{k}  _{m}  _{q}} \tensor{R}{^{q}   ^{i}  _{n}  _{p}} + 
        \frac{1}{2} \tensor{R}{_{i}  _{j}  _{k}  _{l}} \tensor{R}{^{j}  ^{m}  ^{l}  ^{n}} \tensor{R}{_{m}  _{q}  _{n}  _{p}} \tensor{R}{^{q}  ^{i}  ^{p}  ^{k}}   -\frac{1}{2}  \tensor{R}{_{i}  _{j}  _{k}  _{l}} \tensor{R}{^{j}  ^{m}  ^{k}  ^{l}} \tensor{R}{_{m}  _{q}  ^{n}  _{p}} \tensor{R}{^{q}  ^{i}  ^{n}  ^{p}} \right. \nonumber\\
        &&- 
        \left.\frac{1}{4}  \tensor{R}{_{i}  _{j}  _{k}  _{l}} \tensor{R}{^{j}  ^{m}  ^{n}  ^{p}} \tensor{R}{^{k}  ^{l}  _{m}  _{q}} \tensor{R}{^{q}  ^{i}  _{n}  _{p}}  +\frac{1}{16} \tensor{R}{_{i}  _{j}  _{k}  _{l}} \tensor{R}{^{j}  ^{i}  ^{n}  ^{p}} \tensor{R}{^{m}  ^{q}  ^{k}  ^{l}} \tensor{R}{_{q}  _{m}  _{n}  _{p}} + 
        \frac{1}{32}  \tensor{R}{_{i}  _{j}  _{k}  _{l}} \tensor{R}{^{i}  ^{j}  ^{k}  ^{l}} \tensor{R}{_{m}  _{q}  _{n}  _{p}} \tensor{R}{^{m}  ^{q}  ^{n}  ^{p}} \right) \, .\ \ \ \ \ \ \ 
\end{eqnarray}
The $\alpha$ symmetric completion of these terms is rather straightforward to determine, since its $11$ dimensional uplift is obviously
\begin{equation}
    \begin{split}
        {\cal R}^4
        &=3 \cdot 2^7 \left( \mathcal{R}_{\alpha \rho \sigma \delta} \mathcal{R}^{\rho \gamma \delta \beta} \tensor{\mathcal{R}}{^{\mu \sigma}_{\gamma \nu}} \tensor{\mathcal{R}}{^{\nu \alpha} _{\beta \mu}}+\frac{1}{2} \mathcal{R}_{\alpha \delta \beta \sigma}\mathcal{R}^{\delta \gamma \sigma \rho}\mathcal{R}_{\gamma \mu \rho \nu}\mathcal{R}^{\mu \alpha \nu \beta} -\frac{1}{2} \mathcal{R}_{\alpha \delta \rho \sigma} {\mathcal{R}}^{\delta \gamma \rho \sigma} \mathcal{R}_{\gamma \beta \mu \nu} \mathcal{R}^{\beta \alpha \mu \nu}\right.\\ &\left.-\frac{1}{4} \mathcal{R}_{\alpha \rho \sigma \delta} \mathcal{R}^{\rho \gamma \beta \mu} \tensor{\mathcal{R}}{^{\sigma \delta }_{\gamma \nu}} \tensor{\mathcal{R}}{^{\nu \alpha}_{\beta \mu}}+\frac{1}{16} \mathcal{R}_{\rho \sigma \delta \gamma} \mathcal{R}^{\sigma \rho \alpha \beta} \mathcal{R}^{\mu \nu \delta \gamma} \mathcal{R}_{\nu \mu \alpha \beta}+\frac{1}{32} \mathcal{R}_{\gamma \delta \rho \sigma} \mathcal{R}^{\gamma \delta \rho \sigma} \mathcal{R}_{\alpha \beta \mu \nu} \mathcal{R}^{\alpha \beta \mu \nu}\right)\ .
    \end{split}
\end{equation}
The dimensional reduction of these terms includes the following R-R couplings in an $\alpha$ invariant form
\begin{equation}
\begin{split}
        \mathcal{R}^4
        &= R^4+24  \tensor{R}{_{l}  _{m}  _{n}  _{p}}   \tensor{R}{^{l} ^{m} ^{n}^{p}}  \nabla_k\tensor{F}{_{i}  _{j}}\nabla^k\tensor{F}{^{i}^{j}} -192  \tensor{R}{_{j}  ^{m} ^{n}^{p}}   \tensor{R}{_{l}  _{m}  _{n}  _{p}}   \nabla_k\tensor{F}{_{i} ^{l}}\nabla^k\tensor{F}{^{i}^{j}} \\ &-96  \tensor{R}{_{i}  _{l}  ^{n}^{p}}   \tensor{R}{_{j}  _{m}  _{n}  _{p}}   \nabla_k\tensor{F}{^{l}^{m}}\nabla^k\tensor{F}{^{i}^{j}} + 
         48  \tensor{R}{_{i}  _{j}  ^{n}^{p}}   \tensor{R}{_{l}  _{m}  _{n}  _{p}}   \nabla_k\tensor{F}{^{l}^{m}}\nabla^k\tensor{F}{^{i}^{j}} -96  \tensor{R}{_{k}  ^{m} ^{n}^{p}}   \tensor{R}{_{l}  _{m}  _{n}  _{p}}   \nabla^k\tensor{F}{^{i}^{j}}\nabla^l\tensor{F}{_{i}  _{j}} \\
         & +384  \tensor{R}{_{j}  ^{n} _{m} ^{p}}   \tensor{R}{_{k}  _{p}  _{l}  _{n}}   \nabla^k\tensor{F}{^{i}^{j}}\nabla^m\tensor{F}{_{i} ^{l}} +384  \tensor{R}{_{i}  _{l} _{k} ^{p}}   \tensor{R}{_{j}  _{m}  _{n}  _{p}}   \nabla^k\tensor{F}{^{i}^{j}}\nabla^n\tensor{F}{^{l}^{m}} + 384  \tensor{R}{_{j}  ^{n}_{k} ^{p}}   \tensor{R}{_{l}  _{n}  _{m}  _{p}}   \nabla^k\tensor{F}{^{i}^{j}}\nabla^m\tensor{F}{_{i} ^{l}}\\ 
        &- 96  \tensor{R}{_{i}  _{j}  _{n} ^{p}}   \tensor{R}{_{k}  _{p}  _{l}  _{m}}   \nabla^k\tensor{F}{^{i}^{j}}\nabla^n\tensor{F}{^{l}^{m}} - 
         96  \tensor{R}{_{i}  _{j} _{k} ^{p}}   \tensor{R}{_{l}  _{m}  _{n}  _{p}}   \nabla^k\tensor{F}{^{i}^{j}}\nabla^n\tensor{F}{^{l}^{m}}
        +24\nabla_k\tensor{F}{_{l}  _{n}}\nabla^k\tensor{F}{^{i}^{j}}\nabla_m\tensor{F}{_{j} ^{n}}\nabla^m\tensor{F}{_{i} ^{l}} \\
        &+
        48\nabla_k\tensor{F}{_{i} ^{l}}\nabla^k\tensor{F}{^{i}^{j}}\nabla_n\tensor{F}{_{l}  _{m}}\nabla^n\tensor{F}{_{j} ^{m}} -48\nabla^k\tensor{F}{^{i}^{j}}\nabla^l\tensor{F}{_{i}  _{j}}\nabla_n\tensor{F}{_{l}  _{m}}\nabla^n\tensor{F}{_{k} ^{m}}  +48\nabla_j\tensor{F}{_{m}  _{n}}\nabla^k\tensor{F}{^{i}^{j}}\nabla_{l}\tensor{F}{_{k} ^{n}}\nabla^m\tensor{F}{_{i} ^{l}} \\& - 
         12\nabla_k\tensor{F}{^{m}^{n}}\nabla^k\tensor{F}{^{i}^{j}}\nabla_{l}\tensor{F}{_{m}  _{n}}\nabla^l\tensor{F}{_{i}  _{j}}  + 48\nabla_j\tensor{F}{_{k} ^{n}}\nabla^k\tensor{F}{^{i}^{j}}\nabla_{l}\tensor{F}{_{m}  _{n}}\nabla^m\tensor{F}{_{i} ^{l}} \ .
 \end{split}
\end{equation}

\subsection{$\alpha$ symmetric completion of $R^2 \nabla H^2$ terms}
The mixed components (\ref{Alpha3s}) take the following form when written in terms of explicit index contractions
\begin{equation}
    \begin{split}
        R^2 \nabla H^2=
        &-96 \tensor{R}{_{l}  ^{n}  ^{p}  ^{q}}  \tensor{R}{_{m}  _{n}  _{p}  _{q}} \nabla_{i}\tensor{H}{_{j}  _{k}  ^{m}}\nabla^{i}\tensor{H}{^{j}  ^{k}  ^{l}} - 
        192 \tensor{R}{_{k}  ^{p}  _{m}  ^{q}}  \tensor{R}{_{l}  _{q}  _{n}  _{p}} \nabla_{i}\tensor{H}{_{j}  ^{m}  ^{n}}\nabla^{i}\tensor{H}{^{j}  ^{k}  ^{l}} \\
        &+192 \tensor{R}{_{k}  ^{p}  _{l}  ^{q}}  \tensor{R}{_{m}  _{p}  _{n}  _{q}} \nabla_{i}\tensor{H}{_{j}  ^{m}  ^{n}}\nabla^{i}\tensor{H}{^{j}  ^{k}  ^{l}} + 
        192 \tensor{R}{_{j}  _{k}  _{m}  ^{q}}  \tensor{R}{_{l}  _{n}  _{p}  _{q}} \nabla_{i}\tensor{H}{^{m}  ^{n}  ^{p}}\nabla^{i}\tensor{H}{^{j}  ^{k}  ^{l}} \\
        &+24 \tensor{R}{_{m}  _{n}  _{p}  _{q}}  \tensor{R}{^{m}  ^{n}  ^{p}  ^{q}} \nabla^{i}\tensor{H}{^{j}  ^{k}  ^{l}}\nabla_{j}\tensor{H}{_{i}  _{k}  _{l}} - 
        32 \tensor{R}{_{i}  ^{n}  ^{p}  ^{q}}  \tensor{R}{_{m}  _{n}  _{p}  _{q}} \nabla^{i}\tensor{H}{^{j}  ^{k}  ^{l}}\nabla^{m}\tensor{H}{_{j}  _{k}  _{l}} \\ 
        &+96 \tensor{R}{_{i}  _{n}  ^{p}  ^{q}}  \tensor{R}{_{l}  _{m}  _{p}  _{q}} \nabla^{i}\tensor{H}{^{j}  ^{k}  ^{l}}\nabla^{m}\tensor{H}{_{j}  _{k}  ^{n}} - 
        192 \tensor{R}{_{i}  _{m}  ^{p}  ^{q}}  \tensor{R}{_{l}  _{n}  _{p}  _{q}} \nabla^{i}\tensor{H}{^{j}  ^{k}  ^{l}}\nabla^{m}\tensor{H}{_{j}  _{k}  ^{n}} \\
        &+192 \tensor{R}{_{i}  ^{p}  _{m}  ^{q}}  \tensor{R}{_{l}  _{p}  _{n}  _{q}} \nabla^{i}\tensor{H}{^{j}  ^{k}  ^{l}}\nabla^{m}\tensor{H}{_{j}  _{k}  ^{n}} +
        192 \tensor{R}{_{i}  ^{p}  _{l}  ^{q}}  \tensor{R}{_{m}  _{p}  _{n}  _{q}} \nabla^{i}\tensor{H}{^{j}  ^{k}  ^{l}}\nabla^{m}\tensor{H}{_{j}  _{k}  ^{n}} \\ 
        &+192 \tensor{R}{_{i}  ^{q}  _{n}  _{p}}  \tensor{R}{_{k}  _{l}  _{m}  _{q}} \nabla^{i}\tensor{H}{^{j}  ^{k}  ^{l}}\nabla^{m}\tensor{H}{_{j}  ^{n}  ^{p}} +
        384 \tensor{R}{_{i}  _{n}  _{k}  ^{q}}  \tensor{R}{_{l}  _{m}  _{p}  _{q}} \nabla^{i}\tensor{H}{^{j}  ^{k}  ^{l}}\nabla^{m}\tensor{H}{_{j}  ^{n}  ^{p}} \\
        &-768 \tensor{R}{_{i}  _{k}  _{m}  ^{q}}  \tensor{R}{_{l}  _{n}  _{p}  _{q}} \nabla^{i}\tensor{H}{^{j}  ^{k}  ^{l}}\nabla^{m}\tensor{H}{_{j}  ^{n}  ^{p}} - 
        768 \tensor{R}{_{i}  _{m}  _{k}  ^{q}}  \tensor{R}{_{l}  _{n}  _{p}  _{q}} \nabla^{i}\tensor{H}{^{j}  ^{k}  ^{l}}\nabla^{m}\tensor{H}{_{j}  ^{n}  ^{p}} \\ 
        &-384 \tensor{R}{_{i}  _{k}  _{l}  ^{q}}  \tensor{R}{_{m}  _{n}  _{p}  _{q}} \nabla^{i}\tensor{H}{^{j}  ^{k}  ^{l}}\nabla^{m}\tensor{H}{_{j}  ^{n}  ^{p}} + 
        192 \tensor{R}{_{i}  _{n}  _{j}  _{m}}  \tensor{R}{_{k}  _{p}  _{l}  _{q}} \nabla^{i}\tensor{H}{^{j}  ^{k}  ^{l}}\nabla^{m}\tensor{H}{^{n}  ^{p}  ^{q}} \\
        &-192 \tensor{R}{_{i}  _{j}  _{k}  _{n}}  \tensor{R}{_{l}  _{m}  _{p}  _{q}} \nabla^{i}\tensor{H}{^{j}  ^{k}  ^{l}}\nabla^{m}\tensor{H}{^{n}  ^{p}  ^{q}} \ .
        \label{TR2dH2}
    \end{split}
\end{equation}
 To find its $\alpha$ symmetric completion, we propose the most general combination of 11 dimensional terms of the form ${\cal R}^2 \nabla {\cal F}^2$. It turns out that there is a basis $\{B_i\}$ of $i=1,\dots,24$ independent terms of that form \cite{Peeters}, which we recall in Appendix \ref{b}. Each $B_i$ reduces to an $\alpha$ invariant expression in 10 dimensions. We propose a generic linear combination 
\begin{equation}
        \mathcal{R}^2 \nabla \mathcal{F}^2 
        = \sum_{i=1}^{24} b_i B_i \ . \label{genericR2NH2}
\end{equation}
Compactifying this to 10 dimensions, setting  the R-R fields to zero,  eliminating all ambiguities due to Bianchi identities, and forcing the result to coincide with (\ref{TR2dH2}) fixes almost all the coefficients to
\begin{equation}\label{bisR2NH2}
\begin{split}
        &b_1 = b_2 = - b_5 = b_9 = -b_{10} = 192 \ , \ \ \  b_3 = b_7 = b_{15} = b_{18} = b_{19} = 0   \ , \ \ \ b_{6} = b_8 =-384\, , \\ 
        &b_{12}= b_{13} = - b_{14} = -96 \ , \ \ \ 
        b_{16} = b_{17} = - b_{20} = 64  \ , \ \ \ b_{21}= b_{23} = 32 \ , \ \ \ b_{22}= b_{24} = 8  \ .\ \ \ \  
\end{split} 
\end{equation}
The coefficients $b_4$ and $b_{11}$ are free because $B_4$ and $B_{11}$ compactify to $\alpha$ invariant interactions in 10 dimensions that vanish when the R-R fields are set to zero. 

We must compare this with the corresponding terms in the M-theory effective action computed in \cite{Peeters} from the two-graviton/two-three form amplitude, explicitly 
\begin{equation}
\begin{split}
  \mathcal{R}^2 \nabla \mathcal{F}^2 
    &=z (-24 B_5 - 48 B_8 - 24 B_{10} - 6 B_{12}- 12 B_{13} + 12 B_{14} + 8 B_{16} - 4 B_{20} \\ &+
    B_{22} + 4 B_{23} + B_{24}) 
    +z_1  {Z}_1 + z_2  {Z}_2 + z_3  {Z}_3 + z_4  {Z}_4 + z_5  {Z}_5 + z_6  {Z}_6 \, .
\end{split}\label{PetersR2NH2}
\end{equation}
The terms $\{Z_i\}$ with $i = 1,\dots,6$  represent a basis of terms that vanish at quartic order in a background field expansion (see \eqref{z1}-\eqref{z6}). Using (\ref{bisR2NH2}), we find that (\ref{genericR2NH2})  coincides with (\ref{PetersR2NH2}) when
\begin{equation}
    \begin{split}
        z = 16 \ , \ \ \  z_1 = -8 \ , \ \ \  &z_2 = -8 \ , \ \ \  z_3 =16 \ , \ \ \ z_4 =0 \ , \ \ \  z_5 =0\ ,  \\
        &b_4 = 48 + z_6 \ , \ \ \ b_{11} = 2 z_6 \ .
    \end{split}
\end{equation}
Hence, we conclude that the 11 dimensional combination leading exactly (as opposed to up to five-point terms) to (\ref{TR2dH2}) is determined modulo a single coefficient $z_6$. Setting it to zero for simplicity, we end with  
\begin{equation}
    \begin{split}
         \mathcal{R}^2 \nabla \mathcal{F}^2=
         & -192  \tensor{\mathcal{R}}{^{\alpha}  ^{\delta}  ^{\beta}  ^{\gamma}}  \tensor{\mathcal{R}}{^{\iota}  ^{\kappa}  ^{\varepsilon}  ^{\lambda}} \nabla_{\alpha}\tensor{\mathcal{F}}{_{\beta}  _{\varepsilon}  _{\iota}  ^{\mu}}\nabla_{\gamma}\tensor{\mathcal{F}}{_{\delta}  _{\kappa}  _{\lambda}  _{\mu}}  -384  \tensor{\mathcal{R}}{^{\alpha}  ^{\delta}  ^{\beta}  ^{\gamma}}     \tensor{\mathcal{R}}{^{\varepsilon}  ^{\iota}  _{\alpha}  ^{\kappa}} \nabla_{\beta}\tensor{\mathcal{F}}{_{\delta}  _{\varepsilon}  ^{\lambda}  ^{\mu}}\nabla_{\gamma}\tensor{\mathcal{F}}{_{\iota}  _{\kappa}  _{\lambda}  _{\mu}}  \\ 
        &-64  \tensor{\mathcal{R}}{^{\alpha}  ^{\delta}  ^{\beta}  ^{\gamma}}     \tensor{\mathcal{R}}{^{\varepsilon}  ^{\iota}  _{\alpha}  _{\delta}} \nabla_{\beta}\tensor{\mathcal{F}}{_{\varepsilon}  ^{\kappa}  ^{\lambda}  ^{\mu}}\nabla_{\gamma}\tensor{\mathcal{F}}{_{\iota}  _{\kappa}  _{\lambda}  _{\mu}}  -384  \tensor{\mathcal{R}}{^{\alpha}  ^{\delta}  ^{\beta}  ^{\gamma}}     \tensor{\mathcal{R}}{^{\varepsilon}  ^{\iota}  _{\alpha}  ^{\kappa}} \nabla_{\beta}\tensor{\mathcal{F}}{_{\gamma}  _{\varepsilon}  ^{\lambda}  ^{\mu}}\nabla_{\delta}\tensor{\mathcal{F}}{_{\iota}  _{\kappa}  _{\lambda}  _{\mu}}  \\ 
        &+64  \tensor{\mathcal{R}}{^{\alpha}  ^{\delta}  ^{\beta}  ^{\gamma}}     \tensor{\mathcal{R}}{_{\beta}  ^{\varepsilon}  _{\alpha}  ^{\iota}} \nabla_{\gamma}\tensor{\mathcal{F}}{_{\varepsilon}  ^{\kappa}  ^{\lambda}  ^{\mu}}\nabla_{\delta}\tensor{\mathcal{F}}{_{\iota}  _{\kappa}  _{\lambda}  _{\mu}}  -192  \tensor{\mathcal{R}}{^{\alpha}  ^{\delta}  ^{\beta}  ^{\gamma}}     \tensor{\mathcal{R}}{^{\varepsilon}  ^{\iota}  _{\alpha}  ^{\kappa}} \nabla_{\beta}\tensor{\mathcal{F}}{_{\gamma}  _{\delta}  ^{\lambda}  ^{\mu}}\nabla_{\varepsilon}\tensor{\mathcal{F}}{_{\iota}  _{\kappa}  _{\lambda}  _{\mu}}  \\ 
        &+64  \tensor{\mathcal{R}}{^{\alpha}  ^{\delta}  ^{\beta}  ^{\gamma}}     \tensor{\mathcal{R}}{_{\beta}  ^{\varepsilon}  _{\alpha}  ^{\iota}} \nabla_{\gamma}\tensor{\mathcal{F}}{_{\delta}  ^{\kappa}  ^{\lambda}  ^{\mu}}\nabla_{\varepsilon}\tensor{\mathcal{F}}{_{\iota}  _{\kappa}  _{\lambda}  _{\mu}}  -8  \tensor{\mathcal{R}}{^{\alpha}  ^{\delta}  ^{\beta}  ^{\gamma}}     \tensor{\mathcal{R}}{_{\beta}  _{\gamma}  _{\alpha}  ^{\varepsilon}} \nabla_{\delta}\tensor{\mathcal{F}}{^{\iota}  ^{\kappa}  ^{\lambda}  ^{\mu}}\nabla_{\varepsilon}\tensor{\mathcal{F}}{_{\iota}  _{\kappa}  _{\lambda}  _{\mu}}  \\ 
         &+192  \tensor{\mathcal{R}}{^{\alpha}  ^{\delta}  ^{\beta}  ^{\gamma}}     \tensor{\mathcal{R}}{^{\varepsilon}  ^{\iota}  _{\alpha}  ^{\kappa}} \nabla_{\beta}\tensor{\mathcal{F}}{_{\delta}  _{\varepsilon}  ^{\lambda}  ^{\mu}}\nabla_{\iota}\tensor{\mathcal{F}}{_{\gamma}  _{\kappa}  _{\lambda}  _{\mu}}  +32  \tensor{\mathcal{R}}{^{\alpha}  ^{\delta}  ^{\beta}  ^{\gamma}}     \tensor{\mathcal{R}}{^{\varepsilon}  ^{\iota}  _{\alpha}  _{\delta}} \nabla_{\beta}\tensor{\mathcal{F}}{_{\varepsilon}  ^{\kappa}  ^{\lambda}  ^{\mu}}\nabla_{\iota}\tensor{\mathcal{F}}{_{\gamma}  _{\kappa}  _{\lambda}  _{\mu}}  \\ 
         &-192  \tensor{\mathcal{R}}{^{\alpha}  ^{\delta}  ^{\beta}  ^{\gamma}}     \tensor{\mathcal{R}}{^{\iota}  ^{\kappa}  ^{\varepsilon}  ^{\lambda}} \nabla_{\alpha}\tensor{\mathcal{F}}{_{\beta}  _{\gamma}  _{\varepsilon}  ^{\mu}}\nabla_{\iota}\tensor{\mathcal{F}}{_{\delta}  _{\kappa}  _{\lambda}  _{\mu}}  +8  \tensor{\mathcal{R}}{^{\alpha}  ^{\delta}  ^{\beta}  ^{\gamma}}     \tensor{\mathcal{R}}{_{\beta}  _{\gamma}  _{\alpha}  _{\delta}} \nabla^{\varepsilon}\tensor{\mathcal{F}}{^{\iota}  ^{\kappa}  ^{\lambda}  ^{\mu}}\nabla_{\iota}\tensor{\mathcal{F}}{_{\varepsilon}  _{\kappa}  _{\lambda}  _{\mu}}  \\ 
         &-32  \tensor{\mathcal{R}}{^{\alpha}  ^{\delta}  ^{\beta}  ^{\gamma}}     \tensor{\mathcal{R}}{_{\beta}  _{\gamma}  _{\alpha}  ^{\varepsilon}} \nabla_{\iota}\tensor{\mathcal{F}}{_{\varepsilon}  _{\kappa}  _{\lambda}  _{\mu}}\nabla^{\iota}\tensor{\mathcal{F}}{_{\delta}  ^{\kappa}  ^{\lambda}  ^{\mu}}  +96  \tensor{\mathcal{R}}{^{\alpha}  ^{\delta}  ^{\beta}  ^{\gamma}}     \tensor{\mathcal{R}}{^{\varepsilon}  ^{\iota}  _{\alpha}  ^{\kappa}} \nabla_{\delta}\tensor{\mathcal{F}}{_{\varepsilon}  _{\iota}  ^{\lambda}  ^{\mu}}\nabla_{\kappa}\tensor{\mathcal{F}}{_{\beta}  _{\gamma}  _{\lambda}  _{\mu}}  \\ 
        &+96  \tensor{\mathcal{R}}{^{\alpha}  ^{\delta}  ^{\beta}  ^{\gamma}}     \tensor{\mathcal{R}}{_{\beta}  ^{\varepsilon}  _{\alpha}  ^{\iota}} \nabla_{\kappa}\tensor{\mathcal{F}}{_{\varepsilon}  _{\iota}  _{\lambda}  _{\mu}}\nabla^{\kappa}\tensor{\mathcal{F}}{_{\gamma}  _{\delta}  ^{\lambda}  ^{\mu}}  -96  \tensor{\mathcal{R}}{^{\alpha}  ^{\delta}  ^{\beta}  ^{\gamma}}     \tensor{\mathcal{R}}{_{\beta}  ^{\varepsilon}  _{\alpha}  ^{\iota}} \nabla_{\kappa}\tensor{\mathcal{F}}{_{\delta}  _{\varepsilon}  _{\lambda}  _{\mu}}\nabla^{\kappa}\tensor{\mathcal{F}}{_{\gamma}  _{\iota}  ^{\lambda}  ^{\mu}}  \\ 
         &+192  \tensor{\mathcal{R}}{^{\alpha}  ^{\delta}  ^{\beta}  ^{\gamma}}     \tensor{\mathcal{R}}{^{\varepsilon}  ^{\iota}  _{\alpha}  ^{\kappa}} \nabla_{\lambda}\tensor{\mathcal{F}}{_{\delta}  _{\iota}  _{\kappa}  _{\mu}}\nabla^{\lambda}\tensor{\mathcal{F}}{_{\beta}  _{\gamma}  _{\varepsilon}  ^{\mu}}  +48  \tensor{\mathcal{R}}{^{\alpha}  ^{\delta}  ^{\beta}  ^{\gamma}}     \tensor{\mathcal{R}}{^{\iota}  ^{\kappa}  ^{\varepsilon}  ^{\lambda}} \nabla_{\alpha}\tensor{\mathcal{F}}{ _{\delta}  _{\varepsilon}  _{\lambda}^{\mu}}\nabla_{\mu}\tensor{\mathcal{F}}{_{\beta}_{\gamma}  _{\iota}  _{\kappa} }\ .
    \end{split}
\end{equation}

The dimensional reduction of this expression to 10 dimensions can be organized as follows 
\begin{equation}
\mathcal{R}^2 \nabla \mathcal{F}^2= {R^2 \nabla H^2}+ {F_2^2H^2}+ {R^2F_4^2}+ {F_2^2F_4^2}+ {RF_2 H F_4} \ ,
\end{equation}
where ${R^2 \nabla H^2}$ was our starting point (\ref{TR2dH2}), and rest is its $\alpha$ symmetric completion, given by the following contributions
\begin{equation}
    \begin{split}
        {F_2^2H^2} = 
        &96 \nabla_{i}\tensor{H}{_{l}  ^{n}  ^{p}} \nabla^{i}\tensor{F}{^{j}  ^{k}} \nabla_{j}\tensor{F}{^{l}  ^{m}} \nabla_{k}\tensor{H}{_{m}  _{n}  _{p}} + 
        96 \nabla_{i}\tensor{H}{_{k}  ^{n}  ^{p}} \nabla^{i}\tensor{F}{^{j}  ^{k}} \nabla_{j}\tensor{F}{^{l}  ^{m}} \nabla_{l}\tensor{H}{_{m}  _{n}  _{p}} \\
        &-48 \nabla_{i}\tensor{F}{^{l}  ^{m}} \nabla^{i}\tensor{F}{^{j}  ^{k}} \nabla_{j}\tensor{H}{_{k}  ^{n}  ^{p}} \nabla_{l}\tensor{H}{_{m}  _{n}  _{p}} - 
        16 \nabla_{i}\tensor{H}{^{m}  ^{n}  ^{p}} \nabla^{i}\tensor{F}{^{j}  ^{k}} \nabla_{l}\tensor{H}{_{m}  _{n}  _{p}} \nabla^{l}\tensor{F}{_{j}  _{k}}  \\
        &+48 \nabla_{i}\tensor{H}{_{k}  ^{n}  ^{p}} \nabla^{i}\tensor{F}{^{j}  ^{k}} \nabla_{l}\tensor{H}{_{m}  _{n}  _{p}} \nabla^{l}\tensor{F}{_{j}  ^{m}} - 
        192 \nabla_{i}\tensor{H}{_{l}  _{m}  ^{p}} \nabla^{i}\tensor{F}{^{j}  ^{k}} \nabla_{j}\tensor{H}{_{k}  _{n}  _{p}} \nabla^{l}\tensor{F}{^{m}  ^{n}}  \\
        &+192 \nabla_{i}\tensor{H}{_{j}  _{m}  ^{p}} \nabla^{i}\tensor{F}{^{j}  ^{k}} \nabla_{k}\tensor{H}{_{l}  _{n}  _{p}} \nabla^{l}\tensor{F}{^{m}  ^{n}} + 
        192 \nabla^{i}\tensor{F}{^{j}  ^{k}} \nabla_{j}\tensor{H}{_{i}  _{m}  ^{p}} \nabla_{k}\tensor{H}{_{l}  _{n}  _{p}} \nabla^{l}\tensor{F}{^{m}  ^{n}}  \\
        &+48 \nabla_{i}\tensor{H}{_{m}  _{n}  ^{p}} \nabla^{i}\tensor{F}{^{j}  ^{k}} \nabla_{l}\tensor{H}{_{j}  _{k}  _{p}} \nabla^{l}\tensor{F}{^{m}  ^{n}} -
        8 \nabla_{i}\tensor{F}{_{j}  _{k}} \nabla^{i}\tensor{F}{^{j}  ^{k}} \nabla_{l}\tensor{H}{_{m}  _{n}  _{p}} \nabla^{l}\tensor{H}{^{m}  ^{n}  ^{p}}  \\
        &+96 \nabla^{i}\tensor{F}{^{j}  ^{k}} \nabla_{k}\tensor{H}{_{l}  ^{n}  ^{p}} \nabla^{l}\tensor{F}{_{j}  ^{m}} \nabla_{m}\tensor{H}{_{i}  _{n}  _{p}} - 
        96 \nabla^{i}\tensor{F}{^{j}  ^{k}} \nabla_{j}\tensor{F}{^{l}  ^{m}} \nabla_{l}\tensor{H}{_{i}  ^{n}  ^{p}} \nabla_{m}\tensor{H}{_{k}  _{n}  _{p}} \\
        &+96 \nabla^{i}\tensor{F}{^{j}  ^{k}} \nabla_{j}\tensor{H}{_{i}  _{l}  ^{p}} \nabla^{l}\tensor{F}{^{m}  ^{n}} \nabla_{m}\tensor{H}{_{k}  _{n}  _{p}} +
        48 \nabla^{i}\tensor{F}{^{j}  ^{k}} \nabla_{k}\tensor{H}{_{i}  ^{n}  ^{p}} \nabla^{l}\tensor{F}{_{j}  ^{m}} \nabla_{m}\tensor{H}{_{l}  _{n}  _{p}} \\
        &-48 \nabla_{i}\tensor{H}{_{j}  _{k}  ^{p}} \nabla^{i}\tensor{F}{^{j}  ^{k}} \nabla^{l}\tensor{F}{^{m}  ^{n}} \nabla_{m}\tensor{H}{_{l}  _{n}  _{p}} - 
        96 \nabla^{i}\tensor{F}{^{j}  ^{k}} \nabla_{j}\tensor{H}{_{i}  _{k}  ^{p}} \nabla^{l}\tensor{F}{^{m}  ^{n}} \nabla_{m}\tensor{H}{_{l}  _{n}  _{p}}  \\
        &+24 \nabla_{i}\tensor{F}{_{j}  _{k}} \nabla^{i}\tensor{F}{^{j}  ^{k}} \nabla^{l}\tensor{H}{^{m}  ^{n}  ^{p}} \nabla_{m}\tensor{H}{_{l}  _{n}  _{p}} -
        24 \nabla^{i}\tensor{F}{^{j}  ^{k}} \nabla^{l}\tensor{F}{_{j}  _{k}} \nabla_{m}\tensor{H}{_{l}  _{n}  _{p}} \nabla^{m}\tensor{H}{_{i}  ^{n}  ^{p}} \\
        &+48 \nabla_{i}\tensor{F}{_{j}  ^{l}} \nabla^{i}\tensor{F}{^{j}  ^{k}} \nabla_{m}\tensor{H}{_{l}  _{n}  _{p}} \nabla^{m}\tensor{H}{_{k}  ^{n}  ^{p}} - 
        48 \nabla^{i}\tensor{F}{^{j}  ^{k}} \nabla_{j}\tensor{F}{_{i}  ^{l}} \nabla_{m}\tensor{H}{_{l}  _{n}  _{p}} \nabla^{m}\tensor{H}{_{k}  ^{n}  ^{p}}  \\
        &-96 \nabla_{i}\tensor{H}{_{j}  _{m}  ^{p}} \nabla^{i}\tensor{F}{^{j}  ^{k}} \nabla^{l}\tensor{F}{^{m}  ^{n}} \nabla_{n}\tensor{H}{_{k}  _{l}  _{p}} + 
        96 \nabla^{i}\tensor{F}{^{j}  ^{k}} \nabla_{j}\tensor{H}{_{i}  _{m}  ^{p}} \nabla^{l}\tensor{F}{^{m}  ^{n}} \nabla_{n}\tensor{H}{_{k}  _{l}  _{p}} \\
        &-48 \nabla^{i}\tensor{F}{^{j}  ^{k}} \nabla_{j}\tensor{F}{^{l}  ^{m}} \nabla_{n}\tensor{H}{_{l}  _{m}  _{p}} \nabla^{n}\tensor{H}{_{i}  _{k}  ^{p}} + 
        96 \nabla^{i}\tensor{F}{^{j}  ^{k}} \nabla^{l}\tensor{F}{_{j}  ^{m}} \nabla_{n}\tensor{H}{_{l}  _{m}  _{p}} \nabla^{n}\tensor{H}{_{i}  _{k}  ^{p}}  \\
        &-96 \nabla^{i}\tensor{F}{^{j}  ^{k}} \nabla_{j}\tensor{F}{^{l}  ^{m}} \nabla_{n}\tensor{H}{_{k}  _{m}  _{p}} \nabla^{n}\tensor{H}{_{i}  _{l}  ^{p}} + 
        96 \nabla^{i}\tensor{F}{^{j}  ^{k}} \nabla^{l}\tensor{F}{_{j}  ^{m}} \nabla_{n}\tensor{H}{_{k}  _{m}  _{p}} \nabla^{n}\tensor{H}{_{i}  _{l}  ^{p}}  \\
        &-96 \nabla^{i}\tensor{F}{^{j}  ^{k}} \nabla^{l}\tensor{F}{_{j}  ^{m}} \nabla_{n}\tensor{H}{_{k}  _{l}  _{p}} \nabla^{n}\tensor{H}{_{i}  _{m}  ^{p}} -
        48 \nabla^{i}\tensor{F}{^{j}  ^{k}} \nabla_{j}\tensor{H}{_{k}  _{l}  ^{p}} \nabla^{l}\tensor{F}{^{m}  ^{n}} \nabla_{p}\tensor{H}{_{i}  _{m}  _{n}} \\
        &-24 \nabla_{i}\tensor{H}{_{m}  _{n}  ^{p}} \nabla^{i}\tensor{F}{^{j}  ^{k}} \nabla^{l}\tensor{F}{^{m}  ^{n}} \nabla_{p}\tensor{H}{_{j}  _{k}  _{l}} + 
        48 \nabla^{i}\tensor{F}{^{j}  ^{k}} \nabla^{l}\tensor{F}{^{m}  ^{n}} \nabla_{p}\tensor{H}{_{k}  _{m}  _{n}} \nabla^{p}\tensor{H}{_{i}  _{j}  _{l}} \ ,
    \end{split}
\end{equation}

\begin{equation}
    \begin{split}
        {R^2F_4^2} =& 
        -384  \tensor{R}{_{i}  ^{m}  ^{n}  ^{p}}   \tensor{R}{^{i}  ^{j}  ^{k}  ^{l}} \nabla_{j}\tensor{F}{_{m}  _{n}  ^{q}  ^{r}}\nabla_{k}\tensor{F}{_{l}  _{p}  _{q}  _{r}} +
        192  \tensor{R}{^{i}  ^{j}  ^{k}  ^{l}}   \tensor{R}{^{m}  ^{n}  ^{p}  ^{q}} \nabla_{i}\tensor{F}{_{k}  _{m}  _{p}  ^{r}}\nabla_{l}\tensor{F}{_{j}  _{n}  _{q}  _{r}}  \\
        &+384  \tensor{R}{_{i}  ^{m}  ^{n}  ^{p}}   \tensor{R}{^{i}  ^{j}  ^{k}  ^{l}} \nabla_{k}\tensor{F}{_{j}  _{n}  ^{q}  ^{r}}\nabla_{l}\tensor{F}{_{m}  _{p}  _{q}  _{r}} + 
        64  \tensor{R}{_{i}  ^{m}  _{k}  ^{n}}   \tensor{R}{^{i}  ^{j}  ^{k}  ^{l}} \nabla_{j}\tensor{F}{_{m}  ^{p}  ^{q}  ^{r}}\nabla_{l}\tensor{F}{_{n}  _{p}  _{q}  _{r}} \\
        &-64  \tensor{R}{_{i}  _{j}  ^{m}  ^{n}}   \tensor{R}{^{i}  ^{j}  ^{k}  ^{l}} \nabla_{k}\tensor{F}{_{m}  ^{p}  ^{q}  ^{r}}\nabla_{l}\tensor{F}{_{n}  _{p}  _{q}  _{r}} +
        96  \tensor{R}{_{i}  ^{m}  ^{n}  ^{p}}   \tensor{R}{^{i}  ^{j}  ^{k}  ^{l}} \nabla_{j}\tensor{F}{_{n}  _{p}  ^{q}  ^{r}}\nabla_{m}\tensor{F}{_{k}  _{l}  _{q}  _{r}} \\
        &+64  \tensor{R}{_{i}  ^{m}  _{k}  ^{n}}   \tensor{R}{^{i}  ^{j}  ^{k}  ^{l}} \nabla_{j}\tensor{F}{_{l}  ^{p}  ^{q}  ^{r}}\nabla_{m}\tensor{F}{_{n}  _{p}  _{q}  _{r}} - 
        8  \tensor{R}{_{i}  _{j}  _{k}  ^{m}}   \tensor{R}{^{i}  ^{j}  ^{k}  ^{l}} \nabla_{l}\tensor{F}{^{n}  ^{p}  ^{q}  ^{r}}\nabla_{m}\tensor{F}{_{n}  _{p}  _{q}  _{r}} \\
        &-192  \tensor{R}{^{i}  ^{j}  ^{k}  ^{l}}   \tensor{R}{^{m}  ^{n}  ^{p}  ^{q}} \nabla_{i}\tensor{F}{_{j}  _{k}  _{m}  ^{r}}\nabla_{n}\tensor{F}{_{l}  _{p}  _{q}  _{r}} +
        32  \tensor{R}{_{i}  _{j}  ^{m}  ^{n}}   \tensor{R}{^{i}  ^{j}  ^{k}  ^{l}} \nabla_{k}\tensor{F}{_{m}  ^{p}  ^{q}  ^{r}}\nabla_{n}\tensor{F}{_{l}  _{p}  _{q}  _{r}} \\
        &-192  \tensor{R}{_{i}  ^{m}  ^{n}  ^{p}}   \tensor{R}{^{i}  ^{j}  ^{k}  ^{l}} \nabla_{k}\tensor{F}{_{j}  _{l}  ^{q}  ^{r}}\nabla_{n}\tensor{F}{_{m}  _{p}  _{q}  _{r}} + 
        8  \tensor{R}{_{i}  _{j}  _{k}  _{l}}   \tensor{R}{^{i}  ^{j}  ^{k}  ^{l}} \nabla^{m}\tensor{F}{^{n}  ^{p}  ^{q}  ^{r}}\nabla_{n}\tensor{F}{_{m}  _{p}  _{q}  _{r}}  \\
        &-32  \tensor{R}{_{i}  _{j}  _{k}  ^{m}}   \tensor{R}{^{i}  ^{j}  ^{k}  ^{l}} \nabla_{n}\tensor{F}{_{m}  _{p}  _{q}  _{r}}\nabla^{n}\tensor{F}{_{l}  ^{p}  ^{q}  ^{r}} + 
        192  \tensor{R}{_{i}  ^{m}  ^{n}  ^{p}}   \tensor{R}{^{i}  ^{j}  ^{k}  ^{l}} \nabla_{k}\tensor{F}{_{j}  _{n}  ^{q}  ^{r}}\nabla_{p}\tensor{F}{_{l}  _{m}  _{q}  _{r}} \\
        &+96  \tensor{R}{_{i}  ^{m}  _{k}  ^{n}}   \tensor{R}{^{i}  ^{j}  ^{k}  ^{l}} \nabla_{p}\tensor{F}{_{m}  _{n}  _{q}  _{r}}\nabla^{p}\tensor{F}{_{j}  _{l}  ^{q}  ^{r}} - 
        96  \tensor{R}{_{i}  ^{m}  _{k}  ^{n}}   \tensor{R}{^{i}  ^{j}  ^{k}  ^{l}} \nabla_{p}\tensor{F}{_{l}  _{m}  _{q}  _{r}}\nabla^{p}\tensor{F}{_{j}  _{n}  ^{q}  ^{r}}  \\
        &+192  \tensor{R}{_{i}  ^{m}  ^{n}  ^{p}}   \tensor{R}{^{i}  ^{j}  ^{k}  ^{l}} \nabla_{q}\tensor{F}{_{l}  _{n}  _{p}  _{r}}\nabla^{q}\tensor{F}{_{j}  _{k}  _{m}  ^{r}} + 
        48  \tensor{R}{^{i}  ^{j}  ^{k}  ^{l}}   \tensor{R}{^{m}  ^{n}  ^{p}  ^{q}} \nabla_{i}\tensor{F}{_{j}  _{m}  _{n}  ^{r}}\nabla_{r}\tensor{F}{_{k}  _{l}  _{p}  _{q}} \ ,
    \end{split}
\end{equation}

\begin{equation}
    \begin{split}
        {RF_2 H F_4}=
        &-192  \tensor{R}{_{l}  ^{p}  _{m}  ^{q}}  \nabla_{i}\tensor{F}{_{k}  _{n}  _{p}  _{q}} \nabla^{i}\tensor{F}{^{j}  ^{k}} \nabla_{j}\tensor{H}{^{l}  ^{m}  ^{n}} + 
        192  \tensor{R}{_{j}  ^{p}  _{l}  ^{q}}  \nabla_{i}\tensor{H}{^{l}  ^{m}  ^{n}} \nabla^{i}\tensor{F}{^{j}  ^{k}} \nabla_{k}\tensor{F}{_{m}  _{n}  _{p}  _{q}}  \\
        &+192  \tensor{R}{_{i}  ^{p}  _{l}  ^{q}}  \nabla^{i}\tensor{F}{^{j}  ^{k}} \nabla_{j}\tensor{H}{^{l}  ^{m}  ^{n}} \nabla_{k}\tensor{F}{_{m}  _{n}  _{p}  _{q}} - 
        48  \tensor{R}{_{m}  _{n}  ^{p}  ^{q}}  \nabla^{i}\tensor{F}{^{j}  ^{k}} \nabla_{j}\tensor{F}{_{k}  _{l}  _{p}  _{q}} \nabla^{l}\tensor{H}{_{i}  ^{m}  ^{n}} \\
        &-192  \tensor{R}{_{j}  _{l}  ^{p}  ^{q}}  \nabla^{i}\tensor{F}{^{j}  ^{k}} \nabla_{k}\tensor{F}{_{m}  _{n}  _{p}  _{q}} \nabla^{l}\tensor{H}{_{i}  ^{m}  ^{n}} +
        192  \tensor{R}{_{j}  ^{p}  _{m}  ^{q}}  \nabla^{i}\tensor{F}{^{j}  ^{k}} \nabla_{l}\tensor{F}{_{k}  _{n}  _{p}  _{q}} \nabla^{l}\tensor{H}{_{i}  ^{m}  ^{n}} \\
        &-96  \tensor{R}{_{k}  _{m}  ^{p}  ^{q}}  \nabla^{i}\tensor{F}{^{j}  ^{k}} \nabla_{l}\tensor{F}{_{i}  _{n}  _{p}  _{q}} \nabla^{l}\tensor{H}{_{j}  ^{m}  ^{n}} + 
        96  \tensor{R}{_{i}  _{m}  ^{p}  ^{q}}  \nabla^{i}\tensor{F}{^{j}  ^{k}} \nabla_{l}\tensor{F}{_{k}  _{n}  _{p}  _{q}} \nabla^{l}\tensor{H}{_{j}  ^{m}  ^{n}}  \\
        &-192  \tensor{R}{_{i}  ^{p}  _{k}  ^{q}}  \nabla^{i}\tensor{F}{^{j}  ^{k}} \nabla_{l}\tensor{F}{_{m}  _{n}  _{p}  _{q}} \nabla^{l}\tensor{H}{_{j}  ^{m}  ^{n}} -
        96  \tensor{R}{_{l}  _{m}  _{n}  ^{q}}  \nabla_{i}\tensor{F}{_{j}  _{k}  _{p}  _{q}} \nabla^{i}\tensor{F}{^{j}  ^{k}} \nabla^{l}\tensor{H}{^{m}  ^{n}  ^{p}}  \\
        &-192  \tensor{R}{_{j}  _{m}  _{l}  ^{q}}  \nabla_{i}\tensor{F}{_{k}  _{n}  _{p}  _{q}} \nabla^{i}\tensor{F}{^{j}  ^{k}} \nabla^{l}\tensor{H}{^{m}  ^{n}  ^{p}} + 
        32  \tensor{R}{_{j}  _{k}  _{l}  ^{q}}  \nabla_{i}\tensor{F}{_{m}  _{n}  _{p}  _{q}} \nabla^{i}\tensor{F}{^{j}  ^{k}} \nabla^{l}\tensor{H}{^{m}  ^{n}  ^{p}}  \\
        &+96  \tensor{R}{_{l}  ^{q}  _{m}  _{n}}  \nabla^{i}\tensor{F}{^{j}  ^{k}} \nabla_{j}\tensor{F}{_{i}  _{k}  _{p}  _{q}} \nabla^{l}\tensor{H}{^{m}  ^{n}  ^{p}} + 
        96  \tensor{R}{_{j}  ^{q}  _{m}  _{n}}  \nabla^{i}\tensor{F}{^{j}  ^{k}} \nabla_{l}\tensor{F}{_{i}  _{k}  _{p}  _{q}} \nabla^{l}\tensor{H}{^{m}  ^{n}  ^{p}}  \\
        &-96  \tensor{R}{_{i}  _{m}  _{n}  ^{q}}  \nabla^{i}\tensor{F}{^{j}  ^{k}} \nabla_{l}\tensor{F}{_{j}  _{k}  _{p}  _{q}} \nabla^{l}\tensor{H}{^{m}  ^{n}  ^{p}} - 
        192  \tensor{R}{_{i}  ^{q}  _{j}  _{m}}  \nabla^{i}\tensor{F}{^{j}  ^{k}} \nabla_{l}\tensor{F}{_{k}  _{n}  _{p}  _{q}} \nabla^{l}\tensor{H}{^{m}  ^{n}  ^{p}} \\
        &+32  \tensor{R}{_{i}  ^{q}  _{j}  _{k}}  \nabla^{i}\tensor{F}{^{j}  ^{k}} \nabla_{l}\tensor{F}{_{m}  _{n}  _{p}  _{q}} \nabla^{l}\tensor{H}{^{m}  ^{n}  ^{p}} - 
        96  \tensor{R}{_{l}  ^{n}  ^{p}  ^{q}}  \nabla^{i}\tensor{F}{^{j}  ^{k}} \nabla_{j}\tensor{H}{_{k}  ^{l}  ^{m}} \nabla_{n}\tensor{F}{_{i}  _{m}  _{p}  _{q}} \\
        &+24  \tensor{R}{_{l}  _{m}  ^{p}  ^{q}}  \nabla_{i}\tensor{H}{^{l}  ^{m}  ^{n}} \nabla^{i}\tensor{F}{^{j}  ^{k}} \nabla_{n}\tensor{F}{_{j}  _{k}  _{p}  _{q}} -
        48  \tensor{R}{_{l}  _{m}  ^{p}  ^{q}}  \nabla^{i}\tensor{F}{^{j}  ^{k}} \nabla^{l}\tensor{H}{_{i}  ^{m}  ^{n}} \nabla_{n}\tensor{F}{_{j}  _{k}  _{p}  _{q}} \\
        &-96  \tensor{R}{_{l}  ^{n}  ^{p}  ^{q}}  \nabla_{i}\tensor{H}{_{j}  ^{l}  ^{m}} \nabla^{i}\tensor{F}{^{j}  ^{k}} \nabla_{n}\tensor{F}{_{k}  _{m}  _{p}  _{q}} - 
        192  \tensor{R}{_{k}  _{l}  ^{p}  ^{q}}  \nabla^{i}\tensor{F}{^{j}  ^{k}} \nabla^{l}\tensor{H}{_{j}  ^{m}  ^{n}} \nabla_{p}\tensor{F}{_{i}  _{m}  _{n}  _{q}}  \\
        &+48  \tensor{R}{_{m}  _{n}  ^{p}  ^{q}}  \nabla^{i}\tensor{F}{^{j}  ^{k}} \nabla^{l}\tensor{H}{_{i}  ^{m}  ^{n}} \nabla_{p}\tensor{F}{_{j}  _{k}  _{l}  _{q}} - 
        192  \tensor{R}{_{l}  ^{n}  ^{p}  ^{q}}  \nabla^{i}\tensor{F}{^{j}  ^{k}} \nabla_{j}\tensor{H}{_{i}  ^{l}  ^{m}} \nabla_{p}\tensor{F}{_{k}  _{m}  _{n}  _{q}} \\
        &+192  \tensor{R}{_{i}  _{l}  ^{p}  ^{q}}  \nabla^{i}\tensor{F}{^{j}  ^{k}} \nabla^{l}\tensor{H}{_{j}  ^{m}  ^{n}} \nabla_{p}\tensor{F}{_{k}  _{m}  _{n}  _{q}} - 
        192  \tensor{R}{_{k}  ^{n}  ^{p}  ^{q}}  \nabla_{i}\tensor{H}{_{j}  ^{l}  ^{m}} \nabla^{i}\tensor{F}{^{j}  ^{k}} \nabla_{p}\tensor{F}{_{l}  _{m}  _{n}  _{q}}  \\
        &-192  \tensor{R}{_{i}  ^{n}  ^{p}  ^{q}}  \nabla^{i}\tensor{F}{^{j}  ^{k}} \nabla_{j}\tensor{H}{_{k}  ^{l}  ^{m}} \nabla_{p}\tensor{F}{_{l}  _{m}  _{n}  _{q}} +
        384  \tensor{R}{_{l}  ^{p}  _{m}  ^{q}}  \nabla^{i}\tensor{F}{^{j}  ^{k}} \nabla^{l}\tensor{H}{_{j}  ^{m}  ^{n}} \nabla_{q}\tensor{F}{_{i}  _{k}  _{n}  _{p}}  \\
        &-192  \tensor{R}{_{j}  _{m}  _{l}  ^{q}}  \nabla^{i}\tensor{F}{^{j}  ^{k}} \nabla^{l}\tensor{H}{^{m}  ^{n}  ^{p}} \nabla_{q}\tensor{F}{_{i}  _{k}  _{n}  _{p}} - 
        192  \tensor{R}{_{j}  ^{q}  _{l}  _{m}}  \nabla^{i}\tensor{F}{^{j}  ^{k}} \nabla^{l}\tensor{H}{^{m}  ^{n}  ^{p}} \nabla_{q}\tensor{F}{_{i}  _{k}  _{n}  _{p}} \\
        &+192  \tensor{R}{_{k}  ^{p}  _{l}  ^{q}}  \nabla^{i}\tensor{F}{^{j}  ^{k}} \nabla^{l}\tensor{H}{_{j}  ^{m}  ^{n}} \nabla_{q}\tensor{F}{_{i}  _{m}  _{n}  _{p}} + 
        64  \tensor{R}{_{j}  _{k}  _{l}  ^{q}}  \nabla^{i}\tensor{F}{^{j}  ^{k}} \nabla^{l}\tensor{H}{^{m}  ^{n}  ^{p}} \nabla_{q}\tensor{F}{_{i}  _{m}  _{n}  _{p}}  \\
        &+192  \tensor{R}{_{i}  ^{p}  _{l}  ^{q}}  \nabla^{i}\tensor{F}{^{j}  ^{k}} \nabla_{j}\tensor{H}{^{l}  ^{m}  ^{n}} \nabla_{q}\tensor{F}{_{k}  _{m}  _{n}  _{p}} -
        192  \tensor{R}{_{i}  ^{p}  _{l}  ^{q}}  \nabla^{i}\tensor{F}{^{j}  ^{k}} \nabla^{l}\tensor{H}{_{j}  ^{m}  ^{n}} \nabla_{q}\tensor{F}{_{k}  _{m}  _{n}  _{p}} \\
        &+64  \tensor{R}{_{i}  _{l}  _{j}  ^{q}}  \nabla^{i}\tensor{F}{^{j}  ^{k}} \nabla^{l}\tensor{H}{^{m}  ^{n}  ^{p}} \nabla_{q}\tensor{F}{_{k}  _{m}  _{n}  _{p}} + 
        64  \tensor{R}{_{i}  ^{p}  _{k}  ^{q}}  \nabla^{i}\tensor{F}{^{j}  ^{k}} \nabla_{j}\tensor{H}{^{l}  ^{m}  ^{n}} \nabla_{q}\tensor{F}{_{l}  _{m}  _{n}  _{p}} \ .
    \end{split}
\end{equation}

\begin{equation}
\begin{split}
        {F_2^2F_4^2}=
        &-32 \nabla_{i}\tensor{F}{^{l}  ^{m}} \nabla^{i}\tensor{F}{^{j}  ^{k}} \nabla_{j}\tensor{F}{_{l}  ^{n}  ^{p}  ^{q}} \nabla_{k}\tensor{F}{_{m}  _{n}  _{p}  _{q}} - 
        4 \nabla_{i}\tensor{F}{_{j}  ^{l}} \nabla^{i}\tensor{F}{^{j}  ^{k}} \nabla_{k}\tensor{F}{^{m}  ^{n}  ^{p}  ^{q}} \nabla_{l}\tensor{F}{_{m}  _{n}  _{p}  _{q}} \\
        &-2 \nabla_{i}\tensor{F}{^{m}  ^{n}  ^{p}  ^{q}} \nabla^{i}\tensor{F}{^{j}  ^{k}} \nabla_{l}\tensor{F}{_{m}  _{n}  _{p}  _{q}} \nabla^{l}\tensor{F}{_{j}  _{k}} + 
        32 \nabla_{i}\tensor{F}{_{l}  ^{n}  ^{p}  ^{q}} \nabla^{i}\tensor{F}{^{j}  ^{k}} \nabla_{k}\tensor{F}{_{m}  _{n}  _{p}  _{q}} \nabla^{l}\tensor{F}{_{j}  ^{m}} \\
        &+16 \nabla_{i}\tensor{F}{_{k}  ^{n}  ^{p}  ^{q}} \nabla^{i}\tensor{F}{^{j}  ^{k}} \nabla_{l}\tensor{F}{_{m}  _{n}  _{p}  _{q}} \nabla^{l}\tensor{F}{_{j}  ^{m}} - 
        96 \nabla_{i}\tensor{F}{_{l}  _{m}  ^{p}  ^{q}} \nabla^{i}\tensor{F}{^{j}  ^{k}} \nabla_{j}\tensor{F}{_{k}  _{n}  _{p}  _{q}} \nabla^{l}\tensor{F}{^{m}  ^{n}} \\
        &+96 \nabla^{i}\tensor{F}{^{j}  ^{k}} \nabla_{j}\tensor{F}{_{i}  _{m}  ^{p}  ^{q}} \nabla_{k}\tensor{F}{_{l}  _{n}  _{p}  _{q}} \nabla^{l}\tensor{F}{^{m}  ^{n}} +
        24 \nabla_{i}\tensor{F}{_{m}  _{n}  ^{p}  ^{q}} \nabla^{i}\tensor{F}{^{j}  ^{k}} \nabla_{l}\tensor{F}{_{j}  _{k}  _{p}  _{q}} \nabla^{l}\tensor{F}{^{m}  ^{n}} \\
        &+16 \nabla_{i}\tensor{F}{^{l}  ^{m}} \nabla^{i}\tensor{F}{^{j}  ^{k}} \nabla_{j}\tensor{F}{_{l}  ^{n}  ^{p}  ^{q}} \nabla_{m}\tensor{F}{_{k}  _{n}  _{p}  _{q}} + 
        16 \nabla^{i}\tensor{F}{^{j}  ^{k}} \nabla_{k}\tensor{F}{_{i}  ^{n}  ^{p}  ^{q}} \nabla^{l}\tensor{F}{_{j}  ^{m}} \nabla_{m}\tensor{F}{_{l}  _{n}  _{p}  _{q}}  \\
        &-48 \nabla^{i}\tensor{F}{^{j}  ^{k}} \nabla_{j}\tensor{F}{_{i}  _{k}  ^{p}  ^{q}} \nabla^{l}\tensor{F}{^{m}  ^{n}} \nabla_{m}\tensor{F}{_{l}  _{n}  _{p}  _{q}} +
        8 \nabla_{i}\tensor{F}{_{j}  _{k}} \nabla^{i}\tensor{F}{^{j}  ^{k}} \nabla^{l}\tensor{F}{^{m}  ^{n}  ^{p}  ^{q}} \nabla_{m}\tensor{F}{_{l}  _{n}  _{p}  _{q}} \\
        &-8  \nabla^{i}\tensor{F}{^{j}  ^{k}} \nabla^{l}\tensor{F}{_{j}  _{k}} \nabla_{m}\tensor{F}{_{l}  _{n}  _{p}  _{q}} \nabla^{m}\tensor{F}{_{i}  ^{n}  ^{p}  ^{q}} -
        16 \nabla_{i}\tensor{F}{_{j}  ^{l}} \nabla^{i}\tensor{F}{^{j}  ^{k}} \nabla_{m}\tensor{F}{_{l}  _{n}  _{p}  _{q}} \nabla^{m}\tensor{F}{_{k}  ^{n}  ^{p}  ^{q}}  \\
        &+48 \nabla^{i}\tensor{F}{^{j}  ^{k}} \nabla_{j}\tensor{F}{_{i}  _{m}  ^{p}  ^{q}} \nabla^{l}\tensor{F}{^{m}  ^{n}} \nabla_{n}\tensor{F}{_{k}  _{l}  _{p}  _{q}} +
        48 \nabla^{i}\tensor{F}{^{j}  ^{k}} \nabla^{l}\tensor{F}{_{j}  ^{m}} \nabla_{n}\tensor{F}{_{l}  _{m}  _{p}  _{q}} \nabla^{n}\tensor{F}{_{i}  _{k}  ^{p}  ^{q}}  \\
        &-48 \nabla^{i}\tensor{F}{^{j}  ^{k}} \nabla^{l}\tensor{F}{_{j}  ^{m}} \nabla_{n}\tensor{F}{_{k}  _{l}  _{p}  _{q}} \nabla^{n}\tensor{F}{_{i}  _{m}  ^{p}  ^{q}} +
        48 \nabla^{i}\tensor{F}{^{j}  ^{k}} \nabla^{l}\tensor{F}{^{m}  ^{n}} \nabla_{p}\tensor{F}{_{k}  _{m}  _{n}  _{q}} \nabla^{p}\tensor{F}{_{i}  _{j}  _{l}  ^{q}} \ ,
\end{split}
\end{equation}

\subsection{$\alpha$ symmetric completion of $\nabla H^4$ terms}
The terms with quartic powers of the three form (\ref{NH4}) have the following expression when written in terms of explicit index contractions
\begin{equation}
\begin{split}
       \nabla H^4=
        &-144\nabla_{i}\tensor{H}{^{m} ^{n} ^{p}}\nabla^{i}\tensor{H}{^{j} ^{k} ^{l}}\nabla_{j}\tensor{H}{_{k}  _{m} ^{q}}\nabla_{l}\tensor{H}{_{n}  _{p}  _{q}} -
        144\nabla_{i}\tensor{H}{_{j} ^{m} ^{n}}\nabla^{i}\tensor{H}{^{j} ^{k} ^{l}}\nabla_{k}\tensor{H}{_{m} ^{p} ^{q}}\nabla_{l}\tensor{H}{_{n}  _{p}  _{q}} \\ 
        &-48\nabla_{i}\tensor{H}{_{j} ^{m} ^{n}}\nabla^{i}\tensor{H}{^{j} ^{k} ^{l}}\nabla_{k}\tensor{H}{_{l} ^{p} ^{q}}\nabla_{m}\tensor{H}{_{n}  _{p}  _{q}} -
        24\nabla_{i}\tensor{H}{_{j}  _{k} ^{m}}\nabla^{i}\tensor{H}{^{j} ^{k} ^{l}}\nabla_{l}\tensor{H}{^{n} ^{p} ^{q}}\nabla_{m}\tensor{H}{_{n}  _{p}  _{q}} \\ 
        &+4/3\nabla_{i}\tensor{H}{_{j}  _{k}  _{l}}\nabla^{i}\tensor{H}{^{j} ^{k} ^{l}}\nabla_{m}\tensor{H}{_{n}  _{p}  _{q}}\nabla^{m}\tensor{H}{^{n} ^{p} ^{q}} - 
        48\nabla^{i}\tensor{H}{^{j} ^{k} ^{l}}\nabla_{j}\tensor{H}{_{k} ^{m} ^{n}}\nabla_{l}\tensor{H}{_{m} ^{p} ^{q}}\nabla_{n}\tensor{H}{_{i}  _{p}  _{q}} \\
        &+96\nabla_{i}\tensor{H}{_{j} ^{m} ^{n}}\nabla^{i}\tensor{H}{^{j} ^{k} ^{l}}\nabla_{k}\tensor{H}{_{m} ^{p} ^{q}}\nabla_{n}\tensor{H}{_{l}  _{p}  _{q}} +
        24\nabla_{i}\tensor{H}{_{j}  _{k} ^{m}}\nabla^{i}\tensor{H}{^{j} ^{k} ^{l}}\nabla_{n}\tensor{H}{_{m}  _{p}  _{q}}\nabla^{n}\tensor{H}{_{l} ^{p} ^{q}} \\ 
        &+48\nabla^{i}\tensor{H}{^{j} ^{k} ^{l}}\nabla_{j}\tensor{H}{_{k} ^{m} ^{n}}\nabla_{m}\tensor{H}{_{n} ^{p} ^{q}}\nabla_{p}\tensor{H}{_{i}  _{l}  _{q}} +
        96\nabla^{i}\tensor{H}{^{j} ^{k} ^{l}}\nabla_{j}\tensor{H}{_{k} ^{m} ^{n}}\nabla_{l}\tensor{H}{_{m} ^{p} ^{q}}\nabla_{p}\tensor{H}{_{i}  _{n}  _{q}} \\ 
        &-96\nabla^{i}\tensor{H}{^{j} ^{k} ^{l}}\nabla_{j}\tensor{H}{_{k} ^{m} ^{n}}\nabla_{m}\tensor{H}{_{l} ^{p} ^{q}}\nabla_{p}\tensor{H}{_{i}  _{n}  _{q}} - 
        48\nabla_{i}\tensor{H}{_{j}  _{k} ^{m}}\nabla^{i}\tensor{H}{^{j} ^{k} ^{l}}\nabla^{n}\tensor{H}{_{l} ^{p} ^{q}}\nabla_{p}\tensor{H}{_{m}  _{n}  _{q}} \\ 
        &-48\nabla^{i}\tensor{H}{^{j} ^{k} ^{l}}\nabla_{j}\tensor{H}{_{i}  _{k} ^{m}}\nabla^{n}\tensor{H}{_{l} ^{p} ^{q}}\nabla_{p}\tensor{H}{_{m}  _{n}  _{q}} + 
        24\nabla_{i}\tensor{H}{_{j} ^{m} ^{n}}\nabla^{i}\tensor{H}{^{j} ^{k} ^{l}}\nabla_{p}\tensor{H}{_{m}  _{n}  _{q}}\nabla^{p}\tensor{H}{_{k}  _{l} ^{q}} \\
        &+96\nabla_{i}\tensor{H}{_{j} ^{m} ^{n}}\nabla^{i}\tensor{H}{^{j} ^{k} ^{l}}\nabla_{p}\tensor{H}{_{l}  _{n}  _{q}}\nabla^{p}\tensor{H}{_{k}  _{m} ^{q}} -
        48\nabla_{i}\tensor{H}{_{j} ^{m} ^{n}}\nabla^{i}\tensor{H}{^{j} ^{k} ^{l}}\nabla^{p}\tensor{H}{_{k}  _{m} ^{q}}\nabla_{q}\tensor{H}{_{l}  _{n}  _{p}} \\
        &-24\nabla^{i}\tensor{H}{^{j} ^{k} ^{l}}\nabla_{j}\tensor{H}{_{i} ^{m} ^{n}}\nabla^{p}\tensor{H}{_{k}  _{m} ^{q}}\nabla_{q}\tensor{H}{_{l}  _{n}  _{p}} +
        24\nabla_{i}\tensor{H}{^{m} ^{n} ^{p}}\nabla^{i}\tensor{H}{^{j} ^{k} ^{l}}\nabla_{q}\tensor{H}{_{l}  _{n}  _{p}}\nabla^{q}\tensor{H}{_{j}  _{k}  _{m}} \ .
\end{split}
\label{Tnabla^4}
\end{equation}
 To find the $\alpha$ symmetric completion, we propose the most general combination of 11 dimensional terms of the form $\nabla {\cal F}^4$. It turns out that there is a basis $\{C_i\}$ of $i=1,\dots,24$ independent terms of that form \cite{Peeters}, which we recall in Appendix \ref{b}. Each $C_i$ leads to an $\alpha$ invariant expression in 10 dimensions. We propose the generic linear combination 
\begin{equation}
        \nabla \mathcal{F}^4 
        = \sum_{i=1}^{24} c_i C_i \ . \label{genericNF4}
\end{equation}
Compactifying this to 10 dimensions, setting the R-R fields to zero, eliminating all ambiguities due to Bianchi identities, and forcing the result to coincide with (\ref{Tnabla^4}), constrains the coefficients as follows 
\begin{eqnarray}
        &&c_3 = \frac 4 3 - 16 \, c_1 \ , \ \ \  c_{11} = -96\, c_1 + \frac {72} 3 + \frac 1 3 c_{10} + 12\,  c_2 + c_4 - \frac 1 3 c_7 \ , \nonumber \\
&& c_8 = 4 + 48\,  c_1 - 6 \, c_6 \ , \ \ \  c_{17}= 192 c_1 - 56 - 2c_{10} + 12 c_6 \ ,  \nonumber        \\
        &&c_{13} = -96 + 288\, c_1 - c_{10} - 36 \, c_2 - 3\,  c_4 + c_7  \ , \ \ \ c_{15} = -12 + 144 c_1 - c_{10} + 3 c_5 + c_7 - 9 c_9 \ , \nonumber \\
        &&c_{14} = 36 - 144\,  c_1 + 18\,  c_2 + 3\, c_4 + \frac 3 2  c_5  + \frac 9 2 c_6 - c_7 \ , \nonumber \\
        &&c_{18}= -16 + 96 c_1 - 18 c_2 - 3 c_4 - \frac 3 2 c_5 - \frac {15} 2 c_6 + c_7 \ , \nonumber \\
        &&c_{19} = -96 + 576 c_1 - 72 c_2 - 12 c_4 - 6 c_5 - 54 c_6 + 8 c_7 \ , \nonumber \\
        &&c_{20} = -1152 c_1 + 192 +2 c_{10} - 36 c_{16} + 144 c_2 + 24 c_4 + 12 c_5 + 36 c_6 - 8 c_7 \ , \nonumber \\
        &&c_{22}= \frac 4 3 - 16 c_1 - c_{16} + 4 c_2 + c_4 + c_5 - c_9 \ , \nonumber \\
        &&c_{23} = -18 c_2 - 6 c_4 - 
        \frac {15} 2  c_5 + \frac 9 2 c_6 - c_7 + 9 c_9 \ , \nonumber \\
        &&c_{24} = -60 + 432 c_1 - c_{10} + 9 c_{16} - 54 c_2 + c_{21} - 9 c_4 - \frac 9  2 c_5 - \frac {27} 2 c_6 + 3 c_7 \ . \label{Cparams}
\end{eqnarray}
While in the previous subsection there was only a two-parameter freedom left, here we find far more ambiguities. This is because for this particular kind of couplings there are many 11 dimensional combinations of terms that vanish when the R-R fields are set to zero in the 10 dimensional $\alpha$ invariant interactions. 

Now we must compare this with the results  in \cite{Peeters}, computed directly from the four-three form scattering amplitudes in 11 dimensions, specifically
\begin{equation}\label{PetersYs}
\begin{split}
  \nabla \mathcal{F}^4
    =& y \left(3 C_5 + C_6 - 9 C_8 + C_9 - 72 C_{12} + 9 C_{14} + 18 C_{17} - 9 C_{18} - 72 C_{19} - 
    C_{22}\right) \\
    &+y_1 Y_1 + \dots + y_9 Y_9 \, .
\end{split}
\end{equation}
The terms $\{Y_i\}$ with $i = 1,\dots,9$ represent a basis of terms that vanish at quartic order in a background field expansion. We list them in \eqref{y1}-\eqref{y9} for completeness. We find that (\ref{PetersYs}) reduces to (\ref{Tnabla^4}) when
\begin{eqnarray}
        &&y = \frac 8 3 \ , \ \ \  y_4 = 18 y_1 + 18 y_2 + 288 y_3 \ , \ \ \ y_6 = -12 + 36 y_1 + 288 y_3\, ,\nonumber\\
        &&y_7 = \frac 1{12} + \frac 1 {16}y_1 - \frac 1{16}y_2 - y_3 \ , \ \ \ y_8 = \frac 1 3 - y_1 + \frac 12 y_2 + 8 y_3 - \frac 1 {36}y_5\, , \label{Yparams}\\ 
        &&y_9 = -12 - 18 y_1 + 288 y_3 - y_5 \ . \nonumber
\end{eqnarray}
Using the constraints (\ref{Cparams})  and (\ref{Yparams}), we find that (\ref{genericNF4})   coincides with  (\ref{PetersYs})  when the unfixed coefficients satisfy 
\begin{equation}
    \begin{split}
        &c_1 = \frac 1 {12} + \frac 1 {16} y_1 - \frac 1 {16} y_2 \ , \ \ \ c_2 = -1 + \frac 1 2 y_1 + \frac 1 {} 36 y_5 \ , \ \ \ c_4 = -4 + 24 y_1 - 6 y_2 + \frac 1 3 y_5 \ , \\
        &c_5 = 12 - 18 y_1 - \frac 1 3 y_5 \ , \ \ \ c_6 = \frac {16}3 + 2 y_1 - 2 y_2\ , \ \ \ c_7 = 12 + 36 y_1 - 18 y_2 + y_5 \ ,\\
        &c_9 = 4 + 5 y_1 - 3 y_2 + \frac 1 9 y_5 \ , \ \ \ c_{10} = -18 y_1 - 18 y_2 - 288 y_3 \ , \\
        & c_{16} = \frac 8 3 + 2 y_1 - 2 y_2 - 32 y_3 \ , \ \ \ c_{21} = 18 y_1 \ .
    \end{split}
\end{equation}

Selecting $y_1=0$, $y_2=\frac 8 3$, $y_3=-\frac{1}{12}$ and $y_5=36$, the  uplift of (\ref{Tnabla^4}) to 11 dimensions is given by
\begin{equation}
    \begin{split}
        \nabla \mathcal{F}^4
        &=-72 \nabla_{\alpha}\tensor{\mathcal{F}}{_{\beta}  ^{\iota}  ^{\kappa}  ^{\lambda}} \nabla^{\alpha}\tensor{\mathcal{F}}{^{\beta}  ^{\gamma}  ^{\delta}  ^{\varepsilon}} \nabla_{\gamma}\tensor{\mathcal{F}}{_{\delta}  _{\iota}  ^{\mu}  ^{\tau}} \nabla_{\varepsilon}\tensor{\mathcal{F}}{_{\kappa}  _{\lambda}  _{\mu}  _{\tau}} - 
        8 \nabla_{\alpha}\tensor{\mathcal{F}}{_{\beta}  _{\gamma}  ^{\iota}  ^{\kappa}} \nabla^{\alpha}\tensor{\mathcal{F}}{^{\beta}  ^{\gamma}  ^{\delta}  ^{\varepsilon}} \nabla_{\delta}\tensor{\mathcal{F}}{_{\varepsilon}  ^{\lambda}  ^{\mu}  ^{\tau}} \nabla_{\iota}\tensor{\mathcal{F}}{_{\kappa}  _{\lambda}  _{\mu}  _{\tau}} \\ 
        &- \frac 1 {12} \nabla_{\alpha}\tensor{\mathcal{F}}{_{\beta}  _{\gamma}  _{\delta}  _{\varepsilon}}   \nabla^{\alpha}\tensor{\mathcal{F}}{^{\beta}  ^{\gamma}  ^{\delta}  ^{\varepsilon}} \nabla_{\iota}\tensor{\mathcal{F}}{_{\kappa}  _{\lambda}  _{\mu}  _{\tau}} \nabla^{\iota}\tensor{\mathcal{F}}{^{\kappa}  ^{\lambda}  ^{\mu}  ^{\tau}} + 
        48 \nabla^{\alpha}\tensor{\mathcal{F}}{^{\beta}  ^{\gamma}  ^{\delta}  ^{\varepsilon}} \nabla_{\beta}\tensor{\mathcal{F}}{_{\gamma}  ^{\iota}  ^{\kappa}  ^{\lambda}} \nabla_{\delta}\tensor{\mathcal{F}}{_{\varepsilon}  _{\iota}  ^{\mu}  ^{\tau}} \nabla_{\kappa}\tensor{\mathcal{F}}{_{\alpha}  _{\lambda}  _{\mu}  _{\tau}} \\ 
        &+16 \nabla_{\alpha}\tensor{\mathcal{F}}{_{\beta}  _{\gamma}  ^{\iota}  ^{\kappa}} \nabla^{\alpha}\tensor{\mathcal{F}}{^{\beta}  ^{\gamma}  ^{\delta}  ^{\varepsilon}} \nabla_{\delta}\tensor{\mathcal{F}}{_{\iota}  ^{\lambda}  ^{\mu}  ^{\tau}} \nabla_{\kappa}\tensor{\mathcal{F}}{_{\varepsilon}  _{\lambda}  _{\mu}  _{\tau}} + 
        \frac 8 3 \nabla_{\alpha}\tensor{\mathcal{F}}{_{\beta}  _{\gamma}  _{\delta}  ^{\iota}} \nabla^{\alpha}\tensor{\mathcal{F}}{^{\beta}  ^{\gamma}  ^{\delta}  ^{\varepsilon}} \nabla_{\kappa}\tensor{\mathcal{F}}{_{\iota}  _{\lambda}  _{\mu}  _{\tau}} \nabla^{\kappa}\tensor{\mathcal{F}}{_{\varepsilon}  ^{\lambda}  ^{\mu}  ^{\tau}} \\ 
        &-48 \nabla^{\alpha}\tensor{\mathcal{F}}{^{\beta}  ^{\gamma}  ^{\delta}  ^{\varepsilon}} \nabla_{\beta}\tensor{\mathcal{F}}{_{\gamma}  ^{\iota}  ^{\kappa}  ^{\lambda}} \nabla_{\iota}\tensor{\mathcal{F}}{_{\delta}  _{\kappa}  ^{\mu}  ^{\tau}} \nabla_{\mu}\tensor{\mathcal{F}}{_{\alpha}  _{\varepsilon}  _{\lambda}  _{\tau}} + 
        48 \nabla^{\alpha}\tensor{\mathcal{F}}{^{\beta}  ^{\gamma}  ^{\delta}  ^{\varepsilon}} \nabla_{\beta}\tensor{\mathcal{F}}{_{\gamma}  ^{\iota}  ^{\kappa}  ^{\lambda}} \nabla_{\iota}\tensor{\mathcal{F}}{_{\delta}  _{\varepsilon}  ^{\mu}  ^{\tau}} \nabla_{\mu}\tensor{\mathcal{F}}{_{\alpha}  _{\kappa}  _{\lambda}  _{\tau}} \\ 
        &-24 \nabla_{\alpha}\tensor{\mathcal{F}}{_{\beta}  _{\gamma}  ^{\iota}  ^{\kappa}} \nabla^{\alpha}\tensor{\mathcal{F}}{^{\beta}  ^{\gamma}  ^{\delta}  ^{\varepsilon}} \nabla^{\lambda}\tensor{\mathcal{F}}{_{\delta}  _{\iota}  ^{\mu}  ^{\tau}} \nabla_{\mu}\tensor{\mathcal{F}}{_{\varepsilon}  _{\kappa}  _{\lambda}  _{\tau}} - 
        24 \nabla^{\alpha}\tensor{\mathcal{F}}{^{\beta}  ^{\gamma}  ^{\delta}  ^{\varepsilon}} \nabla_{\beta}\tensor{\mathcal{F}}{_{\alpha}  _{\gamma}  ^{\iota}  ^{\kappa}} \nabla^{\lambda}\tensor{\mathcal{F}}{_{\delta}  _{\iota}  ^{\mu}  ^{\tau}} \nabla_{\mu}\tensor{\mathcal{F}}{_{\varepsilon}  _{\kappa}  _{\lambda}  _{\tau}} \\ 
        &+24 \nabla_{\alpha}\tensor{\mathcal{F}}{_{\beta}  ^{\iota}  ^{\kappa}  ^{\lambda}} \nabla^{\alpha}\tensor{\mathcal{F}}{^{\beta}  ^{\gamma}  ^{\delta}  ^{\varepsilon}} \nabla_{\mu}\tensor{\mathcal{F}}{_{\varepsilon}  _{\kappa}  _{\lambda}  _{\tau}} \nabla^{\mu}\tensor{\mathcal{F}}{_{\gamma}  _{\delta}  _{\iota}  ^{\tau}} - 
        \frac {16} 3 \nabla^{\alpha}\tensor{\mathcal{F}}{^{\beta}  ^{\gamma}  ^{\delta}  ^{\varepsilon}} \nabla_{\beta}\tensor{\mathcal{F}}{^{\iota}  ^{\kappa}  ^{\lambda}  ^{\mu}} \nabla_{\iota}\tensor{\mathcal{F}}{_{\gamma}  _{\delta}  _{\varepsilon}  ^{\tau}} \nabla_{\tau}\tensor{\mathcal{F}}{_{\alpha}  _{\kappa}  _{\lambda}  _{\mu}} \ .
    \end{split}
\end{equation}
The dimensional reduction of this expression to 10 dimensions leads to the $\alpha$ symmetric completion of (\ref{Tnabla^4}), which is given by
\begin{equation}
    \begin{split}
      \nabla {\cal F}^4  &={\nabla H^4}+H^2F_4^2+F_4^4\, ,
\end{split}
\end{equation}
where ${\nabla H^4}$ was our starting point \eqref{Tnabla^4} and the rest is its $\alpha$ completion, given by
 \begin{equation}
    \begin{split}
 H^2F_4^2&=     -72 \nabla_{i}\tensor{H}{^{m} ^{n}^{p}} \nabla^{i}\tensor{H}{^{j} ^{k}^{l}} \nabla_{j}\tensor{F}{_{k}  _{m}  ^{q}^{r}} \nabla_{l}\tensor{F}{_{n}  _{p}  _{q}  _{r}} + 
        72 \nabla_{i}\tensor{F}{_{k}  _{m}  ^{q}^{r}} \nabla^{i}\tensor{H}{^{j} ^{k}^{l}} \nabla_{j}\tensor{H}{^{m} ^{n}^{p}} \nabla_{l}\tensor{F}{_{n}  _{p}  _{q}  _{r}} \\ 
        &-16 \nabla_{i}\tensor{H}{_{j}  ^{m}^{n}} \nabla^{i}\tensor{H}{^{j} ^{k}^{l}} \nabla_{k}\tensor{F}{_{l}  ^{p} ^{q}^{r}} \nabla_{m}\tensor{F}{_{n}  _{p}  _{q}  _{r}} -\frac 2 3 \nabla_{i}\tensor{H}{_{j}  _{k}  _{l}} \nabla^{i}\tensor{H}{^{j} ^{k}^{l}} \nabla_{m}\tensor{F}{_{n}  _{p}  _{q}  _{r}} \nabla^{m}\tensor{F}{^{n} ^{p} ^{q}^{r}} \\ 
        & +
        96 \nabla_{i}\tensor{F}{_{k}  _{n}  ^{q}^{r}} \nabla^{i}\tensor{H}{^{j} ^{k}^{l}} \nabla_{m}\tensor{F}{_{l}  _{p}  _{q}  _{r}} \nabla^{m}\tensor{H}{_{j}  ^{n}^{p}}-32 \nabla_{i}\tensor{F}{_{j}  _{k}  _{m} ^{r}} \nabla^{i}\tensor{H}{^{j} ^{k}^{l}} \nabla_{l}\tensor{F}{_{n}  _{p}  _{q}  _{r}} \nabla^{m}\tensor{H}{^{n} ^{p}^{q}}  \\
        &+
        48 \nabla_{i}\tensor{F}{_{j}  _{n}  _{p} ^{r}} \nabla^{i}\tensor{H}{^{j} ^{k}^{l}} \nabla_{m}\tensor{F}{_{k}  _{l}  _{q}  _{r}} \nabla^{m}\tensor{H}{^{n} ^{p}^{q}} +\frac {16} 3 \nabla_{i}\tensor{F}{_{j}  _{k}  _{l} ^{r}} \nabla^{i}\tensor{H}{^{j} ^{k}^{l}} \nabla_{m}\tensor{F}{_{n}  _{p}  _{q}  _{r}} \nabla^{m}\tensor{H}{^{n} ^{p}^{q}}\\
        & - 
        48 \nabla^{i}\tensor{H}{^{j} ^{k}^{l}} \nabla_{j}\tensor{H}{^{m} ^{n}^{p}} \nabla_{k}\tensor{F}{_{l}  _{m}  ^{q}^{r}} \nabla_{n}\tensor{F}{_{i}  _{p}  _{q}  _{r}} +32 \nabla^{i}\tensor{H}{^{j} ^{k}^{l}} \nabla_{l}\tensor{F}{_{m}  ^{p} ^{q}^{r}} \nabla^{m}\tensor{H}{_{j}  _{k} ^{n}} \nabla_{n}\tensor{F}{_{i}  _{p}  _{q}  _{r}} \\
        &+ 
        96 \nabla^{i}\tensor{H}{^{j} ^{k}^{l}} \nabla_{k}\tensor{F}{_{l}  _{m}  ^{q}^{r}} \nabla^{m}\tensor{H}{_{j}  ^{n}^{p}} \nabla_{n}\tensor{F}{_{i}  _{p}  _{q}  _{r}} +32 \nabla_{i}\tensor{H}{_{j}  ^{m}^{n}} \nabla^{i}\tensor{H}{^{j} ^{k}^{l}} \nabla_{k}\tensor{F}{_{m}  ^{p} ^{q}^{r}} \nabla_{n}\tensor{F}{_{l}  _{p}  _{q}  _{r}} \\ 
        &+16 \nabla_{i}\tensor{H}{_{j}  _{k} ^{m}} \nabla^{i}\tensor{H}{^{j} ^{k}^{l}} \nabla_{n}\tensor{F}{_{m}  _{p}  _{q}  _{r}} \nabla^{n}\tensor{F}{_{l}  ^{p} ^{q}^{r}} +96 \nabla^{i}\tensor{H}{^{j} ^{k}^{l}} \nabla_{k}\tensor{F}{_{m}  _{n}  ^{q}^{r}} \nabla^{m}\tensor{H}{_{j}  ^{n}^{p}} \nabla_{p}\tensor{F}{_{i}  _{l}  _{q}  _{r}}  \\ 
        &+
        96 \nabla^{i}\tensor{H}{^{j} ^{k}^{l}} \nabla_{j}\tensor{F}{_{k}  _{m}  _{n} ^{r}} \nabla^{m}\tensor{H}{^{n} ^{p}^{q}} \nabla_{p}\tensor{F}{_{i}  _{l}  _{q}  _{r}} -
        96 \nabla_{i}\tensor{F}{_{j}  _{k}  _{n} ^{r}} \nabla^{i}\tensor{H}{^{j} ^{k}^{l}} \nabla^{m}\tensor{H}{^{n} ^{p}^{q}} \nabla_{p}\tensor{F}{_{l}  _{m}  _{q}  _{r}} \\
        &-96 \nabla^{i}\tensor{H}{^{j} ^{k}^{l}} \nabla_{j}\tensor{F}{_{i}  _{k}  _{n} ^{r}} \nabla^{m}\tensor{H}{^{n} ^{p}^{q}} \nabla_{p}\tensor{F}{_{l}  _{m}  _{q}  _{r}} -24 \nabla^{i}\tensor{H}{^{j} ^{k}^{l}} \nabla^{m}\tensor{H}{_{j}  _{k} ^{n}} \nabla_{p}\tensor{F}{_{m}  _{n}  _{q}  _{r}} \nabla^{p}\tensor{F}{_{i}  _{l}  ^{q}^{r}} \\
        &+
        48 \nabla^{i}\tensor{H}{^{j} ^{k}^{l}} \nabla^{m}\tensor{H}{_{j}  _{k} ^{n}} \nabla_{p}\tensor{F}{_{l}  _{m}  _{q}  _{r}} \nabla^{p}\tensor{F}{_{i}  _{n}  ^{q}^{r}} -24 \nabla^{i}\tensor{H}{^{j} ^{k}^{l}} \nabla_{j}\tensor{H}{_{i}  ^{m}^{n}} \nabla_{p}\tensor{F}{_{l}  _{n}  _{q}  _{r}} \nabla^{p}\tensor{F}{_{k}  _{m}  ^{q}^{r}} \\
        &+192 \nabla^{i}\tensor{H}{^{j} ^{k}^{l}} \nabla_{j}\tensor{H}{^{m} ^{n}^{p}} \nabla_{m}\tensor{F}{_{k}  _{n}  ^{q}^{r}} \nabla_{q}\tensor{F}{_{i}  _{l}  _{p}  _{r}} -
        96 \nabla^{i}\tensor{H}{^{j} ^{k}^{l}} \nabla_{k}\tensor{F}{_{m}  _{n}  ^{q}^{r}} \nabla^{m}\tensor{H}{_{j}  ^{n}^{p}} \nabla_{q}\tensor{F}{_{i}  _{l}  _{p}  _{r}} \\
        &-48 \nabla^{i}\tensor{H}{^{j} ^{k}^{l}} \nabla_{j}\tensor{H}{^{m} ^{n}^{p}} \nabla_{k}\tensor{F}{_{l}  _{m}  ^{q}^{r}} \nabla_{q}\tensor{F}{_{i}  _{n}  _{p}  _{r}} -
        96 \nabla^{i}\tensor{H}{^{j} ^{k}^{l}} \nabla_{j}\tensor{H}{^{m} ^{n}^{p}} \nabla_{m}\tensor{F}{_{k}  _{l}  ^{q}^{r}} \nabla_{q}\tensor{F}{_{i}  _{n}  _{p}  _{r}} \\
        &-48 \nabla^{i}\tensor{H}{^{j} ^{k}^{l}} \nabla_{j}\tensor{H}{^{m} ^{n}^{p}} \nabla_{m}\tensor{F}{_{i}  _{k}  ^{q}^{r}} \nabla_{q}\tensor{F}{_{l}  _{n}  _{p}  _{r}} +
        144 \nabla_{i}\tensor{F}{_{k}  _{m}  ^{q}^{r}} \nabla^{i}\tensor{H}{^{j} ^{k}^{l}} \nabla^{m}\tensor{H}{_{j}  ^{n}^{p}} \nabla_{q}\tensor{F}{_{l}  _{n}  _{p}  _{r}} \\
        &-48 \nabla_{i}\tensor{H}{_{j}  ^{m}^{n}} \nabla^{i}\tensor{H}{^{j} ^{k}^{l}} \nabla^{p}\tensor{F}{_{k}  _{m}  ^{q}^{r}} \nabla_{q}\tensor{F}{_{l}  _{n}  _{p}  _{r}} -
        48 \nabla^{i}\tensor{H}{^{j} ^{k}^{l}} \nabla_{j}\tensor{H}{_{i}  ^{m}^{n}} \nabla^{p}\tensor{F}{_{k}  _{m}  ^{q}^{r}} \nabla_{q}\tensor{F}{_{l}  _{n}  _{p}  _{r}} \\ 
        &-16 \nabla_{i}\tensor{F}{_{k}  _{l}  ^{q}^{r}} \nabla^{i}\tensor{H}{^{j} ^{k}^{l}} \nabla_{j}\tensor{H}{^{m} ^{n}^{p}} \nabla_{q}\tensor{F}{_{m}  _{n}  _{p}  _{r}} -
        144 \nabla^{i}\tensor{H}{^{j} ^{k}^{l}} \nabla^{m}\tensor{H}{_{j}  ^{n}^{p}} \nabla_{q}\tensor{F}{_{l}  _{n}  _{p}  _{r}} \nabla^{q}\tensor{F}{_{i}  _{k}  _{m} ^{r}} \\ 
        &+48 \nabla_{i}\tensor{H}{^{m} ^{n}^{p}} \nabla^{i}\tensor{H}{^{j} ^{k}^{l}} \nabla_{q}\tensor{F}{_{l}  _{n}  _{p}  _{r}} \nabla^{q}\tensor{F}{_{j}  _{k}  _{m} ^{r}} -
        48 \nabla^{i}\tensor{H}{^{j} ^{k}^{l}} \nabla_{j}\tensor{F}{_{k}  _{n}  _{p} ^{r}} \nabla^{m}\tensor{H}{^{n} ^{p}^{q}} \nabla_{r}\tensor{F}{_{i}  _{l}  _{m}  _{q}} \\
        &-72 \nabla_{i}\tensor{F}{_{j}  _{m}  _{n} ^{r}} \nabla^{i}\tensor{H}{^{j} ^{k}^{l}} \nabla^{m}\tensor{H}{^{n} ^{p}^{q}} \nabla_{r}\tensor{F}{_{k}  _{l}  _{p}  _{q}} \, .
    \end{split}
\end{equation}
and
\begin{equation}
    \begin{split}
 F_4^4&=    
        72 \nabla_{i}\tensor{F}{_{j}  ^{n} ^{p}^{q}} \nabla^{i}\tensor{F}{^{j} ^{k} ^{l}^{m}} \nabla_{k}\tensor{F}{_{l}  _{n}  ^{r}^{s}} \nabla_{m}\tensor{F}{_{p}  _{q}  _{r}  _{s}} -
        8 \nabla_{i}\tensor{F}{_{j}  _{k}  ^{n}^{p}} \nabla^{i}\tensor{F}{^{j} ^{k} ^{l}^{m}} \nabla_{l}\tensor{F}{_{m}  ^{q} ^{r}^{s}} \nabla_{n}\tensor{F}{_{p}  _{q}  _{r}  _{s}} \\ 
        &   -
        \frac 1 {12} \nabla_{i}\tensor{F}{_{j}  _{k}  _{l}  _{m}} \nabla^{i}\tensor{F}{^{j} ^{k} ^{l}^{m}} \nabla_{n}\tensor{F}{_{p}  _{q}  _{r}  _{s}} \nabla^{n}\tensor{F}{^{p} ^{q} ^{r}^{s}} +48 \nabla^{i}\tensor{F}{^{j} ^{k} ^{l}^{m}} \nabla_{j}\tensor{F}{_{k}  ^{n} ^{p}^{q}} \nabla_{l}\tensor{F}{_{m}  _{n}  ^{r}^{s}} \nabla_{p}\tensor{F}{_{i}  _{q}  _{r}  _{s}}\\ 
        & +
        16 \nabla_{i}\tensor{F}{_{j}  _{k}  ^{n}^{p}} \nabla^{i}\tensor{F}{^{j} ^{k} ^{l}^{m}} \nabla_{l}\tensor{F}{_{n}  ^{q} ^{r}^{s}} \nabla_{p}\tensor{F}{_{m}  _{q}  _{r}  _{s}}  +
        \frac 8 3 \nabla_{i}\tensor{F}{_{j}  _{k}  _{l} ^{n}} \nabla^{i}\tensor{F}{^{j} ^{k} ^{l}^{m}} \nabla_{p}\tensor{F}{_{n}  _{q}  _{r}  _{s}} \nabla^{p}\tensor{F}{_{m}  ^{q} ^{r}^{s}} \\
        &-48 \nabla^{i}\tensor{F}{^{j} ^{k} ^{l}^{m}} \nabla_{j}\tensor{F}{_{k}  ^{n} ^{p}^{q}} \nabla_{n}\tensor{F}{_{l}  _{p}  ^{r}^{s}} \nabla_{r}\tensor{F}{_{i}  _{m}  _{q}  _{s}} +
        48 \nabla^{i}\tensor{F}{^{j} ^{k} ^{l}^{m}} \nabla_{j}\tensor{F}{_{k}  ^{n} ^{p}^{q}} \nabla_{n}\tensor{F}{_{l}  _{m}  ^{r}^{s}} \nabla_{r}\tensor{F}{_{i}  _{p}  _{q}  _{s}} \\
        & - 
        24 \nabla_{i}\tensor{F}{_{j}  _{k}  ^{n}^{p}} \nabla^{i}\tensor{F}{^{j} ^{k} ^{l}^{m}} \nabla^{q}\tensor{F}{_{l}  _{n}  ^{r}^{s}} \nabla_{r}\tensor{F}{_{m}  _{p}  _{q}  _{s}} -24 \nabla^{i}\tensor{F}{^{j} ^{k} ^{l}^{m}} \nabla_{j}\tensor{F}{_{i}  _{k}  ^{n}^{p}} \nabla^{q}\tensor{F}{_{l}  _{n}  ^{r}^{s}} \nabla_{r}\tensor{F}{_{m}  _{p}  _{q}  _{s}}\\ 
        & +
        24 \nabla_{i}\tensor{F}{_{j}  ^{n} ^{p}^{q}} \nabla^{i}\tensor{F}{^{j} ^{k} ^{l}^{m}} \nabla_{r}\tensor{F}{_{m}  _{p}  _{q}  _{s}} \nabla^{r}\tensor{F}{_{k}  _{l}  _{n} ^{s}} -\frac {16} 3 \nabla^{i}\tensor{F}{^{j} ^{k} ^{l}^{m}} \nabla_{j}\tensor{F}{^{n} ^{p} ^{q}^{r}} \nabla_{n}\tensor{F}{_{k}  _{l}  _{m} ^{s}} \nabla_{s}\tensor{F}{_{i}  _{p}  _{q}  _{r}} \ .
    \end{split}
\end{equation}

\section{Conclusions}\label{6}

We have seen that in KK reductions on tori, diffeomorphisms in the higher dimensional theory admit a linear dependence on the internal coordinates $\xi^i = \alpha_{\#}{}^i y^{\#}$, that descends to a global symmetry in the lower dimensional theory, with constrained parameters $\alpha_{\#}{}^i \partial_i = 0$. We called this $\alpha$ symmetry and showed that it constitutes a symmetry principle that fixes couplings in the lower dimensional action.

The role of $\alpha$ symmetry is in fact the opposite to that of $\beta$ symmetry discussed in  \cite{Beta1,Beta2}. In the latter, enhanced (T-duality) symmetries of the lower dimensional theory can be  non-linearly realized in higher dimensions due to the use of constrained parameters. In $\alpha$ symmetry, instead, certain higher dimensional diffeomorphisms, which are typically broken in the KK truncation, are kept and realized in lower dimensions, also due to the use of constrained parameters. While $\beta$ symmetry is a lower dimensional symmetry realized in higher dimensions, $\alpha$ symmetry is a higher dimensional symmetry realized in lower dimensions. Both of them give rise to non-linear global symmetry principles with constrained parameters, that fix couplings.

We showed that $\alpha$ symmetry is sufficient to exactly determine all interactions  at the two-derivative level for KK theories descending from pure Einstein gravity and half-maximal supergravity in $10$ dimensions, as well as in the circle reduction of maximal supergravity in $11$ dimensions to Type IIA supergravity in $10$ dimensions. In the last two cases, $\alpha$ symmetry must be complemented with $\beta$ symmetry to fully fix all couplings. 

An application of $\alpha$ symmetry that we started exploring here is the prediction of higher derivative  R-R couplings  in Type IIA, as an $\alpha$ symmetric completion of the NS-NS sector. The idea is that $\alpha$ symmetry mixes NS-NS and R-R fields, so that if  one sector is known, the other can be predicted. We investigated to what extent this is the case, for the specific terms contributing to four-point scattering amplitudes at order $\zeta(3)\, \alpha'{}^3\,  t_8 t_8 R^{(-) 4}$. Although $\alpha$ symmetry is highly constraining, there exist $\alpha$ invariants that vanish when the R-R fields are set to zero, giving rise to ambiguities in the procedure, that we have classified in this case. Eliminating them by cross checking with explicit $11$ dimensional four-point scattering amplitude computations, we proposed in Section 5 the full R-R completion of the  $\zeta(3)\, \alpha'{}^3\,  t_8 t_8 R^{(-) 4}$ terms for Type IIA superstring theory in $10$ dimensions.

An obvious extension of our work would be the study of the R-R completion of the  NS-NS $\alpha'{}^3$ couplings in Type IIA involving more than four fields  \cite{Liu2}-\cite{Wulff2}, or even of the full eight-derivative NS-NS effective action computed in \cite{Garousi}. 
Other cases that would be worth exploring are the $\alpha$ symmetries of maximal supergravities in lower than 10 dimensions, which could be analyzed from the point of view of partial truncations of Exceptional Field Theories \cite{EFT}.

Unlike other symmetries (e.g. supersymmetry or $\beta$ symmetry), we do not expect $\alpha$ symmetry to be deformed by higher derivative corrections, due to its origin from compactified diffeomorphisms. More specifically, we expect the existence of a scheme in which $\alpha$ transformations receive no $\alpha'$ corrections. However, if $\alpha$ symmetry receives genuine higher derivative corrections, it might become an interesting tool to assess quantum corrections to diffeomorphisms.

All theories descending through dimensional reduction from a higher dimensional theory that is invariant under general coordinate transformations  are expected to be $\alpha$ symmetric. Hence, 
the presence of this kind of symmetry in a phenomenological model of cosmology or particle physics might  indicate the existence of extra dimensions.

\bigskip 

\noindent{\bf Acknowledgements:} We warmly thank W. Baron for useful comments and discussions. D.M. thanks J. J. Fernandez-Melgarejo and Universidad de Murcia for hospitality. Support by Consejo Nacional de Investigaciones Cient\'ificas y T\'ecnicas (CONICET) and Universidad de Buenos Aires (UBA) is also gratefully acknowledged. 

\bigskip

\begin{appendix}
\section{Conventions and definitions}\label{a}
In this Appendix we introduce the notation used throughout the paper. The indices labeling the original $D$ dimensional  and the reduced $n$ dimensional external and $d$ dimensional internal space-time and tangent coordinates are defined in table \ref{tab:tab}.
\begin{table}[ht]
\centering 
\begin{tabular}{|c|c|c|c|c|} 
\hline 
\ \ Dimension \ \  &\ \    Type \ \   & \ \   Index \ \    \\  
\hline 
\multirow{2}{*}{original $D$}
    & space-time    & $\mu,\nu,\rho,\dots$  \\ 
    & tangent       & $\alpha,\beta,\gamma,\dots$  \\
\hline 
\multirow{2}{*}{external $n$}
    & space-time    & $i,j,k,\dots$  \\
    & tangent       & $a,b,c,\dots$  \\
\hline 
\multirow{2}{*}{internal $d$ }
    & space-time    & $m,n,p,\dots$  \\
    & tangent       & ${\overline a}, {\overline b}, {\overline c},\dots$ \\
\hline 
\multirow{2}{*}{double internal $d+d$}
    & space-time    & $M,N,P,\dots$  \\
    & tangent       & $A,B,C,\dots$  \\
\hline 
\end{tabular}
\caption{Labels of space-time and tangent original $D$ dimensional, external  $n$ dimensional and internal $d$ and double $d+d$ dimensional coordinates}\label{tab:tab}
\end{table}

The Lie derivative of an $n$ dimensional tensor is given by
\be
L_\xi V_i{}^j=\xi^k\partial_k V_i{}^j+\partial_i\xi^k V_k{}^j-\partial_k\xi^j V_i{}^k\, .
\ee
The Christoffel connection is defined in terms of the metric as
\be
\Gamma_{ij}^k=\frac12g^{kl}\left(\partial_i g_{lj}+\partial_j g_{il}-\partial_l g_{ij}\right)\, .
\ee
It transforms anomalously under infinitesimal diffeomorphisms
\be
\delta_\xi\Gamma_{ij}^k=L_\xi\Gamma_{ij}^k+\partial_i\partial_j\xi^k\, ,
\ee
and allows to define the covariant derivative
\be
\nabla_k V_i{}^j=\partial_kV_i{}^j-\Gamma_{ki}^l V_l{}^j +\Gamma_{kl}^j V_i{}^l\, .\ee

The Riemann tensor
\be
R^k{}_{lij}=\partial_i\Gamma_{lj}^k-\partial_j\Gamma^k_{li}+\Gamma^k_{in}\Gamma^n_{lj}-\Gamma^k_{jn}\Gamma^n_{li}
\ee
has the following symmetries and Bianchi identities
\be
R_{klij}=g_{kn}R^n{}_{lij}=R_{([kl][ij])}\, ,\qquad R^k{}_{[lij]}=0\, ,\qquad \nabla_{[i}R_{jk]}{}^l{}_m=0\, .
\ee
The Ricci tensor and scalar are defined as
\be
R_{ij}=R^k{}_{ikj}\qquad {\rm and}\qquad R=g^{ij}R_{ij}\, .
\ee

The metric and its inverse can be written in terms of a frame field and its inverse as
\be
g_{ij}=e_i{}^a g_{ab}e_j{}^b\, \qquad{\rm and}\qquad g^{ij}=e^i{}_a g^{ab}e^j{}_b\, ,
\ee
where $g_{ab}$ is the Minkowski metric. The frame field and its inverse satisfy the following identities
\be
e^i{}_a e_i{}^b=\delta_a^b\, , \qquad e_i{}^a e^j{}_a=\delta_i^j\, ,\qquad e^i{}_a=g^{ij}e_j{}^b g_{ab}\, ,
\ee
and they change under infinitesimal diffeomorphisms and Lorentz transformations as 
\be
\delta e_i{}^a=L_\xi e_i{}^a+e_i{}^b\Lambda_{b}{}^{a}\, ,\qquad \delta e^i{}_a=L_\xi e^i{}_a-\Lambda_a{}^b e^i{}_b\, ,\qquad  \Lambda_{ab}=\Lambda_a{}^c g_{cb}=-\Lambda_{ba}\, .
\ee
The spin connection 
\be
w_{ab}{}^c=D_a e_i{}^c e^i{}_b-e^i{}_a\Gamma_{ij}^k e_k{}^c e^j{}_b\, , \label{spinconnection}
\ee
transforms as 
\be
\delta_\Lambda w_{ab}{}^c=D_a\Lambda_{bc}+w_{db}{}^c\Lambda^d{}_a+w_{adc}\Lambda^d{}_b + w_{abd}\Lambda^d{}_c
\ee
and turns tangent space partial derivatives $D_a=e^i{}_a\partial_i$ into Lorentz covariant derivatives
\be
\nabla_a T_b=D_a T_b +w_{ab}{}^c T_c\, .
\ee
 The Riemann tensor can be written in terms of the spin connection as
\be
R_{abcd} = 2 D_{[a}w_{ b] cd} + 2w_{[ab]}{}^e w_{ecd} + 2 w_{[\underline{a} c}{}^e w_{\underline{b}] ed}\, 
\ee
and its symmetries and Bianchi identities are
\be
R_{abcd} = R_{[ab] [cd]}\, ,\qquad R_{abcd} = R_{cdab} \ , \ \ \  \ R_{[abc] d} = 0 \, .\  \label{BI2}
\ee
The Ricci tensor and scalar curvature are given by the traces
\be
R_{ab} = R^c{}_{acb} \ , \ \ \  R = R_a{}^a \ ,  \ \ \ 
R_{[ab]} = 0\ . \label{BI3}
\ee

We define an internal vielbein, transforming as a scalar under diffeomorphisms, and as follows under the internal Lorentz symmetry
\be
\delta_\Lambda \nu_m{}^{\bar a} = \nu_m{}^{\bar b} \Lambda_{\bar b}{}^{\bar a} \ .
\ee
Its associated spin connection
\be
\Omega_{a\bar b\bar c}=\nu_{m\bar b}D_a \nu_{\bar c}{}^m \ , \label{om}
\ee
transforms as
\be\delta_\Lambda\Omega_{a\bar b\bar c}=D_a\Lambda_{\bar b\bar c}+\Lambda^d{}_a\Omega_{d\bar b\bar c}+\Lambda^{\bar d}{}_{\bar b}\Omega_{a\bar d\bar c}+\Lambda^{\bar d}{}_{\bar c}\Omega_{a\bar b\bar d} \ ,\ee
turning tangent space partial derivatives $D_a$ into covariant derivatives
\be
\nabla_aT_b{}^{\bar c}=D_aT_b{}^{\bar c}+w_{ab}{}^dT_d{}^{\bar c}+\Omega_{a}{}^{\bar c}{}_{\bar d}T_{b}{}^{\bar d}\ .
\ee

In the paper we also consider an internal double vielben transforming as a scalar under diffeomorphisms and as  follows with respect to Lorentz and gauge symmetries 
\be
\delta\nu_M{}^A={\cal L}_\xi{\nu}_{M}{}^A+\nu_{M}{}^{B}\Lambda_B{}^{A}+f^A{}_{BC}\Lambda^B\nu_M{}^C\, .
\ee
We defined a gauge invariant  internal connection
\be
\Omega_{a B C}=\nu_{M B}D_a\nu^M{}_{ C}-f_{ B CD}A_a{}^D\ ,
\ee
transforming as
\ba
\delta\Omega_{a B C}&=&\Lambda^d{}_a\Omega_{dBC}+ D_a \Lambda_{BC} + 2\Omega_{a D[ C}\Lambda^{ D}{}_{ B]}\ , \label{noncovariantOmega}
\ea
and defining the following covariant derivatives on Lorentz covariant tensors
\be
\nabla_a T_{b}{}^C = D_a T_b{}^C + w_{ab}{}^d T_d{}^C + \Omega_{a}{}^C{}_D T_b{}^D \ .
\ee
 \section{Basis of terms in 11 dimensions}\label{b}
For completeness we write here the basis of independent terms defined in \cite{Peeters}. 

For the terms of the form ${\cal R}^2 \nabla {\cal F}^2$ we have
\ba
        B_1&=&\tensor{\mathcal{R}}{_{\alpha}  _{\beta}  _{\gamma}  _{\delta}}   \tensor{\mathcal{R}}{_{\kappa}  _{\lambda}  _{\mu}  _{\varepsilon}}  \nabla^{\gamma}\tensor{\mathcal{F}}{^{\beta}  ^{\delta}  ^{\lambda}  ^{\iota}} \nabla^{\kappa}\tensor{\mathcal{F}}{^{\alpha}  ^{\mu}  ^{\varepsilon}  _{\iota}}\qquad\qquad B_{13}=\tensor{\mathcal{R}}{_{\alpha}  _{\beta}  _{\gamma}  _{\delta}}   \tensor{\mathcal{R}}{_{\kappa}  ^{\alpha}  _{\lambda}  ^{\gamma}}  \nabla_{\iota}\tensor{\mathcal{F}}{^{\beta}  ^{\lambda}  _{\mu}  _{\varepsilon}} \nabla^{\iota}\tensor{\mathcal{F}}{^{\delta}  ^{\kappa}  ^{\mu}  ^{\varepsilon}}\nn\\
        B_2&=&\tensor{\mathcal{R}}{_{\alpha}  _{\beta}  _{\gamma}  _{\delta}}   \tensor{\mathcal{R}}{_{\kappa}  _{\lambda}  _{\mu}  _{\varepsilon}}  \nabla^{\varepsilon}\tensor{\mathcal{F}}{^{\beta}  ^{\delta}  ^{\lambda}  ^{\iota}} \nabla^{\kappa}\tensor{\mathcal{F}}{^{\alpha}  ^{\gamma}  ^{\mu}  _{\iota}}\qquad\qquad B_{14}=\tensor{\mathcal{R}}{_{\alpha}  _{\beta}  _{\gamma}  _{\delta}}   \tensor{\mathcal{R}}{_{\kappa}  ^{\alpha}  _{\lambda}  ^{\gamma}}  \nabla_{\iota}\tensor{\mathcal{F}}{^{\beta}  ^{\delta}  _{\mu}  _{\varepsilon}} \nabla^{\iota}\tensor{\mathcal{F}}{^{\kappa}  ^{\lambda}  ^{\mu}  ^{\varepsilon}}\nn\\
        B_3&=&\tensor{\mathcal{R}}{_{\alpha}  _{\beta}  _{\gamma}  _{\delta}}   \tensor{\mathcal{R}}{_{\kappa}  _{\lambda}  _{\mu}  _{\varepsilon}}  \nabla^{\kappa}\tensor{\mathcal{F}}{^{\alpha}  ^{\gamma}  ^{\mu}  _{\iota}} \nabla^{\lambda}\tensor{\mathcal{F}}{^{\beta}  ^{\delta}  ^{\varepsilon}  ^{\iota}}\qquad\qquad B_{15}=\tensor{\mathcal{R}}{_{\alpha}  _{\beta}  _{\gamma}  _{\delta}}   \tensor{\mathcal{R}}{_{\kappa}  ^{\alpha}  _{\lambda}  ^{\gamma}}  \nabla^{\beta}\tensor{\mathcal{F}}{^{\lambda}  _{\mu}  _{\varepsilon}  _{\iota}} \nabla^{\kappa}\tensor{\mathcal{F}}{^{\delta}  ^{\mu}  ^{\varepsilon}  ^{\iota}}\nn\\
        B_4&=&\tensor{\mathcal{R}}{_{\alpha}  _{\beta}  _{\gamma}  _{\delta}}   \tensor{\mathcal{R}}{_{\kappa}  _{\lambda}  _{\mu}  _{\varepsilon}}  \nabla_{\iota}\tensor{\mathcal{F}}{^{\gamma}  ^{\delta}  ^{\mu}  ^{\varepsilon}} \nabla^{\lambda}\tensor{\mathcal{F}}{^{\iota}  ^{\alpha}  ^{\beta}  ^{\kappa}}\qquad \qquad B_{16}=\tensor{\mathcal{R}}{^{\alpha}  ^{\delta}  ^{\beta}  ^{\gamma}}   \tensor{\mathcal{R}}{_{\beta}  ^{\varepsilon}  _{\alpha}  ^{\iota}}  \nabla_{\delta}\tensor{\mathcal{F}}{_{\gamma}  ^{\kappa}  ^{\lambda}  ^{\mu}} \nabla_{\iota}\tensor{\mathcal{F}}{_{\varepsilon}  _{\kappa}  _{\lambda}  _{\mu}}\nn\\
        B_5&=&\tensor{\mathcal{R}}{_{\alpha}  _{\beta}  _{\gamma}  _{\delta}}   \tensor{\mathcal{R}}{_{\kappa}  _{\lambda}  _{\mu}  ^{\delta}}  \nabla^{\alpha}\tensor{\mathcal{F}}{^{\beta}  ^{\gamma}  _{\varepsilon}  _{\iota}} \nabla^{\kappa}\tensor{\mathcal{F}}{^{\lambda}  ^{\mu}  ^{\varepsilon}  ^{\iota}}\qquad\qquad B_{17}=\tensor{\mathcal{R}}{^{\alpha}  ^{\delta}  ^{\beta}  ^{\gamma}}   \tensor{\mathcal{R}}{_{\beta}  ^{\varepsilon}  _{\alpha}  ^{\iota}}  \nabla_{\varepsilon}\tensor{\mathcal{F}}{_{\gamma}  ^{\kappa}  ^{\lambda}  ^{\mu}} \nabla_{\iota}\tensor{\mathcal{F}}{_{\delta}  _{\kappa}  _{\lambda}  _{\mu}}\nn\\
        B_6&=&\tensor{\mathcal{R}}{_{\alpha}  _{\beta}  _{\gamma}  _{\delta}}   \tensor{\mathcal{R}}{_{\kappa}  _{\lambda}  _{\mu}  ^{\delta}}  \nabla^{\alpha}\tensor{\mathcal{F}}{^{\beta}  ^{\kappa}  _{\varepsilon}  _{\iota}} \nabla^{\gamma}\tensor{\mathcal{F}}{^{\lambda}  ^{\mu}  ^{\varepsilon}  ^{\iota}}\qquad\qquad B_{18}=\tensor{\mathcal{R}}{_{\alpha}  _{\beta}  _{\gamma}  _{\delta}}   \tensor{\mathcal{R}}{_{\kappa}  ^{\alpha}  _{\lambda}  ^{\gamma}}  \nabla^{\delta}\tensor{\mathcal{F}}{^{\beta}  ^{\mu}  ^{\varepsilon}  ^{\iota}} \nabla_{\iota}\tensor{\mathcal{F}}{^{\kappa}  ^{\lambda}  _{\mu}  _{\varepsilon}}\nn\\
        B_7&=&\tensor{\mathcal{R}}{_{\alpha}  _{\beta}  _{\gamma}  _{\delta}}   \tensor{\mathcal{R}}{_{\kappa}  _{\lambda}  _{\mu}  ^{\delta}}  \nabla^{\alpha}\tensor{\mathcal{F}}{^{\beta}  ^{\kappa}  _{\varepsilon}  _{\iota}} \nabla^{\mu}\tensor{\mathcal{F}}{^{\gamma}  ^{\lambda}  ^{\varepsilon}  ^{\iota}}\qquad\qquad B_{19}=\tensor{\mathcal{R}}{_{\alpha}  _{\beta}  _{\gamma}  _{\delta}}   \tensor{\mathcal{R}}{_{\kappa}  _{\lambda}  ^{\gamma}  ^{\delta}}  \nabla_{\iota}\tensor{\mathcal{F}}{^{\alpha}  ^{\kappa}  _{\mu}  _{\varepsilon}} \nabla^{\iota}\tensor{\mathcal{F}}{^{\beta}  ^{\lambda}  ^{\mu}  ^{\varepsilon}}\label{bi}\  \  \   \ \\
        B_8&=&\tensor{\mathcal{R}}{_{\alpha}  _{\beta}  _{\gamma}  _{\delta}}   \tensor{\mathcal{R}}{_{\kappa}  _{\lambda}  _{\mu}  ^{\delta}}  \nabla^{\alpha}\tensor{\mathcal{F}}{^{\gamma}  ^{\kappa}  _{\varepsilon}  _{\iota}} \nabla^{\beta}\tensor{\mathcal{F}}{^{\lambda}  ^{\mu}  ^{\varepsilon}  ^{\iota}}\qquad\qquad B_{20}=\tensor{\mathcal{R}}{_{\alpha}  _{\beta}  _{\gamma}  _{\delta}}   \tensor{\mathcal{R}}{_{\kappa}  _{\lambda}  ^{\gamma}  ^{\delta}}  \nabla^{\alpha}\tensor{\mathcal{F}}{^{\kappa}  _{\mu}  _{\varepsilon}  _{\iota}} \nabla^{\beta}\tensor{\mathcal{F}}{^{\lambda}  ^{\mu}  ^{\varepsilon}  ^{\iota}}\nn\\
        B_9&=&\tensor{\mathcal{R}}{_{\alpha}  _{\beta}  _{\gamma}  _{\delta}}   \tensor{\mathcal{R}}{_{\kappa}  _{\lambda}  _{\mu}  ^{\delta}}  \nabla^{\alpha}\tensor{\mathcal{F}}{^{\gamma}  ^{\kappa}  _{\varepsilon}  _{\iota}} \nabla^{\lambda}\tensor{\mathcal{F}}{^{\beta}  ^{\mu}  ^{\varepsilon}  ^{\iota}}\qquad\qquad B_{21}=\tensor{\mathcal{R}}{_{\alpha}  _{\beta}  _{\gamma}  _{\delta}}   \tensor{\mathcal{R}}{_{\kappa}  _{\lambda}  ^{\gamma}  ^{\delta}}  \nabla^{\alpha}\tensor{\mathcal{F}}{^{\kappa}  _{\mu}  _{\varepsilon}  _{\iota}} \nabla^{\lambda}\tensor{\mathcal{F}}{^{\beta}  ^{\mu}  ^{\varepsilon}  ^{\iota}}\nn\\
        B_{10}&=&\tensor{\mathcal{R}}{_{\alpha}  _{\beta}  _{\gamma}  _{\delta}}   \tensor{\mathcal{R}}{_{\kappa}  _{\lambda}  _{\mu}  ^{\delta}}  \nabla_{\iota}\tensor{\mathcal{F}}{^{\gamma}  ^{\kappa}  ^{\mu}  _{\varepsilon}} \nabla^{\iota}\tensor{\mathcal{F}}{^{\alpha}  ^{\beta}  ^{\lambda}  ^{\varepsilon}}\qquad\qquad B_{22}=\tensor{\mathcal{R}}{_{\alpha}  _{\beta}  _{\gamma}  _{\delta}}   \tensor{\mathcal{R}}{_{\kappa}  ^{\alpha}  ^{\gamma}  ^{\delta}}  \nabla^{\beta}\tensor{\mathcal{F}}{_{\lambda}  _{\mu}  _{\varepsilon}  _{\iota}} \nabla^{\kappa}\tensor{\mathcal{F}}{^{\lambda}  ^{\mu}  ^{\varepsilon}  ^{\iota}}\nn\\
        B_{11}&=&\tensor{\mathcal{R}}{_{\alpha}  _{\beta}  _{\gamma}  _{\delta}}   \tensor{\mathcal{R}}{_{\kappa}  _{\lambda}  _{\mu}  ^{\delta}}  \nabla_{\varepsilon}\tensor{\mathcal{F}}{^{\alpha}  ^{\beta}  ^{\lambda}  _{\iota}} \nabla^{\iota}\tensor{\mathcal{F}}{^{\gamma}  ^{\kappa}  ^{\mu}  ^{\varepsilon}}\qquad\qquad B_{23}=\tensor{\mathcal{R}}{_{\alpha}  _{\beta}  _{\gamma}  _{\delta}}   \tensor{\mathcal{R}}{_{\kappa}  ^{\alpha}  ^{\gamma}  ^{\delta}}  \nabla_{\iota}\tensor{\mathcal{F}}{^{\beta}  _{\lambda}  _{\mu}  _{\varepsilon}} \nabla^{\iota}\tensor{\mathcal{F}}{^{\kappa}  ^{\lambda}  ^{\mu}  ^{\varepsilon}}\nn\\
        B_{12}&=&\tensor{\mathcal{R}}{_{\alpha}  _{\beta}  _{\gamma}  _{\delta}}   \tensor{\mathcal{R}}{_{\kappa}  _{\lambda}  _{\mu}  ^{\delta}}  \nabla^{\gamma}\tensor{\mathcal{F}}{^{\kappa}  ^{\lambda}  _{\varepsilon}  _{\iota}} \nabla^{\mu}\tensor{\mathcal{F}}{^{\beta}  ^{\alpha}  ^{\varepsilon}  ^{\iota}}\qquad\qquad B_{24}=\tensor{\mathcal{R}}{_{\alpha}  _{\beta}  _{\gamma}  _{\delta}}   \tensor{\mathcal{R}}{^{\alpha}  ^{\beta}  ^{\gamma}  ^{\delta}}  \nabla_{\kappa}\tensor{\mathcal{F}}{_{\lambda}  _{\mu}  _{\varepsilon}  _{\iota}} \nabla^{\lambda}\tensor{\mathcal{F}}{^{\kappa}  ^{\mu}  ^{\varepsilon}  ^{\iota}}\nn
    \ea
with the following combinations that vanish for four-point amplitudes
\begin{eqnarray}
Z_1&=&48 B_1 + 48 B_2 - 48 B_3 + 36 B_4 + 96 B_6 + 48 B_7 - 48 B_8 + 96 B_{10} +  12 B_{12} \nn\\
        &&+24 B_{13} - 12 B_{14} + 8 B_{15} + 8 B_{16} - 16 B_{17} + 6 B_{19} +  2 B_{22} + B_{24}\label{z1}\\
        Z_2 &=& -48 B_1 - 48 B_2 - 24 B_4 - 24 B_5 + 48 B_6 - 48 B_8 - 24 B_9 - 72 B_{10} - 24 B_{13} \nn\\
        &&+24 B_{14} - B_{22} + 4 B_{23}\\
        Z_3 &=& 12 B_1 + 12 B_2 - 24 B_3 + 9 B_4 + 48 B_6 + 24 B_7 - 24 B_8 + 24 B_{10} + 6 B_{12} + 6 B_{13} \nn \\
        &&+4 B_{15} - 4 B_{17} + 3 B_{19} + 2 B_{21}\\
        Z_4 &=& 12 B_1 + 12 B_2 - 12 B_3 + 9 B_4 + 24 B_6 + 12 B_7 - 12 B_8 + 24 B_{10} +   3 B_{12} + 6 B_{13}  \nn \\
        &&+4 B_{15} - 4 B_{17} + 2 B_{20}\\
        Z_5&=&4 B_3 - 8 B_6 - 4 B_7 + 4 B_8 - B_{12} - 2 B_{14} + 4 B_{18}\\
        Z_6&=&B_4 + 2 B_{11}\label{z6}
\end{eqnarray}

For the terms of the form $\nabla {\cal F}^4$ we have
\ba
        C_1&=&\nabla_{\alpha}\tensor{\mathcal{F}}{_{\beta}  _{\gamma}  _{\delta}  _{\varepsilon}} \nabla^{\alpha}\tensor{\mathcal{F}}{^{\beta}  ^{\gamma}  ^{\delta}  ^{\varepsilon}} \nabla_{\iota}\tensor{\mathcal{F}}{_{\mu}  _{\mu_1}  _{\kappa}  _{\lambda}} \nabla^{\iota}\tensor{\mathcal{F}}{^{\mu}  ^{\mu_1}  ^{\kappa}  ^{\lambda}}\qquad C_{13}=\nabla_{\alpha}\tensor{\mathcal{F}}{_{\beta}  _{\gamma}  _{\delta}  _{\varepsilon}} \nabla^{\alpha}\tensor{\mathcal{F}}{^{\beta}  _{\iota}  _{\mu}  _{\mu_1}} \nabla^{\gamma}\tensor{\mathcal{F}}{^{\iota}  ^{\mu}  _{\kappa}  _{\lambda}} \nabla^{\delta}\tensor{\mathcal{F}}{^{\varepsilon}  ^{\mu_1}  ^{\kappa}  ^{\lambda}}\nn\\
        C_2&=&\nabla_{\alpha}\tensor{\mathcal{F}}{_{\beta}  _{\gamma}  _{\delta}  _{\varepsilon}} \nabla^{\alpha}\tensor{\mathcal{F}}{^{\beta}  ^{\gamma}  ^{\delta}  _{\iota}} \nabla^{\varepsilon}\tensor{\mathcal{F}}{_{\mu}  _{\mu_1}  _{\kappa}  _{\lambda}} \nabla^{\iota}\tensor{\mathcal{F}}{^{\mu}  ^{\mu_1}  ^{\kappa}  ^{\lambda}}\qquad C_{14}=\nabla_{\alpha}\tensor{\mathcal{F}}{_{\beta}  _{\gamma}  _{\delta}  _{\varepsilon}} \nabla^{\alpha}\tensor{\mathcal{F}}{^{\beta}  _{\iota}  _{\mu}  _{\mu_1}} \nabla_{\kappa}\tensor{\mathcal{F}}{^{\gamma}  ^{\delta}  ^{\iota}  _{\lambda}} \nabla^{\kappa}\tensor{\mathcal{F}}{^{\varepsilon}  ^{\mu}  ^{\mu_1}  ^{\lambda}}\nn\\
        C_3&=&\nabla_{\alpha}\tensor{\mathcal{F}}{_{\beta}  _{\gamma}  _{\delta}  _{\varepsilon}} \nabla^{\alpha}\tensor{\mathcal{F}}{^{\beta}  ^{\gamma}  ^{\delta}  _{\iota}} \nabla_{\mu}\tensor{\mathcal{F}}{^{\varepsilon}  _{\mu_1}  _{\kappa}  _{\lambda}} \nabla^{\mu}\tensor{\mathcal{F}}{^{\iota}  ^{\mu_1}  ^{\kappa}  ^{\lambda}}\qquad C_{15}=\nabla_{\alpha}\tensor{\mathcal{F}}{_{\beta}  _{\gamma}  _{\delta}  _{\varepsilon}} \nabla^{\alpha}\tensor{\mathcal{F}}{_{\iota}  _{\mu}  _{\mu_1}  _{\kappa}} \nabla^{\beta}\tensor{\mathcal{F}}{^{\gamma}  ^{\delta}  ^{\iota}  _{\lambda}} \nabla^{\mu}\tensor{\mathcal{F}}{^{\varepsilon}  ^{\mu_1}  ^{\kappa}  ^{\lambda}}\nn\\
        C_4&=&\nabla_{\alpha}\tensor{\mathcal{F}}{_{\beta}  _{\gamma}  _{\delta}  _{\varepsilon}} \nabla^{\alpha}\tensor{\mathcal{F}}{^{\beta}  ^{\gamma}  _{\iota}  _{\mu}} \nabla^{\delta}\tensor{\mathcal{F}}{^{\varepsilon}  _{\mu_1}  _{\kappa}  _{\lambda}} \nabla^{\iota}\tensor{\mathcal{F}}{^{\mu}  ^{\mu_1}  ^{\kappa}  ^{\lambda}}\qquad C_{16}=\nabla_{\alpha}\tensor{\mathcal{F}}{_{\beta}  _{\gamma}  _{\delta}  _{\varepsilon}} \nabla^{\alpha}\tensor{\mathcal{F}}{_{\iota}  _{\mu}  _{\mu_1}  _{\kappa}} \nabla^{\beta}\tensor{\mathcal{F}}{^{\iota}  ^{\mu}  ^{\mu_1}  _{\lambda}} \nabla^{\kappa}\tensor{\mathcal{F}}{^{\gamma}  ^{\delta}  ^{\varepsilon}  ^{\lambda}}\nn\\
        C_5&=&\nabla_{\alpha}\tensor{\mathcal{F}}{_{\beta}  _{\gamma}  _{\delta}  _{\varepsilon}} \nabla^{\alpha}\tensor{\mathcal{F}}{^{\beta}  _{\iota}  _{\mu}  _{\mu_1}} \nabla^{\iota}\tensor{\mathcal{F}}{^{\mu}  ^{\mu_1}  ^{\kappa}  ^{\lambda}} \nabla_{\kappa}\tensor{\mathcal{F}}{^{\gamma}  ^{\delta}  ^{\varepsilon}  _{\lambda}}\qquad C_{17}=\nabla_{\alpha}\tensor{\mathcal{F}}{_{\beta}  _{\gamma}  _{\delta}  _{\varepsilon}} \nabla^{\beta}\tensor{\mathcal{F}}{^{\alpha}  ^{\gamma}  _{\iota}  _{\mu}} \nabla^{\kappa}\tensor{\mathcal{F}}{^{\varepsilon}  ^{\mu}  ^{\mu_1}  ^{\lambda}} \nabla_{\mu_1}\tensor{\mathcal{F}}{^{\delta}  ^{\iota}  _{\kappa}  _{\lambda}}\nn\\
        C_6&=&\nabla_{\alpha}\tensor{\mathcal{F}}{_{\beta}  _{\gamma}  _{\delta}  _{\varepsilon}} \nabla^{\alpha}\tensor{\mathcal{F}}{^{\beta}  _{\iota}  _{\mu}  _{\mu_1}} \nabla_{\kappa}\tensor{\mathcal{F}}{^{\gamma}  ^{\delta}  ^{\varepsilon}  _{\lambda}} \nabla^{\lambda}\tensor{\mathcal{F}}{^{\iota}  ^{\mu}  ^{\mu_1}  ^{\kappa}}\qquad C_{18}=\nabla_{\alpha}\tensor{\mathcal{F}}{_{\beta}  _{\gamma}  _{\delta}  _{\varepsilon}} \nabla^{\beta}\tensor{\mathcal{F}}{^{\alpha}  _{\iota}  _{\mu}  _{\mu_1}} \nabla_{\kappa}\tensor{\mathcal{F}}{^{\gamma}  ^{\delta}  ^{\iota}  _{\lambda}} \nabla^{\lambda}\tensor{\mathcal{F}}{^{\varepsilon}  ^{\mu}  ^{\mu_1}  ^{\kappa}}\nn\\
        C_7&=& \nabla_{\alpha}\tensor{\mathcal{F}}{_{\beta}  _{\gamma}  _{\delta}  _{\varepsilon}} \nabla^{\beta}\tensor{\mathcal{F}}{^{\alpha}  ^{\gamma}  _{\iota}  _{\mu}} \nabla^{\delta}\tensor{\mathcal{F}}{^{\varepsilon}  _{\mu_1}  _{\kappa}  _{\lambda}} \nabla^{\mu_1}\tensor{\mathcal{F}}{^{\iota}  ^{\mu}  ^{\kappa}  ^{\lambda}}\qquad C_{19}=\nabla_{\alpha}\tensor{\mathcal{F}}{_{\beta}  _{\gamma}  _{\delta}  _{\varepsilon}} \nabla^{\beta}\tensor{\mathcal{F}}{^{\gamma}  _{\iota}  _{\mu}  _{\mu_1}} \nabla^{\iota}\tensor{\mathcal{F}}{^{\delta}  ^{\mu}  _{\kappa}  _{\lambda}} \nabla^{\kappa}\tensor{\mathcal{F}}{^{\alpha}  ^{\varepsilon}  ^{\mu_1}  ^{\lambda}}\nn\\
        C_8&=&\nabla_{\alpha}\tensor{\mathcal{F}}{_{\beta}  _{\gamma}  _{\delta}  _{\varepsilon}} \nabla^{\beta}\tensor{\mathcal{F}}{^{\alpha}  ^{\gamma}  _{\iota}  _{\mu}} \nabla^{\kappa}\tensor{\mathcal{F}}{^{\iota}  ^{\mu}  ^{\mu_1}  ^{\lambda}} \nabla_{\mu_1}\tensor{\mathcal{F}}{^{\delta}  ^{\varepsilon}  _{\kappa}  _{\lambda}}\qquad C_{20}=\nabla_{\alpha}\tensor{\mathcal{F}}{_{\beta}  _{\gamma}  _{\delta}  _{\varepsilon}} \nabla^{\beta}\tensor{\mathcal{F}}{^{\gamma}  _{\iota}  _{\mu}  _{\mu_1}} \nabla^{\iota}\tensor{\mathcal{F}}{^{\alpha}  ^{\delta}  _{\kappa}  _{\lambda}} \nabla^{\kappa}\tensor{\mathcal{F}}{^{\varepsilon}  ^{\mu}  ^{\mu_1}  ^{\lambda}}\nn\\
        C_9&=&\nabla_{\alpha}\tensor{\mathcal{F}}{_{\beta}  _{\gamma}  _{\delta}  _{\varepsilon}} \nabla^{\beta}\tensor{\mathcal{F}}{_{\iota}  _{\mu}  _{\mu_1}  _{\kappa}} \nabla^{\iota}\tensor{\mathcal{F}}{^{\mu}  ^{\mu_1}  ^{\kappa}  _{\lambda}} \nabla^{\lambda}\tensor{\mathcal{F}}{^{\alpha}  ^{\gamma}  ^{\delta}  ^{\varepsilon}}\qquad C_{21}=\nabla_{\alpha}\tensor{\mathcal{F}}{_{\beta}  _{\gamma}  _{\delta}  _{\varepsilon}} \nabla^{\beta}\tensor{\mathcal{F}}{^{\gamma}  _{\iota}  _{\mu}  _{\mu_1}} \nabla^{\delta}\tensor{\mathcal{F}}{^{\iota}  ^{\mu}  _{\kappa}  _{\lambda}} \nabla^{\mu_1}\tensor{\mathcal{F}}{^{\alpha}  ^{\varepsilon}  ^{\kappa}  ^{\lambda}}\nn\\
        C_{10}&=&\nabla_{\alpha}\tensor{\mathcal{F}}{_{\beta}  _{\gamma}  _{\delta}  _{\varepsilon}} \nabla^{\alpha}\tensor{\mathcal{F}}{^{\beta}  ^{\gamma}  _{\iota}  _{\mu}} \nabla^{\kappa}\tensor{\mathcal{F}}{^{\varepsilon}  ^{\mu}  ^{\mu_1}  ^{\lambda}} \nabla_{\mu_1}\tensor{\mathcal{F}}{^{\delta}  ^{\iota}  _{\kappa}  _{\lambda}}\qquad C_{22}=\nabla_{\alpha}\tensor{\mathcal{F}}{_{\beta}  _{\gamma}  _{\delta}  _{\varepsilon}} \nabla^{\beta}\tensor{\mathcal{F}}{_{\iota}  _{\mu}  _{\mu_1}  _{\kappa}} \nabla^{\iota}\tensor{\mathcal{F}}{^{\gamma}  ^{\delta}  ^{\varepsilon}  _{\lambda}} \nabla^{\lambda}\tensor{\mathcal{F}}{^{\alpha}  ^{\mu}  ^{\mu_1}  ^{\kappa}}\nn\\
        C_{11}&=&\nabla_{\alpha}\tensor{\mathcal{F}}{_{\beta}  _{\gamma}  _{\delta}  _{\varepsilon}} \nabla^{\alpha}\tensor{\mathcal{F}}{^{\beta}  ^{\gamma}  _{\iota}  _{\mu}} \nabla^{\delta}\tensor{\mathcal{F}}{^{\iota}  _{\mu_1}  _{\kappa}  _{\lambda}} \nabla^{\mu}\tensor{\mathcal{F}}{^{\varepsilon}  ^{\mu_1}  ^{\kappa}  ^{\lambda}}\qquad C_{23}=\nabla_{\alpha}\tensor{\mathcal{F}}{_{\beta}  _{\gamma}  _{\delta}  _{\varepsilon}} \nabla^{\beta}\tensor{\mathcal{F}}{_{\iota}  _{\mu}  _{\mu_1}  _{\kappa}} \nabla^{\iota}\tensor{\mathcal{F}}{^{\gamma}  ^{\delta}  ^{\mu}  _{\lambda}} \nabla^{\lambda}\tensor{\mathcal{F}}{^{\alpha}  ^{\varepsilon}  ^{\mu_1}  ^{\kappa}}\nn\\
        C_{12}&=&\nabla_{\alpha}\tensor{\mathcal{F}}{_{\beta}  _{\gamma}  _{\delta}  _{\varepsilon}} \nabla^{\alpha}\tensor{\mathcal{F}}{^{\beta}  _{\iota}  _{\mu}  _{\mu_1}} \nabla^{\gamma}\tensor{\mathcal{F}}{^{\delta}  ^{\iota}  _{\kappa}  _{\lambda}} \nabla^{\kappa}\tensor{\mathcal{F}}{^{\varepsilon}  ^{\mu}  ^{\mu_1}  ^{\lambda}}\qquad C_{24}=\nabla_{\alpha}\tensor{\mathcal{F}}{_{\beta}  _{\gamma}  _{\delta}  _{\varepsilon}} \nabla^{\beta}\tensor{\mathcal{F}}{_{\iota}  _{\mu}  _{\mu_1}  _{\kappa}} \nabla^{\iota}\tensor{\mathcal{F}}{^{\alpha}  ^{\gamma}  ^{\delta}  _{\lambda}} \nabla^{\mu}\tensor{\mathcal{F}}{^{\varepsilon}  ^{\mu_1}  ^{\kappa}  ^{\lambda}} \nn
       \ea
with the following combinations that vanish for four-point amplitudes
\ba
        Y_1&=&-C_3 + 12 C_4 - 6 C_5 + 72 C_7 - 9 C_8 - C_9 + 54 C_{10} - 6 C_{11} - 144 C_{12} \nn\\
        &&+  18 C_{14} - 27 C_{18} + 18 C_{21}\label{y1}\\
        Y_2&=&C_3 - 6 C_5 - 18 C_7 + 9 C_8 + C_9 + 6 C_{11} + 9 C_{18} + 18 C_{23}\\
        Y_3&=&C_1 + 96 C_4 - 96 C_5 + 32 C_6 + 288 C_7 + 64 C_9 + 32 C_{22}\\
        Y_4&=&-C_{10} + 2 C_{12} + 2 C_{20}\\
        Y_5&=&C_7 + C_{10} + 4 C_{19}\\
        Y_6&=& -C_7 - C_{10} + 2 C_{17}\\
        Y_7&=&C_1 - 8 C_2 + 32 C_6 + 32 C_9 + 32 C_{16}\\
        Y_8&=&-C_2 - 12 C_4 + 12 C_5 - 4 C_9 - 12 C_{11} + 36 C_{15}\\
        Y_9&=&C_{10} - 2 C_{12} + C_{13}\label{y9}
    \ea

\end{appendix}

\end{document}